\documentclass[prb,twocolumn,showpacs,amsmath,amssymb]{revtex4-1}

\usepackage[dvipdfmx]{graphicx}
\usepackage{hyperref}

\usepackage{bm}

\usepackage{color}
\usepackage{braket}
\usepackage{physics}
\usepackage{ulem}

\begin{document}

\title{Entanglement dynamics in the many-body Hatano-Nelson model}

\author{Takahiro Orito}
\affiliation{
Graduate School of Advanced Science and Engineering, Hiroshima University, 739-8530, Japan
\\
Institute for Solid State Physics, The University of Tokyo,
5-1-5 Kashiwanoha, Kashiwa 277-8581, Japan}
\author{Ken-Ichiro Imura}
\affiliation{Institute of Industrial Science, The University of Tokyo,
5-1-5 Kashiwanoha, Kashiwa 277-8574, Japan}

\date{\today}

\begin{abstract}
The entanglement dynamics 
in a non-Hermitian quantum system is studied
numerically 
and
analyzed from the viewpoint of
quasiparticle picture.
As a concrete model,
we consider
a one-dimensional tight-binding model with asymmetric hopping (Hatano-Nelson model)
under onsite disorder and
nearest-neighbor interaction.
As opposed to an assertion of previous studies,
the entanglement dynamics
in this non-Hermitian quantum system
is very different from
the one
in its Hermitian counterpart,
especially in the delocalized regime with weak disorder;
there
the entanglement entropy $S_{\rm ent}(t)$
shows a characteristic non-monotonic time evolution.
We have clarified and quantified the nature of this behavior
in the quasiparticle picture.
In the asymptotic regime of $t\rightarrow\infty$,
the entanglement entropy $S_{\rm ent}(t)$
in this regime
saturates to a much suppressed value,
which increases only logarithmically with respect to the size of
the subsystem.
\end{abstract}

\maketitle

\section{Introduction}



\begin{figure}
\includegraphics[width=80mm]{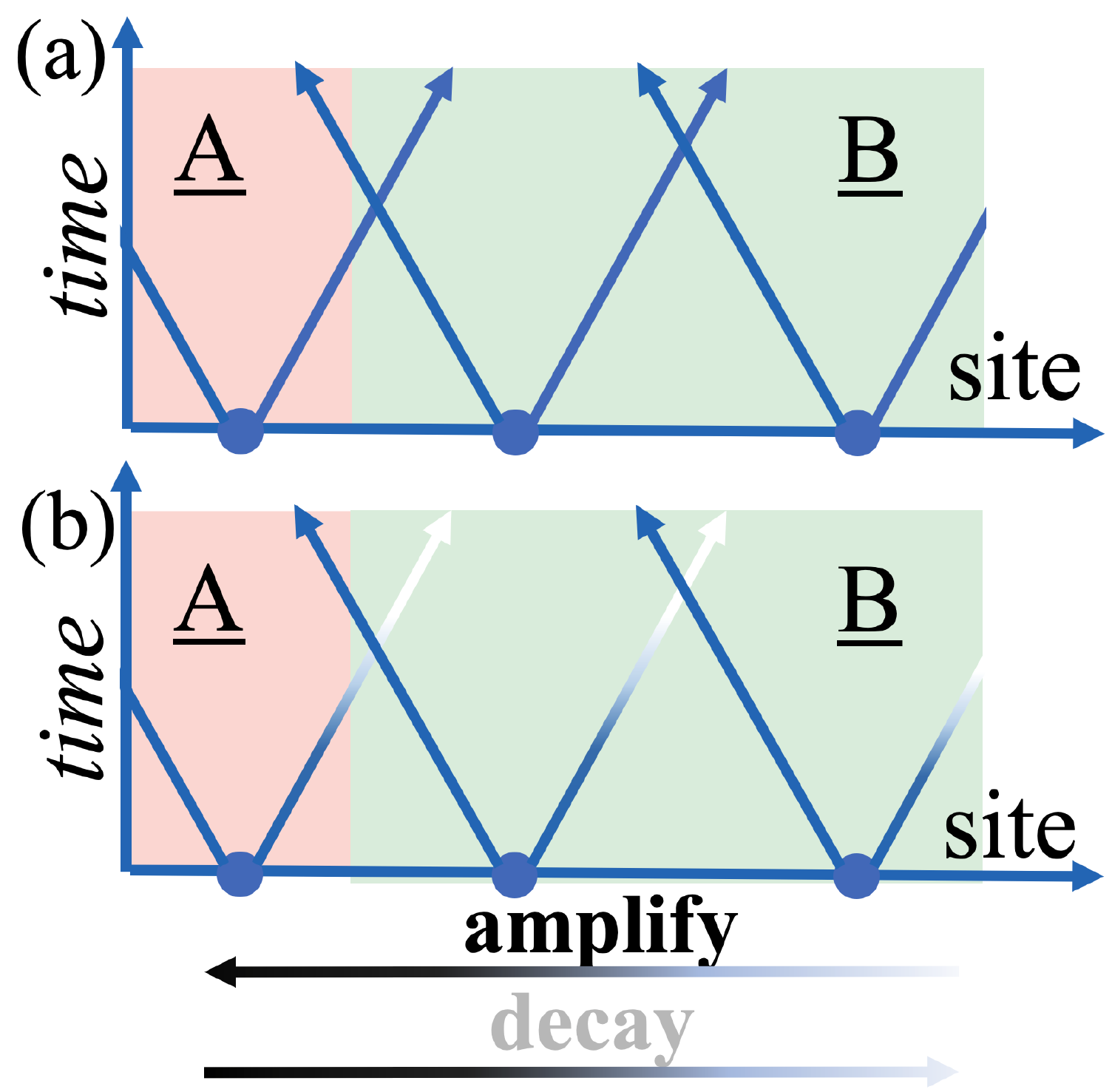}
\caption{
Schematic illustration of the quasiparticle picture:~(a) in the Hermitian case, 
and (b) in the case of the Hatano-Nelson type asymmetric hopping model
(non-Hermitian case).
}
\label{Fig1}
\end{figure}

The entanglement entropy $S_{\rm ent}$
quantifies
non-local correlation between quasiparticles
in a many-body quantum state,
such as the one in an EPR (Einstein-Podolsky-Rosen) pair.
\cite{EPR,nobel_c,nobel_z,aspect,aspect1?,aspect3}
In the process of
quantum thermalization\cite{ETH1,ETH2,ETH3} or relaxation\cite{GGE1,GGE2,GGE3,GGE4}, 
the so-called
quasiparticle picture \cite{Calabrese1,quasi-particle1,quasi-particle2,quasi-particle3,quasi-particle4}
(Fig.~\ref{Fig1})
makes this point explicit.
%
A pair of entangled quasiparticles
generated at $t=0$ move apart, and
as time passes by,
they are more likely found in a
different subsystem; see panel (a) of Fig.~\ref{Fig1}.
This leads to an increase of
the entanglement entropy $S_{\rm ent}$;
cf. its bipartite definition, Eq.~(\ref{EE_def}).
Correspondingly,
the reduced density matrix of the subsystem becomes a {\it mixed} state [cf. Eq.~(\ref{DM_redu})].\cite{MBL_review,MBL5}

In a system in which
this quasiparticle picture is well applicable,
the entanglement entropy $S_{\rm ent}$ 
is an 
extensive quantity,
obeying the volume law\cite{Pagesan}: $S_{\rm ent}\propto V=L^{d}$
($L$: size, $d$: dimension of the system);
and indeed serves a {\it thermodynamic} quantity,
while there are cases in which
$S_{\rm ent}$ obeys the area law scaling $S_{\rm ent}\propto V=L^{d-1}$.
The latter includes the cases of
non-unitary time evolution induced by dissipation,\cite{Marco1} 
projective measurements,\cite{MIP,Fuji-Ashida}
and also some
parameter regime of a PT symmetric system.\cite{PT-quasi-particle,Marco2} 
If one can
manipulate a parameter of the system
to drive the system from one case to the other,
the entanglement entropy $S_{\rm ent}$ is subject to
a transition from volume to area law scaling.\cite{MIP} 
This transition, dubbed as the entanglement transition, 
has been attracting much attention recently,
in theoretical,\cite{MIP1,MIP2,MIP3,MIP4,MIP5,MIP6,MIP7}
experimental,\cite{MIP_experiment,MIP_exp2}
and numerical contexts.\cite{ET1,PuriT1,PuriT2}

We consider, as a concrete example,
the case of a many-body
Hatano-Nelson (HN) model;\cite{HN_PRB97,HN_PRB98,HN_PRL}
a one-dimensional tight-binding model
with asymmetric (non-reciprocal) hopping,
which is specified by a parameter $g$.
In dynamics,
due to the asymmetry in hopping, 
an initial wave packet does not spread as in the Hermitian case,
but rather slides in the direction
specified by the asymmetry of hopping (sign of $g$).\cite{Longhi_AA}
Such a unidirectional motion is
robust against disorder 
and suppresses wave packet spreading.
\footnote{See Supplemental Material at [URL will be inserted by the publisher] for a numerical demonstration and detailed explanation of wave-packet dynamics.
The Supplemental Material also contains Refs.~\onlinecite{SM-revise1,SM-revise2,thouless}.
}
This peculiar wave-packet dynamics
leads to a remarkable
non-monotonic 
time evolution of the entanglement entropy $S_{\rm ent}(t)$.\cite{OI22A}
In the body of the paper,
we provide an intuitive explanation on the increase of $S_{\rm ent}$ 
in the delocalized regime
from the viewpoint of
the quasiparticle picture.

This work is also an outcome of a technical advancement 
we have made in our numerics.
Here, we have successfully employed the Krylov subspace method\cite{Arnoldi}
in our problem, 
which has allowed us to deal with a system of larger size than the previous studies,
e.g., 
the one of our own.\cite{OI22A}
This has been particularly helpful in the study of
the scaling property of $S_{\rm ent}$.

The paper is organized as follows. 
In Sec.~\ref{Sec2}, we introduce the HN model, the numerical conditions, and the definition of entanglement entropy.
In Sec.~\ref{Sec3}, we systematically investigate the effect of disorder on the density dynamics in the real and momentum spaces and entanglement dynamics.
In Sec.~\ref{Sec4}, we point out the difference in the entanglement dynamics between the non-Hermitian and Hermitian systems in the clean limit from the perspective of the quasiparticle picture.
In Sec.~\ref{Sec_scal}, we investigate whether non-Hermiticity induces entanglement transition.
In Sec.~\ref{Sec5}, we examine various aspects of $S_{\rm ent}$, such as its scaling behavior, the effect of interaction, and its relation to the correlation function.
Section \ref{Sec6} is devoted to concluding remarks. 
Some details are left in the Appendixes and Supplemental Materials.

\section{Many-body Hatano-Nelson Model}
\label{Sec2}


Let us first introduce our model,
which is a many-body extension of the so-called Hatano-Nelson (HN) model,
\cite{HN_PRB97,HN_PRB98,HN_PRL}
and reads as 
in the second quantization representation
\begin{eqnarray}
{\cal H}&=& - \sum_{j=0}^{L-1}
\Big(\Gamma_L c_j^\dagger c_{j+1}+\Gamma_R c_{j+1}^\dagger c_{j}\Big)
\nonumber \\
&+&\sum_{j=0}^{L-1}
\Big(V \hat{n}_j \hat{n}_{j+1}+W_j \hat{n}_j\Big),
\label{ham_mp}
\end{eqnarray}
where $c_j^\dagger$ ($c_j$) is a fermionic creation (annihilation) operator of a particle 
at site $j$, 
while $\hat{n}_j=c_j^{\dagger}c_j$ is a number operator which counts
the number $n_j$ of such particles
found at site $j$. 
Here, we choose the boundary conditions to be periodic,
i.e., $c_L=c_0$ and $c_L^\dagger=c_0^\dagger$.
The first two-terms represent the asymmetric hopping, where the degree of non-reciprocity (asymmetry) is specified by the parameter $g$
\begin{equation}
\Gamma_L=e^g\Gamma_0,\ \
\Gamma_R=e^{-g}\Gamma_0.
\end{equation}
In the third term, $V$ represents 
the strength of the nearest neighbor inter-particle interaction,
while
in the last term, $W_j$ represents 
the depth
of an on-site disorder potential
at a site $j$.
Here, unlike in the original
Hatano-Nelson model,
\cite{HN_PRB97,HN_PRB98,HN_PRL}
in which
the random numbers
$W_j$'s obey to a uniform distribution,
we consider the case in which
$W_j$ represents a quasi-periodic potential
(cf. the Aubry-Andr\'e model \cite{AA}):
\begin{equation}
\label{AA}
W_j= W \cos (2\pi \alpha j+\theta),
\end{equation}
where
$\alpha$ should be chosen to be an irrational number,
e.g., $\alpha=(\sqrt{5}-1)/2$.
For $\alpha$ thus chosen,
the quasi-periodic potential $W_j$ mimics a random/disorder potential
as the one in the original Hatano-Nelson model,
$W$ represents the strength of the disorder potential.
%
If an average over different
disorder configurations is necessary,
one can activate the parameter $\theta$ in Eq.~(\ref{AA}),
and take the average over $\theta$.

Although the original Hatano-Nelson model [case of $V=0$ in Eq.~(\ref{ham_mp})]
has first appeared
\cite{HN_PRB97,HN_PRB98,HN_PRL}
as an effective model 
describing the phenomenon of vortex (de)pinning,
it is now considered to be a prototypical non-Hermitian situation,
and readapted in a number of different works.
The aspect of asymmetric hopping: $g\neq 0$
leads (under the open boundary)
to the so-called non-Hermitian skin effect, 
and is much discussed in the context of the idea of
non-Hermitian topological insulator.
\cite{PhysRevX_Gong,YW,YM,imura1,imura2,Ashida_2020,HN_topo1,HN_topo3,HN-skin2,HN_topo2,YH2,HN-skin1} 
The competition between
the effect of asymmetric hopping: $g\neq 0$ and 
that of the disorder potential $W\neq 0$
leads to a typical delocalization-localization transition 
in this non-Hermitian system,
and the model is also much discussed in this context.
\cite{HN-localization,HN-localization2,NonHMBL1,NonHMBL2,nonHMBL-AA,GL-non-herm,non-Hermitian-MBL-arXiv}
If the localization length $\xi$ is known in the Hermitian limit,
the localization transition is expected to occur at
$g=\xi^{-1}$
in the corresponding non-Hermitian model.
\cite{HN_PRB97,HN_PRB98,HN_PRL}
In a non-interacting system, either Hermitian or non-Hermitian, 
$\xi$ can be calculated by the transfer matrix method.
\cite{transfer_matrix1-Herm,transfer_matrix2-Herm,transfer_matrix-nonH1,transfer_matrix-nonH2} 
In an interacting system, this is simply not possible, while
the study of the interacting Hatano-Nelson model 
brings about some information on the many-body localization length,
\cite{Gil,OI22A}
since
the asymmetric hopping $g$
can be interpreted (under the periodic boundary)
as an imaginary flux.
\footnote{Inserting a real flux $\Phi$ (Hermitian system) changes a localized eigenstate $\psi(j,\Phi=0)\sim \exp(-\frac{|j|}{\xi})$ to $\psi(j, \Phi\neq0)\sim\exp(-\frac{|j|}{\xi}+i\Phi j)$. In contrast, inserting an imaginary flux $ig$ (HN-model) modifies a localized eigenstate $\psi(j,g=0)\sim \exp(-\frac{|j|}{\xi})$ to
\[
\begin{aligned}[t]
    \psi(j,g\neq0)\sim&\exp\left(-\frac{(1-g\xi)|j|}{\xi}\right) & \text{if } & j<0 \\
    \psi(j,g\neq0)\sim&\exp\left(-\frac{(1+g\xi)|j|}{\xi}\right) & \text{if } & j\geq0.
\end{aligned}
\]
One can easily observe that the delocalization transition is induced by $ig$ and occurs at $g=\xi^{-1}$, from which we can determine the localization length.}


\subsection{Non-Hermitian many-particle dynamics}


%
In the simulation of many-particle dynamics,
we will typically consider the 
initial state:
\begin{equation}
|\Psi (0)\rangle=|\Psi (t=0)\rangle = |101010\cdots\rangle,
\label{DW}
\end{equation}
i.e.,
the one in the
density wave
form,
or in the
N\'eel form in the spin language.
\cite{panda}
On the right hand side of
Eq.~(\ref{DW}),
we have employed the computational basis $|n_1 n_2 \cdots n_L\rangle$;
$n_j=0,1$ represents occupation of the $j$th site.
At time $t=0$, 
the initial state (\ref{DW}) 
can be expressed as a superposition of eigenstates as
\begin{equation}
|\Psi (0)\rangle=\sum_{\alpha} c_\alpha (0)|\alpha\rangle,
\label{psi_0}
\end{equation}
where
$|\alpha\rangle$ represents a many-body eigenstate of the Hamiltonian (\ref{ham_mp}),
i.e., ${\cal H}|\alpha\rangle=E_\alpha |\alpha\rangle$.
Note that the eigenenergy $E_\alpha$ is generally complex.
In Eq.~(\ref{psi_0}),
$|\alpha\rangle$ represents a 
right eigenstate
corresponding to the eigenenergy $E_\alpha$,
which is generally not identical 
to the Hermitian conjugate of
the corresponding 
left eigenstate $\langle\langle\alpha|$,
where
\begin{equation}
\langle\langle\alpha |{\cal H}=E_\alpha\langle\langle\alpha|,
\end{equation}
or its conjugate
\begin{equation}
{\cal H}^\dagger|\alpha\rangle\rangle=E_\alpha^*|\alpha\rangle\rangle,
\label{left-eigen}
\end{equation}
or rather,
$\langle\langle\alpha |\neq |\alpha\rangle^\dagger$.
To find the coefficients $c_\alpha (0)$ in Eq.~(\ref{psi_0}),
one actually needs to find
such left eigenstates, i.e.,
\begin{equation}
c_\alpha(0) =\langle\langle\alpha|\Psi(0)\rangle.
\end{equation}
Note that
the left and right eigenstates satisfy the biorthogonal condition:\cite{Brody_2013}
\begin{equation}
\langle\langle\alpha |\beta\rangle=\delta_{\alpha,\beta}.
\label{biortho}
\end{equation}

We then let the state Eq.~(\ref{psi_0}) evolve
into $\Psi(t)$, 
in principle,
via the Schr\"odinger equation,
\begin{equation}
i\hbar{\partial\over\partial t}|\Psi(t)\rangle=H|\Psi(t)\rangle,
\label{sh-eq}
\end{equation}
though,
in practice, we let it evolve through a numerical recipe
outlined in the next subsection.
In case of a {\it unitary} time evolution 
driven by a Hermitian Hamiltonian,
the weight of each eigenstate $|\alpha\rangle$ is unchanged 
in the time-evolved wave packet $|\Psi(t)\rangle$;
if one expresses 
$|\Psi(t)\rangle$
as a superposition of eigenstates as in Eq.~(\ref{psi_0}),
or 
\begin{equation}
|\Psi(t)\rangle=\sum_{\alpha} c_\alpha (t)|\alpha\rangle,
\label{psi_t}
\end{equation}
the magnitude of the coefficients
\begin{equation}
c_\alpha (t) = c_\alpha (0) e^{-iE_\alpha t}
\label{c_alpha}
\end{equation}
is 
conserved
in the case of unitary time evolution 
driven by a {\it Hermitian} Hamiltonian;
i.e.,
in the course of time;
$|c_\alpha (t)|^2$ are just constants, or $|c_\alpha (t)|^2=|c_\alpha (0)|^2$.
Here,
in the case of {\it non-unitary} time evolution driven by a {\it non-Hermitian} Hamiltonian,
this is no longer the case;
the coefficients $c_\alpha (t)$ that appear
in
Eqs. (\ref{psi_t}) and (\ref{c_alpha}) 
change constantly their 
amplitudes
in the time evolution.
In such non-unitary time evolution,
the total probability $\langle\Psi(t)|\Psi(t)\rangle$
is {\it a priori} not conserved
\footnote{
In the Hermitian case,
the total probability:
$\sum_\alpha |c_\alpha (t)|^2 (=1)$
is, of course, conserved.
Here, in the non-Hermitian case,
the quantity, $\sum_\alpha |c_\alpha (t)|^2$
itself does not have much meaning, since 
$\langle\Psi(t)|\Psi(t)\rangle\neq \sum_\alpha |c_\alpha (t)|^2$.
If one expands $\langle\Psi(t)|$ into contributions from different {\it left} eigenmodes as,
$\langle\Psi(t)|=\sum_\alpha b_\alpha (t)\langle\langle\alpha|$, and uses the biorthogonal relation
(\ref{biortho}), then one finds, 
$\langle\Psi(t)|\Psi(t)\rangle =\sum_\alpha b_\alpha (t) c_\alpha (t)$.
}
due to post-selection (see Appendix~\ref{GKSL-section}). 
In the actual numerical calculation, 
we renormalize $|\Psi(t)\rangle$ as\cite{Longhi_AA,OI22A} 
\begin{equation}
|\Psi (t)\rangle
\to
|\tilde{\Psi}(t)\rangle=
{|\Psi (t)\rangle\over\sqrt{\langle\Psi(t)|\Psi(t)\rangle}}.
\label{renorm}
\end{equation}
Under this renormalization 
[justified physically, in Appendix~\ref{GKSL-section}, in the context of Lindblad/GKSL (Gorini - Kossakowski - Sudarshan - Lindblad) dynamics\cite{GKSL1,GKSL2}],
the total probability is conserved, 
but the relative importance of $c_\alpha (t)$ in Eq.~(\ref{psi_t}) 
with respect to other $c_\alpha (t)$'s can vary.
In the GKSL/quantum trajectory approach, 
the renormalization factor (the denominator) in Eq. (\ref{renorm})
appears naturally 
as a result of the projection to null outcome; 
here, continuous measurement and post-selection is assumed 
(see Appendix~\ref{GKSL-section} for details).
%

Here, in the case of {\it non-Hermitian}
{\it non-unitary} dynamics,
a remarkable fact is that
as time passes by,
contributions from
those $|\alpha\rangle$'s
whose eigenenergy has a large positive imaginary part
become dominant
in the superposition of many eigenstates $|\alpha\rangle$
in Eq.~(\ref{psi_t});
for
\begin{equation}
{\rm Im}(E_{\alpha_1}) > {\rm Im}(E_{\alpha_2})> \cdots,
\end{equation}
\begin{equation}
|c_{\alpha_1} (t)|^2 \gg |c_{\alpha_2} (t)|^2 \gg \cdots,
\end{equation}
i.e.,
only the first few
$|\alpha_1\rangle,|\alpha_2\rangle,\cdots$
become relevant
in the superposition (\ref{psi_t})
if 
$|\alpha_1\rangle,|\alpha_2\rangle,\cdots$
are labeled 
in the decreasing order of ${\rm Im}(E_{\alpha})$,
and if the maximal
${\rm Im}(E_{\alpha_1})$ is sufficiently larger than the rest.
If ${\rm Im}(E_{\alpha_1}) \gg {\rm Im}(E_{\alpha_2})$,
in the end of the time evolution ($t\to\infty$),
the wave packet
$|\Psi(t)\rangle$ 
will be completely dominated by a single eigenstate
$|\alpha_1\rangle$;
i.e., apart from an unimportant phase factor,
\begin{equation}
\lim_{t\to\infty}|\tilde{\Psi}(t)\rangle
\sim
|\alpha_1\rangle.
\label{alpha_1}
\end{equation}
Thus, in the
non-Hermitian quantum dynamics,
the non-unitarity of the time evolution 
associated with the imaginary part of the eigenenergy
gives rise to
{\it collapse of the superposition}
of an initial wave packet (\ref{psi_0}).
After a long enough non-unitary time evolution,
a generic initial state
composed of many different eigenstates
tends to converge to a single (or to a few) eigenstate(s).
\footnote{Later we will encounter the case in which
some largest
Im($E_{\alpha})$'s are quasi-degenerate:
${\rm Im}(E_{\alpha_1})\simeq {\rm Im}(E_{\alpha_2}) \simeq \cdots$,
and contribute equally to 
$|\Psi (t\to\infty)\rangle$.
Such degeneracy in the imaginary part
becomes indeed relevant
in the long-time dynamics of the non-interacting case;
see Secs.~\ref{Sec3} and~\ref{Sec5} for details.}

\begin{figure*}
\includegraphics[width=150mm]{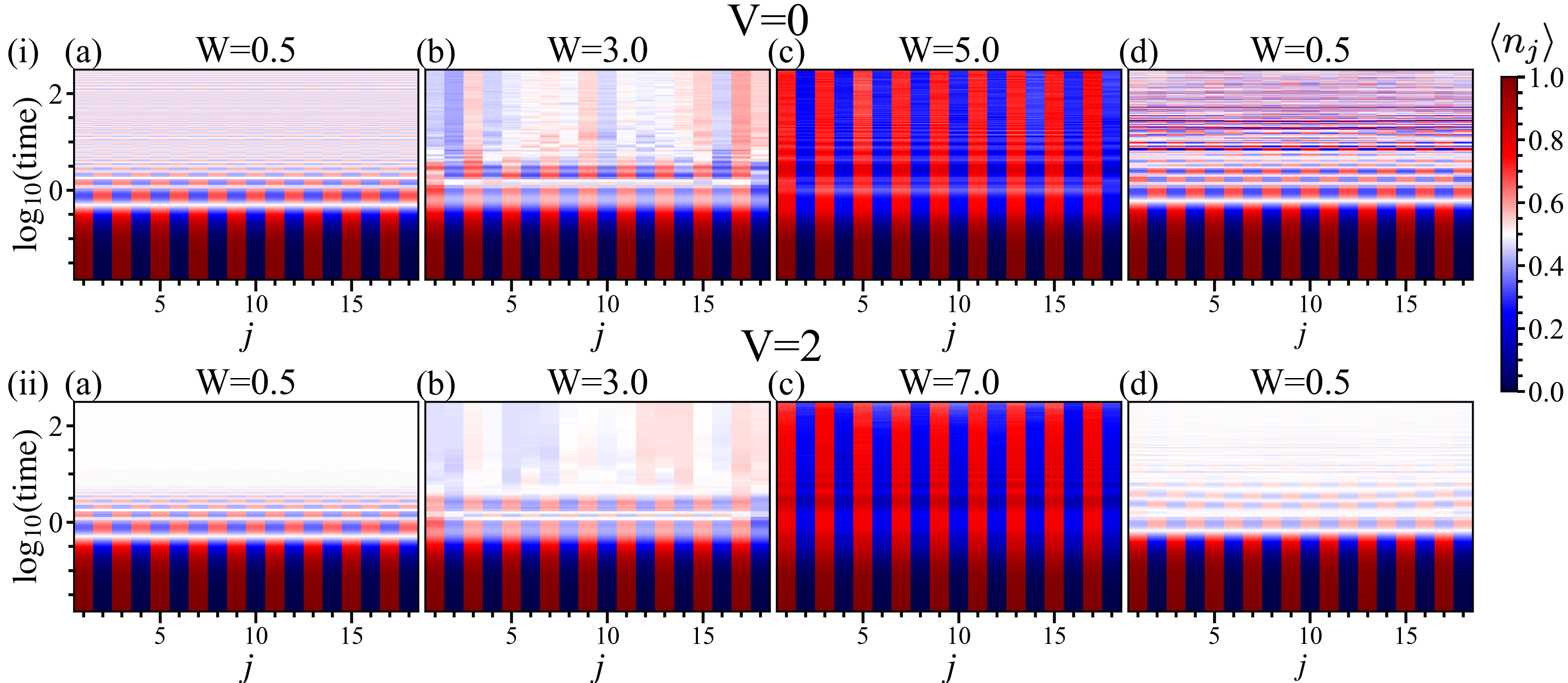}
\caption{
Time evolution of the spatial profile of the density $n_j(t)$. The first row [(i-a)-(i-c)]: non-interacting case ($V=0$), asymmetric hopping ($g=0.5$) and, (i-d): $V=0$, $g=0.0$. The second row [(ii-a)-(ii-c)]: interacting case ($V=2$), $g=0.5$ and, (ii-d): $V=2$, $g=0.0$.
In numerical calculation, for $g\neq0$
we carried out the evaluations with
$80$ (for $V=2$) and $40$ (for $V=0$)
samples.
For the Hermitian case, we used $80$ samples.
}
\label{Fig2}
\end{figure*}

\subsection{Numerical simulation}

Simulating a many-body quantum system is challenging, since
the size of the Hilbert space increases 
exponentially with the increase of size $L$ of the system.
In a simulation of a Hermitian system
using the exact diagonalization method,
$L=18$ may be a typical maximal size
one can handle comfortably
in a present day computer performance.
In a non-Hermitian system, however,
it is necessary to consider not only the eigenenergy and right eigenvector,
but also the left eigenvector.
Consequently, most studies are limited to treating system sizes up to $L=16$.\cite{Hamaz,OI22A,NonHMBL1,NonHMBL2,panda,MBskin-dynamics}
Confronted with this numerical challenge, 
we have decided to employ the Krylov subspace method.
In order to make it compatible with a non-Hermitian matrix, 
we have generated
the orthonormal Krylov subspace $V_M$ 
using the Arnoldi method,
instead of the Lanczos method.\cite{Arnoldi}
The Krylov subspace is given by $K_M=span(|\Psi(t)\rangle, H|\Psi(t)\rangle, \cdots, H^{M-1}|\Psi(t)\rangle)$.
The time evolution of quantum state is described by
\begin{equation}
|\Psi(t+\delta t)\rangle\sim V_M e^{-i\delta tH'}V_M^\dagger|\Psi(t)\rangle=V_M e^{-i\delta tH'}|e_1\rangle,
\label{psit2}
\end{equation}
where 
$|e_1\rangle\equiv(1,0,\cdots,0)^T$ 
and 
$H'=V_M^\dagger H V_M$.
This 
allows us 
to calculate $|\Psi(t+\delta t)\rangle$ 
by dealing with 
matrix $H'$ 
of size $M\times M$
instead of diagonalizing the original Hamiltonian $H$, 
and
eventually
enables us to study a system of larger size than 
those in the previous studies\cite{OI22A}.
In the actual numerical calculations, 
we choose $\delta t=10^{-2}-2\times10^{-1}$ and $M=10-25$.


\subsection{Entanglement entropy: definitions}
In the study of many-body dynamics,
we are not only interested in
how the density spreads
but also how correlation spreads in the system.
To quantify the latter,
we consider the entanglement dynamics.
The entanglement entropy is a 
quantity to characterize the non-locality of a quantum state,
which is often defined
in the sense of {\it bipartite} entanglement entropy:
\begin{equation}
\label{EE_def}
S_{\rm ent} = - {\rm Tr_A}[\Omega_A\log \Omega_A],
\end{equation}
where
\begin{equation}
\label{DM_redu}
\Omega_A={\rm Tr_B}[\Omega]
\end{equation}
is the reduced density matrix of the subsystem A;
we have divided the entire system (of size $L$)
into two subsystems A and B.
%
In practice,
such a division can be done using a many-body basis $\tilde{n}$
represented by a set of 
quantum numbers $\tilde{n}=\{n_1,n_2,\cdots,n_L\}$,
which can be divided into the two parts as
$\tilde{n}=\{\tilde{n}_A,\tilde{n}_B\}$, where
$\tilde{n}_A=\{n_1,n_2,\cdots,n_{\ell}\}$ spans the subsystem A,
while
the remaining part:
$\tilde{n}_B=\{n_{\ell+1},n_{\ell+2},\cdots,n_{L}\}$
spans the subsystem B.
$\ell$ is the size of the subsystem A.
In this basis,
a many-body state $|\Psi\rangle$ may be represented as
\begin{eqnarray}
\label{Psi}
|\Psi\rangle&=&\sum_{\tilde{n}} \psi_{\tilde{n}}|\tilde{n} \rangle=
\sum_{\tilde{n}_A\in A,\tilde{n}_B\in B} \psi_{\tilde{n}_A,\tilde{n}_B}
|\tilde{n}_A\rangle|\tilde{n}_B\rangle.\nonumber\\
\end{eqnarray}
Using this, 
one can explicitly trace out the subsystem B
from the density matrix:
\begin{eqnarray}
\Omega=
|\Psi\rangle\langle\Psi|
=
\sum_{\tilde{n},\tilde{n}'}
\psi_{\tilde{n}}\psi_{\tilde{n}'}^*|\tilde{n}\rangle\langle\tilde{n}'|,
\label{DM}
\end{eqnarray}
i.e., 
the reduced density matrix (\ref{DM_redu}) becomes
\begin{eqnarray}
\Omega_A
&=&
\sum_{\tilde{n}_B''\in B}
\langle\tilde{n}_B''|\Omega |\tilde{n}_B''\rangle
\nonumber \\
&=&
\sum_{\tilde{n}_A,\tilde{n}_A'\in A,\tilde{n}_B\in B}
\psi_{\tilde{n}_A,\tilde{n}_B}\psi_{\tilde{n}_A',\tilde{n}_B}^*|\tilde{n}_A\rangle\langle\tilde{n}_A'|.
\label{DM_redu2}
\end{eqnarray}
Here, we consider the time evolution of 
a many-body density matrix:
$\Omega(t)=|\tilde{\Psi} (t)\rangle\langle\tilde{\Psi}(t)|$,
and the corresponding 
entanglement entropy $S_{\rm ent}(t)$.
Additionally, three definitions of $S_{\rm ent}$ (and $\Omega$) arise in the context of a non-Hermitian system because the system has two eigenvectors, left and right eigenvectors.
In Appendix \ref{choise-of-entanglement}, we provide further explanation regarding these definitions and how they differ from one another.


\begin{figure*}
\includegraphics[width=150mm]{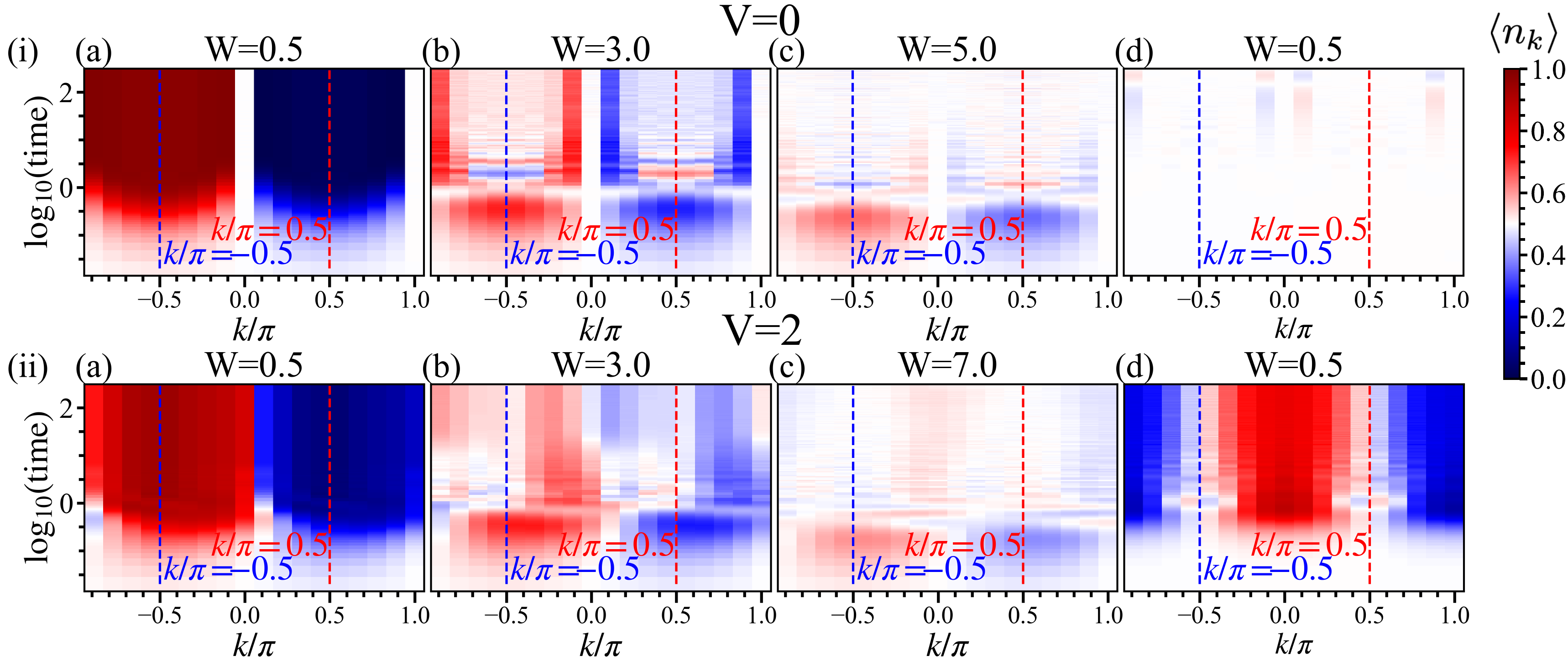}
\caption{Evolution of the density profile in the crystal-momentum ($k$-) space; i.e., $n_k(t)$.
Similarly to Fig. \ref{Fig2},
the first row [(i-a)-(i-c)]: non-interacting case ($V=0$),
asymmetric hopping ($g=0.5$), and (i-d): $V=0$, symmetric hopping ($g=0.0$). 
The second row [(ii-a)-(ii-c)]: interacting case ($V=2.0$), $g=0.5$ and, (ii-d): $V=2.0$, $g=0.0$.
We used the same wave functions as in Fig.~\ref{Fig2}.
}
\label{Fig3}
\end{figure*}


\section{Density and entanglement dynamics}
\label{Sec3}

In this section, 
we sketch the results of our numerical simulation on the (many-particle) density and entanglement dynamics.
We focus on the many-body HN model with the periodic boundary condition in this study,
but it is an interesting direction to study that with the open boundary condition,
where the skin effect that is an intrinsic nature of the non-reciprocal hopping system appears.
Based on previous studies,
we comment on how a skin effect affects
entanglement dynamics as well as density dynamics in Appendix~\ref{boudary-condition}.

\subsection{Density dynamics in real vs. reciprocal spaces}

Let us first focus on the time evolution of the density profile 
in real space:
\begin{equation}
n_j (t)=
\langle\Psi (t)|
c_j^\dagger c_j
|\Psi (t)\rangle,
\end{equation}
where
$|\Psi (t)\rangle$ actually means
$|\tilde{\Psi} (t)\rangle$ in Eq.~(\ref{renorm}), but
to simplify the notation,
here, we have omitted the tilde in $|\tilde{\Psi}(t)\rangle$,
and we will omit it hereafter.
Figure~\ref{Fig2}
shows 
the time evolution of 
$n_j(t)$
for the initial density wave (DW) pattern (\ref{DW})
in the non-interacting ($V=0$)
[first row, panels (i-a)-(i-d)],
and in the interacting ($V=2.0$)
[second row, panels (ii-a)-(ii-d)]
cases.
In both cases
the last panel [(i-d) and (ii-d)]
represents 
the Hermitian case $g=0$ for comparison.
Otherwise,
$g$ is chosen as $g=0.5$ (non-Hermitian).
Different panels correspond to the varying strength of disorder:
$W=0.5$ for panels (i-a) and (ii-a),
$W=3.0$ for panels (i-b) and (ii-b),
$W=5.0$ for panels (i-c), and $W=7.0$ for (ii-c).

In the first column (delocalized phase, $W=0.5$, panel (i-a) and (ii-a)),
the initial density wave pattern tends to be lost in the time evolution,
while in the Hermitian case ($V=0$) [first row, panel (i-d)],
the initial spatial profile does not fade but is replaced with a fast temporally oscillatory pattern
\footnote{
In the thermodynamic and clean limit, the evolution of $\langle n_j\rangle$ exhibits algebraic decay with oscillation and ultimately reaches the homogenous state. This tendency is consistent with the feature of delocalization.}
which is 
a feature reminiscent of an integrable system in which
a perpetual motion 
on a regular ideal orbital 
is ensured by the existence of some integrals of motion
(conserved quantities).
%
In the interacting case ($V=2$) [second row, panel (ii-d)],
scatterings induced by the inter-particle interaction
mix such regular ideal orbitals
and wash out 
the perpetual motion.
After some relaxation time $t_1\sim 10^0=1$
the spatial profile becomes {\it literally} uniform.
The second column [panels (i-b) and (ii-b)]
corresponds to the critical (crossover) regime
so that
the initial density wave pattern remains at least for a relatively 
long time.
As far as these real space features are concerned,
the time evolution of the density profile 
$n_j(t)$ is {\it not so}
different from
the Hermitian case [fourth column, panels (i-d) and (ii-d)].
The third column [panel (i-c) and panel (ii-c)] corresponds to the localized phase,
where the initial density wave pattern remains over time,
effectively similar to the localized phase in the Hermitian case.

Figure~\ref{Fig3} shows
time evolution of the density distribution:
\begin{equation}
n_k (t)=
\langle\Psi (t)|
c_k^\dagger c_k
|\Psi (t)\rangle,
\end{equation}
in the reciprocal 
crystal-momentum space ($k$-space),
where
\begin{equation}
c_k=\sum_j c_j e^{ikj}.
\end{equation}
As in Fig.~\ref{Fig2},
it shows the evolution of $n_k(t)$
both in the non-interacting ($V=0$)
[first row, panels (i-a)-(i-d)]
and in the interacting ($V=2.0$)
[second row, panels (ii-a)-(ii-d)]
cases.
In both cases
the last panel (i-d) and (ii-d)
represent 
the Hermitian case $g=0$ for comparison;
otherwise, $g=0.5$. 
Different panels correspond to the varying strength of disorder:
$W=0.5$ for panels (i-a) and (ii-a),
$W=3.0$ for panels (i-b) and (ii-b),
$W=5.0$ for panels (i-c), and $W=7.0$ for (ii-c).

First,
unlike in the real space (Fig.~\ref{Fig2})
the time evolution of the density profile
shows
very different features
in the Hermitian [column (d)]
and non-Hermitian [especially, first two columns: (a) and (b)]
cases.
In these 
columns,
one can see that
as time evolves,
the density distribution $n_k(t)$ 
in the reciprocal space
tends to 
converge to a certain asymptotic distribution,
implying that
in the regime of sufficiently long time $t\gg 1$,
the many-body wave packet
$|\Psi (t)\rangle$ 
tends to approach to a single eigenstate $|\alpha_1\rangle$
as in Eq.~(\ref{alpha_1});
in the non-interacting case ($V=0$, e.g., in the first row of Fig.~\ref{Fig3})
and also at $W=0$
$|\alpha_1\rangle$ will be given as
\begin{equation}
|\alpha_1\rangle=\left(\prod_{k<0} c_k^\dagger\right) |0\rangle,
\label{alpha_1_k}
\end{equation}
implying
a sharp Fermi-sea like asymptotic density distribution $n_k^{(\infty)}=n_k(t\rightarrow\infty)$ such that
\begin{equation}
n_k^{(\infty)} = \left\{\begin{array}{ll}
1     & {\rm for}\  k<0 \\
0     & {\rm for}\  k>0
\end{array}
\right..
\label{nk_asymp}
\end{equation}
%
Such a density distribution $n_k(t)$ 
localized in the crystal-momentum space
prevails 
in the regime of weak disorder
also
in the case of weak inter-particle interaction;
e.g., case of Fig.~\ref{Fig3} (ii) (a).
As $W$ is increased, e.g., in panel (b)
the distribution is smeared out, and
a sharp signature as in Eq. (\ref{nk_asymp})
becomes no longer visible.
Note that
in the crystal momentum space, both on-site potential $W_j$ and the inter-particle interaction $V$
are sources of scattering.
$V$ corresponds to two-particle scattering process;
two particles with wave number $k$ and $k'$
exchange their momenta.
These are all very different from the Hermitian case [column (d)]
where
$n_k(t)$
remains uniform during the time evolution
in the non-interacting case [panel (i-d)],
while
in the second row [panel (ii-d)],
$n_k(t)$ evolves into 
an equilibrium distribution, which is 
reminiscent of the one realized in the thermodynamic limit
[cf. eigenstate thermalization hypothesis (ETH)].
Inter-particle scatterings induced by a finite $V$
introduces (an effective form of) dissipation in the system
(i.e., in 
the eigenstate), 
bringing it to an effective thermal equilibrium.


\subsection{Entanglement dynamics}

\begin{figure}
\includegraphics[width=85mm]{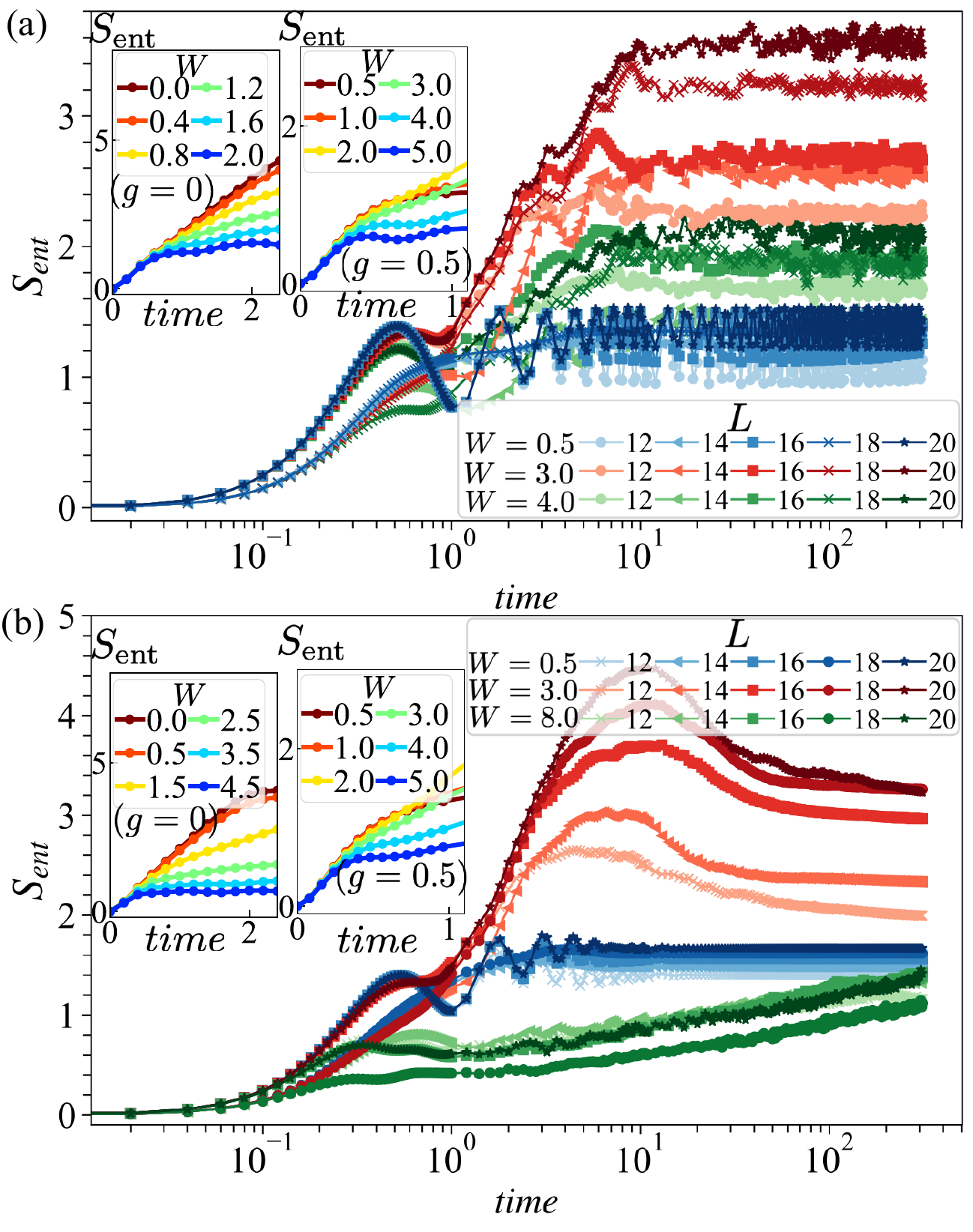}
\caption{
Entanglement dynamics
in three different regimes of disorder strength:
delocalized, critical and localized regimes.
Size dependence of the entanglement entropy is also shown.
(a) non-interacting case: $V=0$,
(b) interacting case: $V=2$.
In numerical calculation, we carried out evaluations for different system sizes using varying sample sizes for $V=2$.
We employed $100, 100, 100, 80$, and $40$ samples for $L=12, 14, 16, 18$, and $20$, respectively.
Similarity, for $V=0$, we used $100, 40, 40, 40$, and $20$ samples for $L=12, 14, 16, 18$, and $20$.
We choose the length of subsystem $\ell$ to be $\ell=L/2$.
In inset panels of panel (a), we conducted evaluations for $L=18$ with $80$ samples (for Hermitian case) and $40$ samples (for non-Hermitian case).
In inset panels of panel (b), we conducted evaluations for $L=18$ with $80$ samples (for both Hermitian and non-Hermitian cases).
}
\label{Fig4}
\end{figure}

Figure~\ref{Fig4} shows examples of entanglement dynamics
at various strength of disorder
and in systems of different size.
The 
asymmetry in hopping is fixed at $g=0.5$.
%
Panel (a) represents the non-interacting case ($V=0$),
while in panel (b)
a moderate strength of inter-particle interaction ($V=2.0$)
is assumed.
In the insets 
of the two panels 
different curves
represent
time evolution of the entanglement entropy $S_{\rm ent}$
at different strengths of disorder $W$
but for a system of size fixed at $L=18$.


In the non-interacting case [panel (a)],
the critical strength of disorder $W_c$ for the localization transition is 
$W_c=2e^g\simeq 3.297...$, so that
\\
(i)
$W=0.5, 1.0, 2.0$ correspond to 
the regime of weak disorder and delocalized wave function,
\\
(ii) $W=3$ roughly corresponds to the critical disorder strength $W_c$,
therefore, may be classified into the critical regime,
while
\\
(iii) $W=4.0, 5.0$ fall on 
regime of strong disorder and localized wave functions.

In the interacting case,
the corresponding values of $W$ 
in each regime
depends on the strength of the interaction $V$,
since in principle, $W_c$ depends on $V$.
In case of panel (b); i.e., at $V=2.0$, the classification may be such that
regime (i): $W=0.5, 1.0, 2.0$,
regime (ii): $W=3.0, 4.0, 5.0, 6.0$,
regime (iii): $W=7.0, 8.0$.
In the main panel,
the size dependence of the entanglement entropy is shown
in each of the three different regimes.

In the non-interacting case;
in panel (a), main panel,
after the initial growth $t>10^0$-$10^1$, 
$S_{\rm ent}(t)$ tends to become saturated;
in regime (i) to a value $\simeq 1.5$,
while in regime (ii) this value is much enhanced,
and
in regime (iii)
the saturated value 
gets back to the ones comparable to those in regime (i).
Thus,
as the strength $W$ of disorder is varied (increased),
the saturated value of the entanglement entropy 
changes non-monotonically;
it is first enhanced by $W$, then suppressed.

In the interacting case [panel (b)],
the behavior of $S_{\rm ent}$ 
in regime (i) is similar to the non-interacting case,
while
the behavior of $S_{\rm ent}$ changes qualitatively
in regimes (ii) and (iii).
In regime (ii),
$S_{\rm ent}(t)$ is much enhanced in the intermediate time range $t\sim 10^0 - 10^1$,
but tends to be suppressed afterwards $t>10^2$;
$S_{\rm ent}(t)$ shows a non-monotonic growth in this regime.
In regime (iii) 
$S_{\rm ent}(t)$ continues to grow after the initial growth:
i.e.,
$S_{\rm ent}(t)\sim \log t$ at $t\gg 10^0$;
behavior characteristic to the many-body localized regime.
\cite{logt0,logt1,logt2,logt3,Huse_FMBL}
Thus,
as the strength $W$ of disorder is varied (increased),
the overall magnitude of $S_{\rm ent}(t)$ in its dynamics
is again non-monotonic
as in the non-interacting case.
This is quite a curious behavior
if we recall that
in the Hermitian case
many-body states 
become less entangled with the increase of $W$.\cite{MBL-EEdisorder1,MBL-EEdisorder2}
Here,
the many-body states tend to become more entangled 
with the increase of $W$, 
i.e., in the weakly disordered regime (i),
while
they tend to become less entangled 
beyond a certain critical value $W>W_{\rm c}$ [in regime (iii)].
%
Such non-monotonic dependence on $W$ is a characteristic non-Hermitian feature.
In the critical regime [regime (ii)], on the other hand,
another
non-monotonic feature is emergent in the entanglement dynamics;
i.e.,
the non-monotonic time 
evolution of the entanglement entropy $S_{\rm ent}(t)$ in time.

A careful 
reader may notice
on top of the above overall feature
that
$S_{\rm ent}$ exhibits also
a rapid oscillation 
typically in the weakly disorder regime.
The oscillation tends to damp in the course of time
in the interacting case [panel (b)],
while
it remains in the non-interacting case [panel (a)].
The oscillation is also conspicuous in
the case of even number of particles $L/2=6, 8, 10$,
while
less pronounced 
in the case of odd number of particles:
$L/2=7,9$.
We show, in Appendix~\ref{system-size},
that the oscillation stems from a 
two-fold degeneracy (in the imaginary part)
of the asymptotic state.

\section{The quasiparticle picture}
\label{Sec4}

The entanglement dynamics 
in a Hermitian system and
in (or close to) the clean limit ($W=0$)
is well described by the quasiparticle picture.
Here, we discuss,
how 
the entanglement dynamics in a non-Hermitian system
we have sketched in the previous section
can or cannot be compatible
with this picture.

The observed behavior of the density dynamics in the crystal momentum space introduced earlier (see Sec.~\ref{Sec3}) is directly relevant to the description of the quasiparticle picture.
In Fig.~\ref{Fig3} and related descriptions, we have seen that
$\langle n_k\rangle$ is almost uniform
($\langle n_k\rangle\simeq 1/2$)
in the Hermitian case,
while
$\langle n_k\rangle$ converges to
$\langle n_k\rangle=1$ for $k<0$, and
$\langle n_k\rangle=0$ for $k>0$
in the non-Hermitian case.
The two panels of Fig.~\ref{Fig1}
are in a sense
a pictorial representation of
these contrasting behaviors,
i.e.,
Fig.~\ref{Fig1}, panel (a) corresponds to Fig.~\ref{Fig3}, panel (i-d);
the Hermitian case,
and
Fig.~\ref{Fig1}, panel (b) corresponds to Fig.~\ref{Fig3}, panel (i-a);
the non-Hermitian case (close to the clean limit).

\subsection{GGE vs. non-unitary dynamics}

Let us focus on the clean and non-interacting limit ($W=0$ and $V=0$).
First, in the Hermitian case, in this case,
$\hat{n}_k=c_k^\dagger c_k$ 
is a conserved quantity: $[\hat{n}_k, H]=0$.
In this integrable system, 
the expectation value,
such as 
$\langle \hat{n}_k\rangle=\langle\Psi|\hat{n}_k|\Psi\rangle$
is also expressed as
a statistical average in the so-called generalized Gibbs ensemble (GGE)
characterized by an infinite number of
Lagrange multipliers $\lambda_k$,
each associated with the conservation of $n_k$;
see Appendix~\ref{Append-GGE} for details.
To be explicit,
$n_k$
can be expressed as
\begin{equation}
\langle \hat{n}_k\rangle=\frac{1}{1+e^{\lambda_k}},
\label{nk_H}
\end{equation}
where
for the density-wave like initial state (\ref{DW})
all $\lambda_k$ are equal to 0, \cite{nk=0.5} 
i.e.,
$\langle \hat{n}_k\rangle=1/2$.
We have seen this in the density dynamics
studied in Sec.\ref{Sec3}.
In Fig.~\ref{Fig3},
in all the panels,
the initial and early time density distribution $n_k(t=0)$ 
shows such a uniform profile
($\langle \hat{n}_k\rangle\simeq 1/2$),
while in the first raw,
panel (i-d), i.e.,
in the non-interacting ($V=0$) and Hermitian ($g=0$) case,
such an initial profile 
is maintained,
though approximately, due to a small but finite $W=0.5$. 
Note that
the (entanglement) entropy
associated with a generalized Gibbs ensemble specified by the distribution (\ref{nk_H})
is given as (see Appendix~\ref{Append-GGE} for its derivation)
\begin{equation}
s(k)=-\langle \hat{n}_k\rangle \log(\langle \hat{n}_k\rangle)-(1-\langle \hat{n}_k\rangle)\log(1-\langle \hat{n}_k\rangle).
\label{sk}
\end{equation}
Note that this takes a maximal value $\log 2$ at $\langle\hat{n}_k\rangle=1/2$ ($\lambda_k = 0$).

\begin{figure}
\includegraphics[width=80mm]{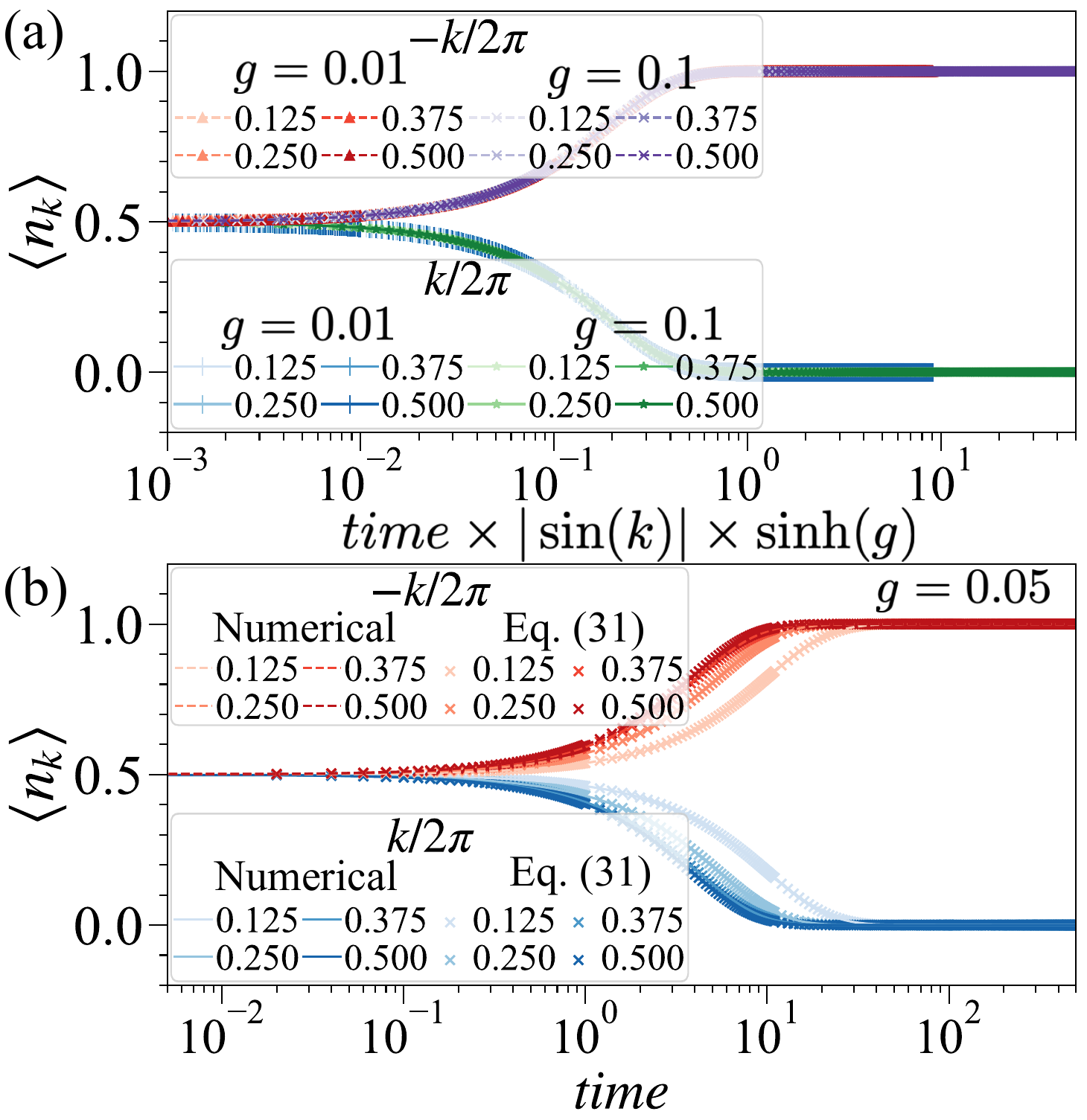}
\caption{Time-dependence of $\langle \hat{n}_k\rangle$:~
(a)~$\langle n_{k}\rangle$ versus $time\times\sin(k)\times\sinh(g)$ with the various values of $g$,
(b)~$\langle n_{k}\rangle$ (scatter plot) and Eq.~(\ref{nk_tilde}) (solid or dashed line) versus $time$. For panel (b), $g$ is fixed at $g=0.05$.
Numerical calculation is conducted by the following parameter: $L=16$, $W=0$, and $V=0$.
}
\label{Fig5}
\end{figure}

In the {\it non-Hermitian} case
with
$\rm{Im}(\epsilon_k)\neq 0$
($\epsilon_k$ is a single particle eigenenergy),
if we repeat the same argument leading to
Eq.~(\ref{nk_H}),
one is left with (see Appendix~\ref{Append-GGE})
\begin{equation}
\langle \hat{n}_k\rangle=\frac{1}{1+e^{\lambda_k-2\rm{Im}(\epsilon_k)t}},
\label{nk_NH}
\end{equation}
i.e.,
$\langle \hat{n}_k\rangle$
is no longer conserved
in this case.
Again,
for the initial DW like pattern (\ref{DW}),
all $\lambda_k$'s are to be set to 0
in Eq.~(\ref{nk_NH}).
%
Noticing that
Im$(\epsilon_k)>0$ for $k<0$,
Im$(\epsilon_k)<0$ for $k>0$,
and thus $\langle \hat{n}_k\rangle$ converges either to 0 ($k>0$)
or to 1 ($k<0$).
In Fig.~\ref{Fig5},
we have plotted the calculated value of $n_k(t)$
against the scaling function
\begin{equation}
\langle \hat{n}_k\rangle=\frac{1}{1+e^{-2\rm{Im}(\tilde{\epsilon}_k)t}},
\label{nk_tilde}
\end{equation}
where
$\tilde{\epsilon}_k=2\epsilon_k$.
Figure~\ref{Fig5} shows
that the numerical data
fit quite well with the scaling function (\ref{nk_tilde})
expected in GGE
except for a factor 2
in the definition of $\tilde{\epsilon}_k$.
\footnote{That is, $\langle \hat{n}_k\rangle$ suggested by GGE (see Appendix~\ref{Append-GGE} for more details) converges either to 0 or to 1 more slowly than Eq.~(\ref{nk_tilde}).
This discrepancy is because we assume superposition consists of various filling to derive Eq.~(\ref{Eq:k,g-dependence-GGE2}), whereas in the actual numerical calculation, we consider the half-filling case.
If the initial state is prepared as a superposition consists various filling $Q=\sum_i k_i/L$, the time dependence of $\langle \hat{n}_k\rangle$ is akin to Eq.~(\ref{Eq:k,g-dependence-GGE2}) (see Appendix~\ref{time-dependence nk}).}

As shown also in a more generic context in Appendix~\ref{Append-GGE}
[see, e.g., Eqs.~(\ref{Eq:k,g-dependence}) and (\ref{Eq:k,g-dependence-GGE})],
the time dependence of $\langle \hat{n}_k \rangle$ 
[here, e.g., Eq. (\ref{nk_NH})] 
is analogous 
to that of the imaginary time evolution driven by a {\it Hermitian} Hamiltonian, 
often employed in 
a numerical recipe 
to find 
the ground state, 
e.g., in 
a path-integral quantum Monte Carlo, or
in a tensor-network method.\cite{Avella2013}
In the imaginary time evolution,
the parameter, $t$ (time) 
corresponds to a ``temperature'' of the statistical ensemble.
Thus,
an evolution driven by a non-Hermitian matrix 
leads to an effective decrease in temperature.
Consequently, as time passes by, 
the temperature 
decreases, and the entropy ($S_{\rm ent}$) also seems to decrease.
In the following subsection, we investigate how 
$S_{\rm ent}$ behaves under a non-unitary dynamics.



\subsection{Entanglement dynamics in the quasiparticle picture in the clean and non-interacting limit ($W=0$ and $V=0$)}

In the quasiparticle picture,
the initial state $|\Psi(t=0)\rangle$ consists of a superposition of a highly excited state, acting as a source of quasiparticle excitations.
Pairs of quasiparticles with opposite momenta $k$ and $-k$ are emitted 
at the same point,
and as times passes by,
they move,
in the Hermitian case,
symmetrically
in opposite directions.
This is schematically
depicted in
panel (a) of Fig.~\ref{Fig1}.
%
Once each quasiparticle is located in the different subsystem, $S_{\rm ent}$ increases.
This process is formulated by 
\begin{equation}
S_{\rm ent}(t)\propto 2t\int_{2v(k)t<\ell} dkv(k)s(k)+\ell\int_{2v(k)t>\ell} dks(k),
\label{qpp}
\end{equation}
where $\ell$ is the subsystem size,
$k$ is a momentum of quasiparticles, $v(k)$ is its velocity, and $s(k)$ determines the production rate of $S_{\rm ent}$.
This production rate tightly relates to the entropy of statistical mechanics because $s(k)=-\langle \hat{n}_k\rangle \log(\langle \hat{n}_k\rangle)-(1-\langle \hat{n}_k\rangle)\log(1-\langle \hat{n}_k\rangle)$.

Entanglement dynamics (\ref{qpp}) has a characteristic time scale $t_c(k)$ determined by $\ell$ and $v(k)$.
When $t_c(k)\equiv\frac{\ell}{2v(k)}>t$, each of the quasiparticles emitted at the same points begin to be located in the different subsystems, contributing to the entanglement production as a function of $s(k)v(k)t$.
While $t_c(k) <t$, most of each of the quasiparticles are located in the different subsystems; therefore, the contribution of $S_{\rm ent}$ from pairs of quasiparticles $s(k)$ becomes constant value $s(k)\ell$.
\footnote{Strictly speaking, since we treat a finite system, each quasiparticles can be located in the same subsystem due to the boundary effect (we later comment on this effect), leading to decay in the $S_{\rm ent}$.
Although this effect can be non-negligible in a finite system, it is already known that the less important this effect, the larger the system size we treat;\cite{EE_revival} therefore, we can interpret $t_c(k)$ as a characteristic time scale.}

\begin{figure}
\includegraphics[width=65mm]{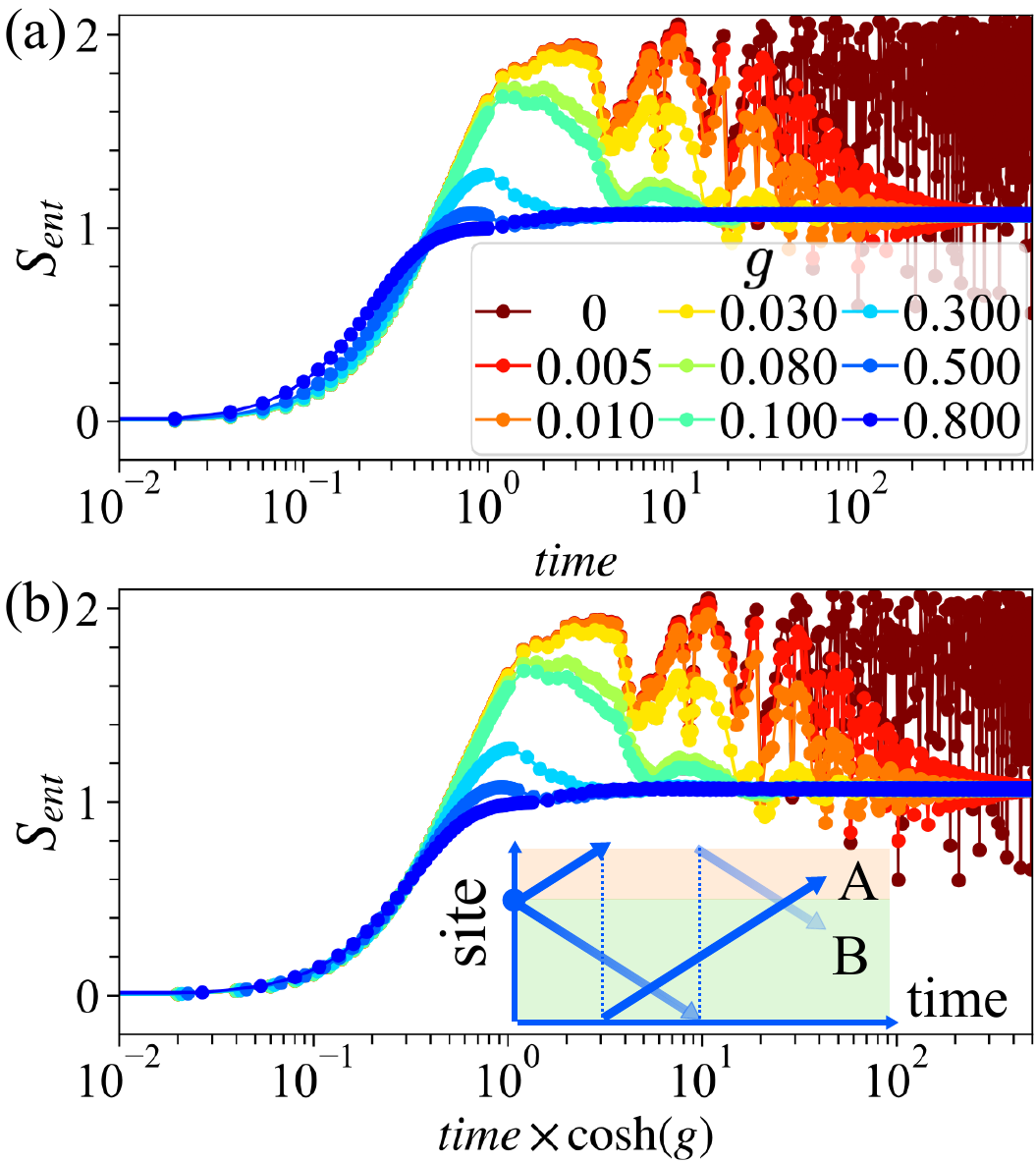}
\caption{Entanglement dynamics of a free particle case ($W=0.0$ and $V=0.0$) with $\ell=3$ with decrease of $g$: (a) $S_{\rm ent}$ versus $time$ and (b) $S_{\rm ent}$ versus $time\times\cosh(g)$ (analogy of quasiparticle picture).
$L=16.$}
\label{Fig6}
\end{figure}

In the non-Hermitian case, as time passes by, one of the quasiparticles is amplified while the other is attenuated due to Im$(E)$, resulting in a unidirectional motion (depicted in panel (b) of Fig.~\ref{Fig1}).
Moreover, this characteristic relaxation of $\langle \hat{n}_k\rangle$ (\ref{nk_tilde}) results in a variation of $\langle \hat{n}_k\rangle$ from $\langle \hat{n}_k\rangle=0.5$, indicating a decrease in $S_{\rm ent}$ as suggested by Eqs.~(\ref{nk_tilde}) and (\ref{qpp}).
Therefore, we investigate how the non-Hermiticity, specifically this characteristic relaxation, modifies the quasiparticle picture and entanglement dynamics.
Panels (a) and (b) of Fig.~\ref{Fig6} show $S_{\rm ent}$ as a function of $time$ and $time\times \cosh(g)$, respectively, with various values of $g$, in the clean and non-interacting limit ($W=0$ and $V=0$). 
We observe distinct behaviors of $S_{\rm ent}(t)$ arising from the quasiparticle picture and non-unitary time evolution.
We first focus on the initial growth of $S_{\rm ent}$.
According to the quasiparticle picture, the initial growth of $S_{\rm ent}$ depends on $v_g t$ ($v_g$ is a group velocity, and see Eq.~(\ref{qpp}) rather than $e^gt$ since $S_{\rm ent}$ is carried by quasiparticles as well as correlation.
In this case, $v_g=-2\cosh(g)\sin(k)$, and thus we expect that the initial growth of $S_{\rm ent}$ depends on $\cosh(g)\sin(k)t$.
We observe that the initial growth of $S_{\rm ent}$ can be approximated by a single curve, as is shown in panel (b), consistent with the quasiparticle picture and implying the validity of the quasiparticle picture at early time scales.
Following the initial growth, $S_{\rm ent}(t)$ depends on $g$.
$S_{\rm ent}(t)$ shows non-monotonic behavior for weak $g$, whereas it only converges to $S_{\rm ent}$ of $|\alpha_1\rangle$ for strong $g$.
This difference in $S_{\rm ent}(t)$ between weak and strong $g$ stems from the relaxation of $\langle \hat{n}_k\rangle$.
For weak $g$, Eq.~(\ref{nk_tilde}) implies that the relaxation of $\langle \hat{n}_k\rangle$ takes a considerable amount of time, causing $S_{\rm ent}(t)$ to resemble the behavior observed in the Hermitian case within this regime, leading to $S_{\rm ent}(t)> S_{\rm ent}(t\to\infty)$. However, Im$(E)$ eventually causes $\langle \hat{n}_k\rangle$ to converge to either 0 or 1, thereby resulting in the convergence of $S_{\rm ent}(t)$ to $S_{\rm ent}(\infty)$ and non-monotonic behavior of $S_{\rm ent}$.
\footnote{In numerical calculation, we choose the length of subsystem size $\ell$ to be small because it may be the simplest way to realize the non-monotonic behavior of $S_{\rm ent}$. 
Since $S_{\rm ent}(t\to\infty)$ decreases with a decrease of $\ell$, in case of small $\ell$, the condition $S_{\rm ent}(t)> S_{\rm ent}(t\to\infty)$, which is required to realize such a behavior, becomes easier to achieve.}
Additionally, we also observe Hermitian-type behavior in which the non-monotonic behavior of $S_{\rm ent}$ is accompanied by oscillations.
These oscillations occur when quasiparticles move through the left or right ends (see inset of panel (b) of Fig.~\ref{Fig6}) and are located within the same subsystem.
This oscillation behavior is known for entanglement revivals\cite{EE_revival}, predicted by the quasiparticle picture.
For large $g$, $\langle \hat{n}_k\rangle$ immediately converges to either 0 or 1, and thus $S_{\rm ent}(t)$ only converges to $S_{\rm ent}(\infty)$.


As we have observed, the interplay between the quasiparticle picture and the relaxation described by Eq.~(\ref{nk_tilde}) qualitatively captures entanglement dynamics of the HN model.
In Appendix~(\ref{Quasi}), we compare the numerical result with $S_{\rm ent}$ suggested by the quasiparticle picture to verify the accuracy of this picture.
While this picture provides a qualitative characterization of $S_{\rm ent}(t)$, we find a quantitative discrepancy between the numerical result and $S_{\rm ent}$ suggested by this picture.
Further work is required to identify the reason why this quantitative discrepancy presents.

\subsection{Disordered case: disorder enhances the entanglement}

The quasiparticle picture also provides us with a natural interpretation on
why disorder enhances the entanglement in the non-Hermitian case.
First,
in the Hermitian case, 
as disorder is introduced to the system, 
here in our analysis,
in the form of a quasi-periodic potential,
the entanglement entropy tends to be suppressed;
see e.g., inset of Fig.~\ref{Fig4}.
This is because
such a quasi-periodic potential
introduces scattering between quasiparticles,
preventing quasiparticle pairs from reaching a different subsystem
[Fig.~\ref{Fig1}, panel (a)].
If the pairs tend to stay in the same subsystem,
the entanglement entropy tends naturally to be decreased.
In the non-Hermitian case, 
scattering between quasiparticles
introduced by the quasi-periodic potential
may lead to quite a different
consequence.
As repeatedly mentioned,
the quasiparticle motion is uni-directional 
in the absence of scattering
[Fig.~\ref{Fig1}, panel (b)],
while
in the presence of scattering
this is expected to be no longer purely uni-directional,
but
become more bi-directional.
As a result,
disorder helps
quasiparticle pairs to reach a different subsystem, 
leading naturally to 
the increase of entanglement entropy.

In the density dynamics (Fig.~\ref{Fig3}),
we have seen that
$\langle n_k\rangle$ 
converges sharply to $0$ or $1$ 
in the clean limit, while
in the presence of disorder, this convergence is relaxed.
This clearly leads to the increase of
thermodynamic entanglement entropy: Eq.~(\ref{sk}) (see Appendix~\ref{Append-GGE} for more details).
In the interacting case, 
the behavior of $\langle n_k\rangle$
is not much different from the non-interacting case
[Fig.~\ref{Fig3}, panel (ii)],
the above reasoning in the non-interacting case
applies also, at least qualitatively,
to the interacting case.

Previously,
we have attributed this enhancement of the entanglement entropy
due to disorder
to
cascade-like spreading of the wave packet
in the single-particle dynamics.
\cite{OI22A,OI2023}
Here,
we have shown that
the quasiparticle picture gives
a more natural explanation of the same phenomenon,
which is
more likely valid in the interacting case.

\section{Logarithmic scaling in the asymptotic regime: $t\rightarrow\infty$}
\label{Sec_scal}

\begin{figure}
\includegraphics[width=85mm]{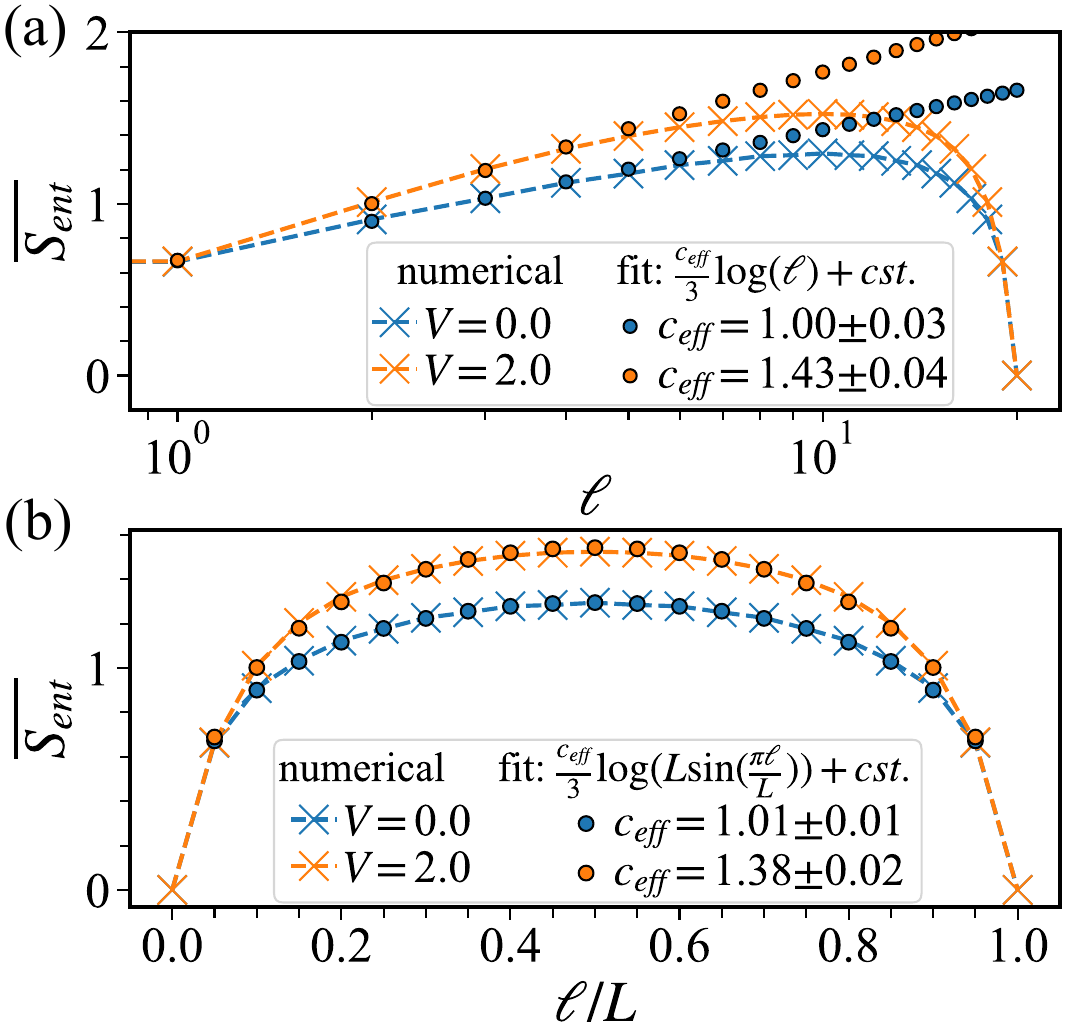}
\caption{Scaling of $S_{\rm ent}$ as a function of $\ell$, where $\ell$ is a length of subsystem: (a) $S_{\rm ent}$ versus $\ell$, (b)$S_{\rm ent}$ versus $\ell/L$.
For panel (b), we take into account the boundary condition,
so that fitting function $\frac{c_{\rm eff}}{3}\log(L\sin(\frac{\pi \ell}{L}))+cst.$ (scatter plot) seems to be fitted the numerical data (dashed line). $cst.$ is corresponding to a constant value.
$g=0.5$. $W=0.0.$}
\label{Fig7}
\end{figure}

In the previous subsection, 
we have seen characteristic behaviors of the entanglement entropy $S_{\rm ent}(t)$, 
which reflects the collapse of the superposition in the time-evolving many-body state $|\Psi (t)\rangle$; i.e., its convergence to a single eigenstate (\ref{alpha_1}).
In the Hermitian system,
the asymptotic value of the entanglement entropy,
\begin{equation}
S_\infty=S_{\rm ent}(t\rightarrow\infty),
\label{S_inf}
\end{equation}
obeys the volume-law scaling.
Here, we address what type scaling 
$S_\infty$ shows in the non-Hermitian case.
For a given entire system of size $L$; here, we fix it at $L=20$,
we vary the bipartite division $\ell$, i.e., the size of the subsystem A, 
and evaluate the entanglement entropy $S_\infty (\ell)$ in the asymptotic regime $t\rightarrow\infty$.
Fig.~\ref{Fig7} (a) shows a result of such analyses 
in the case of $W=0$, $V=0$, in which
$S_\infty (\ell)$
is plotted against $\log \ell$.
One can see that
for $\ell \ll L$
at which size effects are negligible,
the entanglement entropy $S_\infty (\ell)$ is well fit by
the scaling function:
\begin{equation}
\label{log_x}
S_\infty (\ell)={1\over 3}\log \ell + cst.,
\end{equation}
known in the Hermitian case
for a fermionic ground state,
which falls on the case of central charge $c=1$ 
(case of free bosonic excitation spectrum).
\cite{conformal1,conformal2,conformal3,conformal4,conformal5,conformal6}

One can even improve the fitting 
by taking into account
the finite size of the system and the periodic boundary condition;
replacing the length $\ell$ of the subsystem A
in Eq.~(\ref{log_x})
with the corresponding chord distance:
\begin{equation}
d(\ell)=2L\sin (\pi \ell/L)
\end{equation}
of a circle of circumference $L$, one finds
\begin{equation}
\label{log_sin}
S_\infty (\ell)={1\over 3}\log \left[2L\sin (\pi \ell/L)\right] + cst.
\end{equation}
In Fig.~\ref{Fig7} (b) 
the same data of the entanglement entropy $S_{\rm ent}$
is plotted
against the subsystem size $\ell$ 
in linear scale 
and fit by the scaling function (\ref{log_sin}).
One can see
all the data: $\ell=1,2,\cdots,L-1$
of $S_{\rm ent}$
is well fit by this modified scaling function.

In case of the fermionic ground state,
\begin{equation}
|\Psi_G\rangle=\left(\prod_{k s.t. |k|<k_F} c_k^\dagger\right) |0\rangle,
\end{equation}
the logarithmic term in Eq.~(\ref{log_x})
stems from discontinuities 
in the momentum space
at $k=k_F$ and $k=-k_F$,
where
$k_F$ is the Fermi wave number or
Fermi (crystal) momentum
associated with the Fermi energy
$\epsilon_{\rm F}=\hbar^2 k_F^2/(2m)$.
In the asymptotic expansion for large $L$,
the sub-leading logarithmic term becomes relevant
as a result of the vanishing of
the leading linear term $\propto L$ (volume-law term).
In the non-Hermitian dynamics
the many-body wave packet $|\Psi (t)\rangle$ 
may converge to a single eigenstate (\ref{alpha_1_k}).
Then,
the corresponding
momentum distribution (\ref{nk_asymp}) 
exhibits
discontinuities at $k=0$ and $k=-\pi$ 
in case of the half-filling.
These discontinuities lead to 
logarithmic scaling of the entanglement entropy (\ref{log_x}),
known in the 
fermionic ground state.\cite{fisherhartwig1,fisherhartwig2}

In the interacting case,
the entanglement entropy $S_\infty (\ell)$ seems still logarithmic
[Fig.~\ref{Fig7}, panel (b)],
in the sense that
the data are well fit by the following scaling function:
\begin{equation}
\label{log_sin_fit}
S_\infty (\ell)={c_{\rm eff}\over 3}\log \left[2L\sin (\pi \ell/L)\right] + cst.,
\end{equation}
where
$c_{\rm eff}$ is a fitting parameter;
of course, the naming implies that
we are tempted to interpret it as an effective central charge.
Our data clearly shows that
$c_{\rm eff}$ exhibits a deviation from the non-interacting value $c=1$,
which is uncommon in
the Hermitian case.\cite{Nishimoto_2011} 
Another remark is that in the regime of larger $V$
we found a discrepancy of our data
with the fitting function (\ref{log_sin_fit});
see Appendix~\ref{central-charge} for more details.
The discrepancy may be simply due to 
a finite size effect, 
but in any case
a further investigation 
in a system of larger size $L$ will be necessary, employing the methods such as
Bethe ansatz,\cite{Bethe2,Bethe3,DMRG3andBethe} 
the tensor-network,\cite{DMRG1,DMRG2,tensornet3,DMRG3andBethe,tensor} 
and the quantum Monte-Carlo simulation.\cite{QMC1,QMC2,QMC3,QMC4}

\section{Further scaling properties: 
behavior of Im$(E)$ and correlation function}
\label{Sec5}

\begin{figure}
\includegraphics[width=80mm]{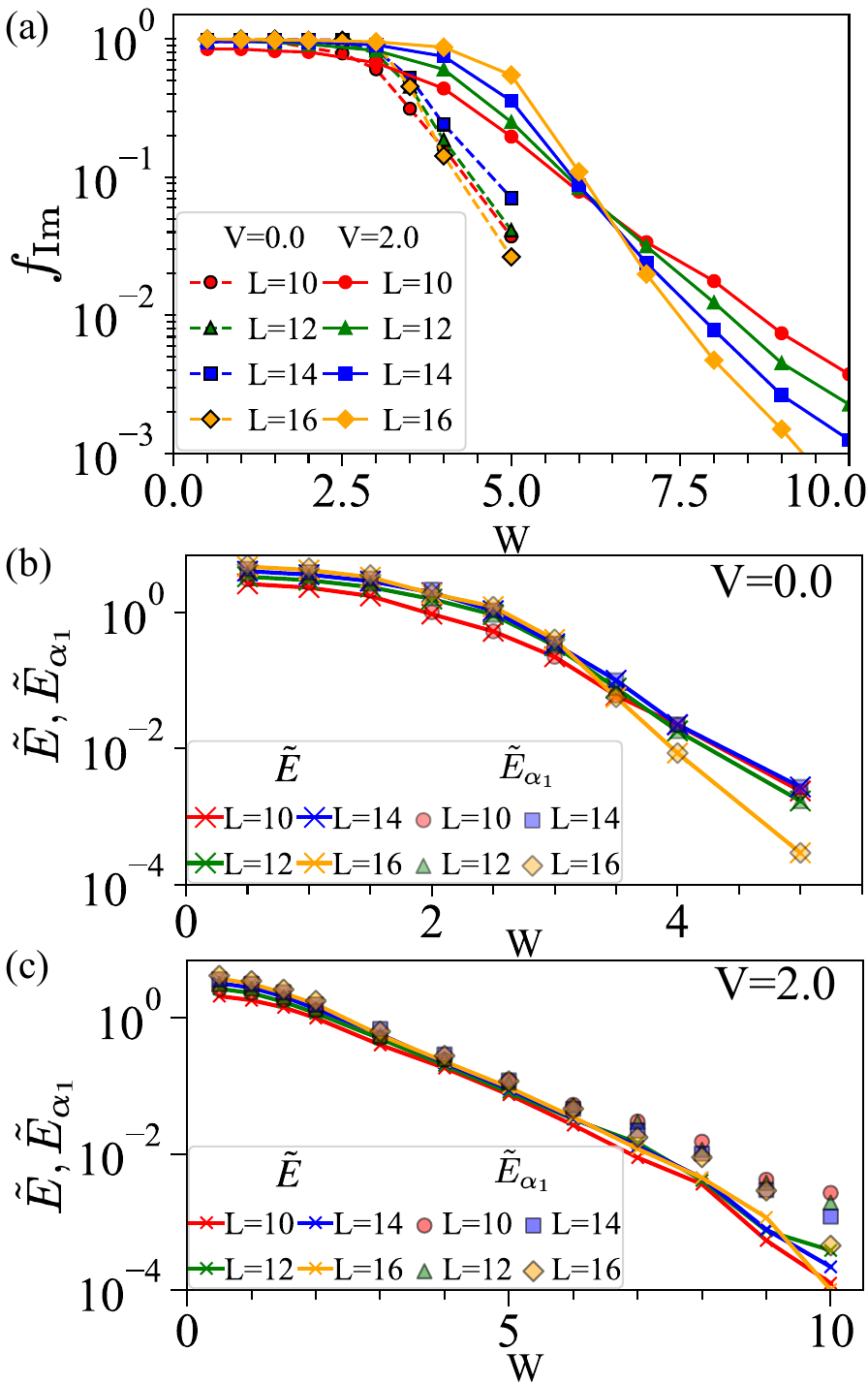}
\caption{Disorder dependence of the quantity to characterize the property of Im$(E)$ for various system sizes $L$. (a) $f_{\rm Im}$ for non-interacting ($V=0$, dashed line) and interacting ($V=2$, solid line) cases, respectively. The largest Im$(E)$, $\tilde{E}_{\alpha_1}\equiv${\rm Max(Im}$(E)$) (scatter plot), and the average value of Im$(E)$ taken from the second to fifth, $\tilde{E}$ (solid line) for non-interacting (panel (b)) and interacting (panel (c)) cases, respectively.
In numerical calculation, we carried out evaluations for various system sizes $L=10, 12, 14$, and $16$ using varying sample sizes: $1000, 500, 500$, and $200$ samples for $L=10, 12, 14$, and $16$.
}
\label{Fig8}
\end{figure}

\subsection{Imaginary part of the eigenenergy; origin of the non-monotonic time evolution}

In Sec.~\ref{Sec3}, we have seen that
the entanglement entropy 
$S_{\rm ent}(t)$ 
exhibits a non-monotonic time evolution, 
typically,
in the regime of 
intermediate disorder
and in the interacting case [Fig.~\ref{Fig4} (b)].
A sensible reader would immediately associate
this intriguing behavior,
unique also to the non-Hermitian case, 
with the complex nature of the spectrum
characteristic to the system, which is indeed the case.
While,
if that is simply the reason,
one may then wonder why
the non-monotonic evolution is specific to the interacting case,
and does not appear in the non-interacting case [Fig.~\ref{Fig4}, panel (a)],
albeit that
the complex spectrum also appears in non-interacting case.
Below, we will carefully focus on the complex nature of the spectrum, highlighting
especially
the degeneracy in the imaginary part of the spectrum Im$(E)$.
The crucial difference 
that also leads to the conspicuous difference in the behavior of entanglement entropy $S_{\rm ent}(t)$ 
in the interacting vs. non-interacting cases
lies in the difference (absence vs. presence) of such degeneracy in Im$(E)$
in the complex spectrum.
After briefly looking into the scaling of the Im ratio $f_{\rm Im}$, 
relevant to the identification of the real-complex transition in spectrum,
we will proceed to a more careful study of such degeneracies in Im$(E)$.

The
fraction $f_{\rm Im}$, which is defined as the ratio of the number of the eigenenergies with non-zero imaginary part ($|{\rm Im}(E)|>10^{-10}$) $D_{\rm Im}$ to the total number of the eigenenergies $D$; i.e., 
\begin{equation}
f_{\rm Im} = D_{\rm Im}/D,
\label{Eq:fraction}
\end{equation}
is often employed 
in the study of real-complex transition.\cite{Hamaz,nonHMBL-AA,non-Hermitian-MBL-arXiv}
$f_{\rm Im}$ is typically averaged within a defined energy range or across the entire spectrum.
Thus, we can consider $f_{\rm Im}$ as a measure to describe the statistical properties of a complex spectrum.
In the delocalized phase, $f_{\rm Im}$ is close or almost equal to $1$, 
whereas it 
practically vanishes
in the localized phase.
$f_{\rm Im}$ in the non-interacting case shown in Panel~(a) of Fig.~\ref{Fig8} (dashed line) take almost constant value $f_{\rm Im}\sim1$ for weak $W$, and as $W$ approaches  $W_c\sim3.3$, $f_{\rm Im}$ 
sharply
decreases.
This tendency becomes more enhanced as $L$ increases, 
and 
in the $(W,$ $f_{\rm Im})$-plane,
different curves for $f_{\rm Im}$ 
calculated at different system size $L$
looks
intersecting 
at a single point, ($W_c$, $f_{\rm Im}(W_c)$) [Fig.~\ref{Fig8}, panel (a)],
implying that this
real-complex transition
at $W=W_c$ is a true phase transition robust until the thermodynamic limit: $L\rightarrow\infty$.
In the interacting case 
(solid line), 
the position of the crossing is shifted to
a regime of larger $W$ compared with the non-interacting case (dashed line),
while the overall behavior is 
unchanged 
from the non-interacting case.
%
Thus,
so far as the scaling analysis of $f_{\rm Im}$ implies,
the real-complex transition of the spectrum occurs
practically in the same way
both in the interacting and non-interacting cases.
Then, how could that be compatible with
a relatively different dynamics of the entanglement entropy
$S_{\rm ent}(t)$
in the interacting and non-interacting cases?

In Sec.~\ref{Sec2} A, we have argued that
in the non-unitary time evolution
the many-body wave packet $|\Psi(t)\rangle$,
which is initially a superposition of many eigenstates,
tends to lose such a superposed nature, and
collapse into a single eigenstate $|\alpha_1\rangle$ [see Eq. (\ref{alpha_1})],
where
$|\alpha_1\rangle$ is such an eigenstate whose eigenenergy $E$ has a maximal imaginary part, Im$(E)$.
This picture demonstrated in Sec.~\ref{Sec2} A is, however,
slightly oversimplified in the sense that
it did not consider the case in which some eigenstate have (practically) the same, or very close Im$(E)$;
the case in which
\begin{equation}
{\rm Im}(E_{\alpha_1}) \simeq {\rm Im}(E_{\alpha_2}) \simeq \cdots.
\end{equation}

Which quantity is relevant for determining 
how quickly the state $|\Psi(t)\rangle$ converges
(or not) to a single eigenstate $|\alpha_1\rangle$?
Im$(E_{\alpha_\nu})>0$ amplifies the amplitude of the coefficient $c_{\alpha_\nu}(t)$, expressed as $|c_{\alpha_\nu}(t)|^2= |c_{\alpha_\nu}(t=0) e^{{\rm Im}(E_{\alpha_\nu}t)}|^2$. 
%
%
The relative importance of the state $|\alpha_1\rangle$
with respect to another state, e.g., $|\alpha_\nu\rangle$ 
in the time-evolving wave packet $|\Psi(t)\rangle$
may be quantified by the ratio:
\begin{equation}
\frac{|c_{\alpha_\nu}(t)|^2}{|c_{\alpha_1}(t)|^2}\propto |e^{{\rm Im}(2(E_{\alpha_\nu}- E_{\alpha_1})t)}|= e^{-\Delta_{\rm Im}^\nu t}.
\label{Eq:Delat-Im}
\end{equation}
%
%
Here, we  
consider the quantity,
$\tilde{E}$ defined as the average of the second to the fifth largest value of Im$(E)$,
and
conjecture that
the difference between $\tilde{E}_{\alpha_1}\equiv {\rm Max(Im(}E))$
would be
a good measure for characterizing how quickly the state $|\Psi(t)\rangle$
converges to (or not to) a hypothetical asymptotic state $|\alpha_1\rangle$.

Panels~(b) and (c) of Fig.~\ref{Fig8} show $\tilde{E}_{\alpha_1}$ and $\tilde{E}$ as a function of $W$ in the non-interacting and interacting case, respectively.
Both $\tilde{E}_{\alpha_1}$ and $\tilde{E}$ decrease with an increase of $W$, leading to a decrease in $\Delta_{\rm Im}^\nu$ (cf. Eq.~(\ref{Eq:Delat-Im})).
Interestingly, in the localized phase of the non-interacting system, $\tilde{E}$ is the same as $\tilde{E}_{\alpha_1}$, which means $\tilde{E}_{\alpha_1}$ is degenerate.
When $\tilde{E}_{\alpha_1}$ is degenerate, the corresponding eigenstates are amplified under time evolution similarly, i.e., $\Delta_{\rm Im}^\nu=0$, and thus superposition $c_{\alpha_\nu}(t)$ is maintained (cf. Eq.~(\ref{Eq:Delat-Im})) even in the non-unitary time evolution. 
Such a degeneracy stems from the fact that Im$(E_{\alpha_\nu})$ is a sum of single particle eigenenergies $\epsilon_\alpha$'s.
In the localized phase, most of $\epsilon_\alpha$ are real spectra, but some $\epsilon_\alpha$ have non-zero imaginary parts of eigenenergies due to the finite system size effect, causing the combination of the sum of the real and complex spectra to leads to the degeneracy of Im$(E)$.
In contrast to the non-interacting case, in an interacting case, $\tilde{E}_{\alpha_1}$ is not the same as the $\tilde{E}$, which means that $|\Psi(t)\rangle$ generally converges to a single eigenstate $|{\alpha_1}\rangle$ and non-monotonic behavior of $S_{\rm ent}$ appears consequently.

\begin{figure}
\includegraphics[width=80mm]{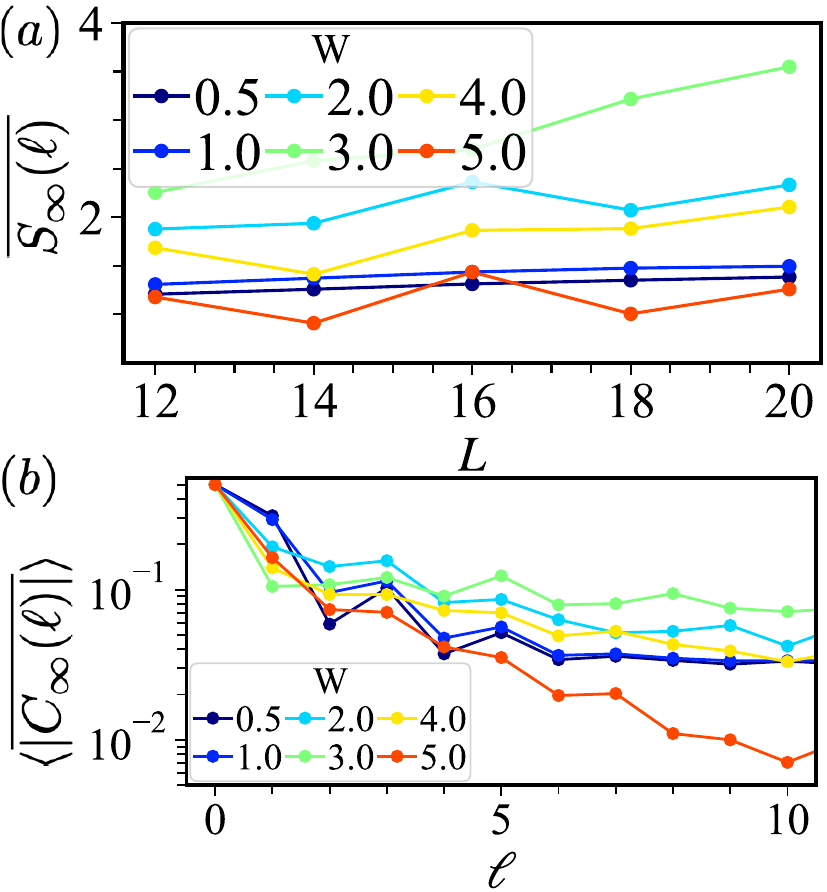}
\caption{
Size dependence of the entanglement entropy $S_\infty(L)$ (the asymptotic value) [panel (a)]
and behavior of the
correlation function $C_\infty (\ell)$ [panel (b)]
both in the non-interacting limit.
(a) $S_\infty=S_{\rm ent}(t\rightarrow\infty)$ is calculated in the system of size $L$
and plotted as function of $L$. 
Partly, the same data as the ones in Fig.~\ref{Fig4} have been replotted.
(b) behavior of the
correlation function $C_\infty (\ell)=C_\infty (\ell,t\rightarrow\infty)=\langle c_j^\dagger c_{j+\ell}\rangle$,
site and sample averaged;
as for precise definitions and conditions, see Eqs. (\ref{corr}), (\ref{corr_av}), and main text.
The same wave functions as those in Fig.~\ref{Fig4} have been used.
}
\label{Fig9}
\end{figure}

\begin{figure}
\includegraphics[width=80mm]{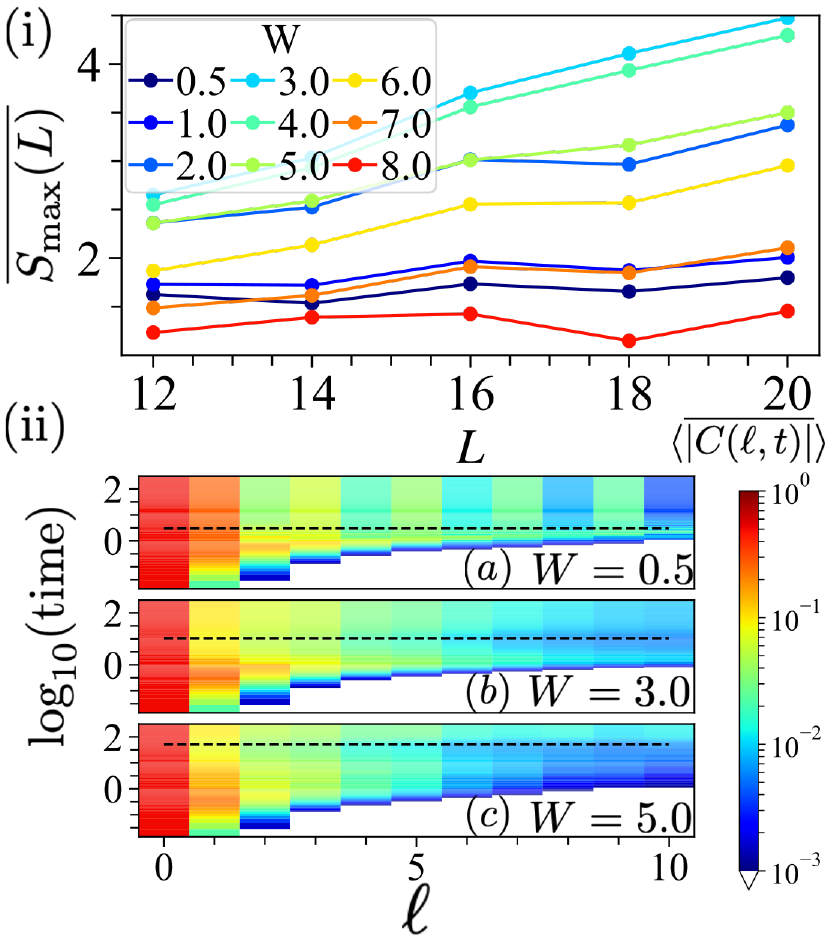}
\caption{
Scaling of $S_{\rm max} (L)$ 
at various strength of disorder $W$ [panel (i)],
and behavior of the correlation function 
$C(l,t)$ in the space time $(l,t)$ [panel (ii)].
(i) $S_{\rm max} (L)$ [introduced in Eq. (\ref{S_max})]
is plotted as a function of $L$.
Partly, the same data as the ones in Fig.~\ref{Fig4} have been used.
(ii) The behavior of the correlation function $C(l,t)$,
introduced in Eq. (\ref{corr}),
is shown as a color map
for three different values of disorder strength $W$: (a) $W=0.5$, (b) $W=3.0$, 
and (c) $W=5.0$ (subpanels).
We focused on
the magnitude of the correlation function $|C(l,t)|$,
which has been also site and ensemble averaged;
cf. Eq. (\ref{corr_av}).
The same wave functions as those in Fig.~\ref{Fig4} have been used.
}
\label{Fig10}
\end{figure}

\subsection{Entanglement entropy and correlation function}

Finally, we focus on how the scaling of the entanglement entropy
$S_{\rm ent}$ 
and 
the
correlation function 
as a function of the strength $W$
of disorder.
In 
Panel (a) of Fig.~\ref{Fig9}, 
$S_\infty=S_{\rm ent}(t\rightarrow\infty)$
[same as in Eq. (\ref{S_inf})],
evaluated in a system of size $L$; so we will also call it $S_\infty (L)$,
ensemble averaged,
is plotted as function of $L$, i.e., how 
$S_\infty (L)$
scales with $L$
at a various strength of disorder $W$ in the non-interacting limit ($V=0$).
%
One remarkable point is that 
the plotted curves $S_\infty (L)$
show
a non-monotonic evolution 
as a function of $W$.
For weak $W$, 
$S_\infty (L)$ obeys the logarithmic scaling, as we saw in the previous section, 
while 
as $W$ approaches the critical value $W_c$,
a sharp distribution of $\langle \hat{n}_k\rangle$
in the crystal momentum space
in the manner
Eq. (\ref{nk_asymp})
tends to be lost, and simultaneously, 
$S_\infty (L)$ 
starts to obey the volume-law. 
In this critical regime,
$\langle \hat{n}_k\rangle$ takes values other than 0 or 1, 
implying
an increase in the thermodynamic entropy [cf. Eq.~(\ref{qpp})].
Once $W$ exceeds $W_c$, scaling of 
$S_\infty (L)$ turns to the area law, 
as is also the case in a Hermitian localized phase. 
The evolution of the scaling behavior of $S_\infty (L)$,
or $S_\infty (L,W)$ may be summarized as,
\begin{equation}
S_\infty (L)\sim\left\{
\begin{array}{ll}
\log L  &(W \ll W_c)   \\
L  &(W\simeq W_c)\\
1  &(W>W_c)      
\end{array}
\right..
\label{S_scaling}
\end{equation}

Another interesting issue is that
the behavior of
the correlation function:
\begin{equation}
C_j(l,t) 
=\langle \psi(t)|
c_j^\dagger c_{j+\ell}
|\psi(t)\rangle
\label{corr}
\end{equation}
shows
a similar ``non-monotonic'' dependence 
on $W$
as the one seen in the entanglement entropy (\ref{S_scaling}).
%
Panel (b) of Fig.~\ref{Fig9} shows
how
this correlation function, 
\begin{equation}
C_\infty(l)= \lim_{t\rightarrow\infty} C(l,t)
\label{corr_inf}
\end{equation}
decays
with the distance $\ell$;
again, in the non-interacting case.
To be precise,
in the panel
the magnitude (absolute value) of the correlation function, 
both ensemble and site averaged:
\begin{equation}
\langle
\overline{
|C_\infty(l)|
}
\rangle
=
\left\langle
\frac{1}{L}\sum_{j} 
|C_j(l,t\rightarrow\infty)|
\right\rangle
\equiv C_\infty(l)
\label{corr_av}
\end{equation}
has been plotted. The brackets $\langle\cdots\rangle$ represents
the ensemble average.
At weak $W$, 
it is 
expected to show 
an algebraic decay: \cite{Skin-EE-transition}
\begin{equation}
C_\infty(l)
\propto \frac{1-e^{i\pi\ell}}{l}
\label{algeb}
\end{equation}
reflecting the sharp distribution of $\langle \hat{n}_k\rangle$
in the crystal momentum space (\ref{nk_asymp}).
%
As $W$ increases, 
the decay of the correlation function $C_\infty(l)$ 
becomes slower;
compare the greenish plots with the bluish ones,
implying that
the non-locality of the system is increased. 
This is in concomitant with
the evolution of the scaling of
the entanglement entropy 
$S_\infty (L)$
from logarithmic to volume law; cf. Eq.~(\ref{S_scaling}).
Once $W$ exceeds $W_c$, 
the correlation function
$C_\infty(l)$ 
decreases exponentially,
reflecting the
localized nature of the wave function.

In the interacting case ($V=2$), we focus on 
the maximal value of $S_{\rm ent}(t)$ in its evolution:
\begin{equation}
S_{\rm max}=
{\rm Max}[S_{\rm ent}(t)] 
\equiv
S_{\rm ent}(t=t_0),
\label{S_max}
\end{equation}
instead of 
$S_\infty=S_{\rm ent}(t\rightarrow\infty)$
[Eq. (\ref{S_inf})].
This is for a practical reason;
the Krylov subspace method employed in this work
is very effective 
for reducing the computational difficulty of 
dealing with a system of large size,
while it does not reduces that of a very long-time dynamics.
\footnote{We note that we can also use $S_{\rm ent}(\infty)$ as a quantity to characterize the delocalization-localization transition.
We expect that $S_{\rm ent}(\infty)$ obeys the volume-law in the delocalized phase and the area-law in the delocalized phase, reflecting the property of $|\alpha_1\rangle$.}
In Fig.~\ref{Fig4}, we have already seen an overall behavior of $S_{\rm ent}(t)$ 
at different values of $W$, i.e., both in the delocalized and localized phases,
and also
at different system sizes $L$.
%
Here, we have focused on 
the size-dependence $S_{\rm max} (L)$ 
[panel (i) of Fig.~\ref{Fig10}]
in the interacting case.
The scaling behavior of $S_{\rm max} (L)$ 
at various values of $W$
shows that 
$S_{\rm max} (L)$
increases (decreases) with $W$ in the delocalized (localized) regime, 
indicating that 
$S_{\rm max} (L)$
is a good measure of delocalization/localization transition/crossover
in this interesting case,
playing a similar role as 
$S_\infty(L)$ 
in the non-interacting case.

We also evaluate the correlation function $C(l,t)$ introduced in Eq.~(\ref{corr})
in the interacting case;
here,
we focus on its time-dependent behavior,
since we are interested in
how the relaxation of a quantum state $|\Psi(t)\rangle$
due to Im$(E)$
is reflected in the behavior of the correlation function.
Three panels of Fig.~\ref{Fig10} (ii)
show the time evolution of the correlation function (\ref{corr})
at disorder strength $W=0.5$ (panel (a)), $3.0$ (panel (b)), and $5.0$ (panel (c)).
The absolute value of the correlation function $|C(\ell,t)|$
is plotted as a color map
in the space of
$\ell$ (the $x$-axis)
and 
$\log_{10}({\rm time})$ (the $y$-axis).
The black dashed line in each panel
represents the time $t_0$ 
when $S_{\rm ent}(t)$ takes the maximal value 
$S_{\rm max}={\rm Max}[S_{\rm ent}(t_0)]$;
see Eq. (\ref{S_max}).
At weak $W$ [panel (a)],
the correlation spreads rapidly;
i.e.,
$C(l,t)$ quickly (i.e., around $t=t_0$) converges to 
an asymptotic distribution $C_\infty(l)$ [see Eq. (\ref{corr_inf})]
which is spatially modulating;
reminiscent of
the algebraic decay (\ref{algeb}) 
in the non-interacting case; 
see also Fig.~\ref{Fig9} (b), e.g., case of $W=0.5$ (blue plots).
As $W$ is increased, 
the spreading of correlation becomes delayed;
$t_0$ becomes larger,
while
the asymptotic distribution
$C_\infty(l)$
becomes a 
monotonically
decreasing function [case of panels (b) and (c)].

Unlike in the non-interacting case,
we have not observed a non-monotonic feature 
with respect to $W$
in the behavior of the correlation function $C(l,t)$
in the interacting case.
Still, 
we have made a notable observation that 
$t_0$ 
corresponds to the time $t$ when 
the behavior of the correlation function changes qualitatively;
i.e.,
from a strongly non-equilibrium type behavior ($t<t_0$)
to that of a steady state type ($t>t_0$).
Once 
$t$ exceeds $t_c$, 
the correlation function $C(l,t)$
tends to become time-independent, indicating that the quantum state $|\Psi(t)\rangle$ 
reaches a steady state.
This observation suggests that $S_{\rm max}$ is a good quantity 
that encodes the transition/crossover of the system or of the state $|\Psi(t)\rangle$
from a non-equilibrium to a steady state.

\section{Concluding remarks}
\label{Sec6}

In this paper, we have highlighted 
the differences in the dynamical behavior between non-Hermitian and Hermitian disordered systems based on the quasiparticle picture.
First, we have systematically studied the dynamical behavior of the many-body HN model, including $n_j(t)$, $n_k(t)$, and $S_{\rm ent}$, using the Krylov subspace method.
Although the difference 
between non-Hermitian and Hermitian systems
are somewhat masked in the behavior of $n_j(t)$ in real space,
we find that it sharply manifest
in $n_k(t)$,
and in $S_{\rm ent}$ as well.
In the non-interacting system, we 
demonstrated both numerically and analytically
the characteristic relaxation of $n_k(t)$, where $n_k(t)$ converges either to 0 (for $k>0$) or to 1 (for $k<0$) in the clean limit.
This behavior stems from the presence of Im$(E)$, which is an intrinsic nature of the non-Hermitian system.
We also discussed the relationship between the relaxation of $n_k(t)$ and $S_{\rm ent}$ based on the quasiparticle picture and provided an intuitive explanation for the non-monotonic behavior of $S_{\rm ent}$ as a function
of $W$.
Interestingly, we found that $S_{\rm ent}$ 
exhibits 
a non-monotonic behavior as a function 
of time 
in the interacting case.
By carefully examining the distribution of Im$(E)$,
especially,
through comparison with the non-interacting, 
we have clarified the nature 
of this non-monotonic time evolution, which is unique to this non-Hermitian
interacting system.

The non-monotonic behavior of $S_{\rm ent}$ with respect to time stems from Im$(E)$, which implies the instability of the many-body localized phase.\cite{Avalanche,Avalanche2}
Recent studies suggest that thermalization symptoms\cite{thermal,thermal2} appear even in a strongly disordered system, resulting in the study of many-body delocalization-localization transition\cite{ETH-MBL1,ETH-MBL2,ETH-MBL3,ETH-MBL4,ETH-MBL5} at a turning point.
They have examined the response of a quantum system to the inclusion of thermal grain\cite{insta_MBL1} and evaluated the imaginary part of eigenenergy,\cite{insta_MBL2} which may relate to the real-complex transition of the HN model.

For the non-interacting case, we have recently noticed that the non-monotonic behavior of $S_{\rm ent}(t\to\infty)$ is also reported in Ref.~\onlinecite{Skin-EE-transition}. They have employed a recently proposed numerical approach\cite{skin-EE} instead of the exact diagonalization and performed calculations in larger system sizes compared to ours.
Their findings indicate that the scaling of $S_{\rm ent}(t\to\infty)$ exhibits logarithmic-area law transition. This result is contradiction with our findings at critical regime ($W\sim W_c$), which may come from the finite size effect.
This discrepancy raises a new question as to whether $S_{\rm ent}$ obeys volume law scaling in an interacting system.
We intend to address this question in future work.


\acknowledgments
Quspin\cite{Quspin1,Quspin2} 
has been employed for generating
the matrix elements of Eq.~(\ref{ham_mp}).
K.-I.I thanks Marco Schiro and Kohei Kawabata for useful discussions, comments and suggestions.
This work was supported by JSPS KAKENHI Grant Numbers JP23KJ0360 (T.O.), JP20K03788 (K.-I.I), and JP21H01005(K.-I.I), and JST SPRING: Grant Number JPMJSP2132 (T.O.).
\appendix

\section{A viewpoint from the Lindblad/GKSL equation}
\label{GKSL-section}
The Lindblad/GKSL equation\cite{GKSL1,GKSL2} is a fundamental equation describing a quantum system coupled to an environment or a measuring apparatus.
GKSL equation is expressed as
\begin{eqnarray}
\frac{\partial\rho(t)}{\partial t}=-i[H_{eff},\rho(t)]+\sum_m L_m\rho(t)L_m^\dagger,
\label{GKSL}    
\end{eqnarray}
where 
$H_{eff}=H-\frac{i}{2}\sum_m L_m^\dagger L_m$, and 
$L_m$ is called the Lindblad operator
that stems from the interaction between the original quantum system and environment or a measuring apparatus.

The purpose of this Appendix is to clarify the relationship between Eq.~(\ref{GKSL}) and Eqs.~(\ref{sh-eq}, \ref{renorm}).
Indeed,
Eqs.~(\ref{sh-eq}, \ref{renorm}) can be derived from Eq.~(\ref{GKSL})
by simply neglecting the second term of Eq.~(\ref{GKSL}).
Without the second term,
the evolution of the density matrix $\rho(t)$
is determined by an effective von Neumann equation
prescribed by the generally non-Hermitian effective Hamiltonian $H_{eff}$.
If a pure state is chosen as an initial state, 
this dynamics coincides with the one obtained 
in the time evolution so that $|\Psi(t)\rangle$ is 
determined by Eqs.~(\ref{sh-eq}, \ref{renorm}).
Conversely,
our non-Hermitian Schr\"odinger dynamics
prescribed by Eqs.~(\ref{sh-eq}, \ref{renorm})
may be thus justified
in the context of
the GKSL description of
an open quantum system.

Of course, why and under what circumstances
the second term of Eq.~(\ref{GKSL}) is negligible
is left to be explained
(one may need also a further justification), 
and
so is
the meaning of neglecting the second term.
This may be best illustrated in the so-called
quantum trajectory 
picture
\cite{Daley_2014}
(cf. also quantum jump\cite{quantum-jump} and 
first-order Monte Carlo methods
\cite{first-monte-1,first-monte-2}).
In this picture
the time evolution of a wave function $|\Psi(t)\rangle$
is regarded as a stochastic process described below,
and 
a series of such a stochastic process
(corresponding to the entire time evolution of the wave function)
is referred to as a quantum trajectory;
in the end, an ensemble average of many trajectories will be taken.
Also, here,
the environment means an ensemble of measuring apparatus,
represented by an operator $L_m$.
After
each time step of $\delta t$, 
a quantum state 
$|\Psi(t)\rangle$
evolves 
with a probability $1-p$
into
\begin{eqnarray}
|\Psi(t+\delta t)\rangle 
=\frac{(1-iH_{eff}\delta t)|\Psi(t)\rangle}{\sqrt{1-p}}
\label{trajectory1}    
\end{eqnarray}
and
with a probability $p_m$
into
\begin{eqnarray}
|\Psi(t+\delta t)\rangle
&=&\frac{L_m|\Psi(t)\rangle}{\sqrt{p_m/\delta t}},
\label{trajectory2}    
\end{eqnarray}
where $p=\sum_m p_m$ and 
$p_m=\langle\Psi(t)| L_m^\dagger L_m |\Psi(t)\rangle \delta t$
.
Equation~(\ref{trajectory2})
describes the case in which 
the measurement apparatus $m$ obtains an outcome, while
Eq.~(\ref{trajectory1}) 
describes the situation in which 
none of the 
measurement apparatus obtains an outcome,
i.e., the case of null outcome.
A series of this stochastic process determines a single trajectory of
the quantum state $|\Psi(t)\rangle$.
Our non-Hermitian Schr\"odinger dynamics
prescribed by Eqs.~(\ref{sh-eq}, \ref{renorm})
is, on the other hand,
obtained by
selecting, after each time step,
the case of null outcome (post-selection);
or,
in other words,
by projecting the quantum state 
onto its subspace of such successive null outcomes
(or by choosing such a trajectory).
Note that
in Eq.~(\ref{trajectory1})
a change in the amplitude of the wave function
in the numerator,
i.e., $(1-iH_{eff}\delta t)|\Psi(t)$,
is precisely compensated by 
the normalization factor $\sqrt{1-p}$
in the denominator,
which is equivalent to the renormalization 
we adopted in Eq.~(\ref{renorm}).


\section{Other definitions of the entanglement entropy}
\label{choise-of-entanglement}
Since a non-Hermitian system has right and left eigenvectors, three possible definitions of $S_{\rm ent}$ and density matrix $\Omega$ have been considered.
In particular, $S_{\rm ent}$ is defined as
\begin{eqnarray}
S_{\rm ent}^{R,R}=-{\rm Tr}[\rho_{R,R}\ln(\rho_{R,R})],
\label{Eq:ent_RR}
\end{eqnarray}
\begin{eqnarray}
S_{\rm ent}^{L,L}=-{\rm Tr}[\rho_{L,L}\ln(\rho_{L,L})],
\label{Eq:ent_LL}
\end{eqnarray}
and
\begin{eqnarray}
S_{\rm ent}^{R,L}=-{\rm Tr}[\rho_{R,L}\ln(\rho_{R,L})],
\label{Eq:ent_RL}
\end{eqnarray}
where $\rho_{R,R}={\rm Tr_B}[|\alpha\rangle\langle\alpha|/\langle \alpha|\alpha\rangle]$,
$\rho_{L,L}={\rm Tr_B}|[\alpha\rangle\rangle\langle\langle\alpha|/\langle\langle \alpha|\alpha\rangle\rangle]$, and
$\rho_{R,L}={\rm Tr_B}[|\alpha\rangle\rangle\langle\alpha|/\langle\langle \alpha|\alpha\rangle]$.
Here, a superscript ($R$ or $L$) represents which eigenstate (left or right) is used to construct the density matrix.
The definition we employ in this work relates to Eq.~(\ref{Eq:ent_RR}), which yields a non-negative value of $S_{\rm ent}$ as well as $S_{\rm ent}$ of a Hermitian system.
This non-negativity of $S_{\rm ent}$ holds, which can be shown by Schmid value decomposition.
Whereas, in the case of Eq.~(\ref{Eq:ent_RL}), the non-negativity of $S_{\rm ent}$ does not have to hold since $\rho_{R,L}$ can become a non-Hermitian matrix, which has been studied in the context of non-unitary CFT.\cite{non-unitary-CFT1,non-unitary-CFT2}
Furthermore, in this case, the qualitative behavior of R\'{e}nyi entropy does not coincide with $S_{\rm ent}$,\cite{negative-c} which is hardly seen in the Hermitian case and Eqs.~(\ref{Eq:ent_RR},\ref{Eq:ent_LL}).\cite{EE-herm-like}
The previous study mainly focused on the static behavior (eigenvector) of  Eq.~(\ref{Eq:ent_RL}), and thus, it may be an interesting direction to investigate the dynamical behavior of  Eq.~(\ref{Eq:ent_RL}).
\section{Choice of the boundary conditions: effect of the skin effect}
\label{boudary-condition}

The Hatano-Nelson model exhibits a so-called non-Hermitian skin effect under the open boundary conditions (OBC).
Skin effect is a localization phenomenon where an extensive number of the eigenstates are at the boundary with real eigenenergy.
Although this feature has already been reported in the original works of Hatano and Nelson, it has now been recognized as a hallmark of topological phases of non-Hermitian physics.
Here, we comment on whether or how the choice of OBC (skin effect) affects features of entanglement dynamics compared to our study (periodic boundary condition (PBC)). 
In the quench dynamics under OBC, 
the non-reciprocal hopping makes 
the density dynamics 
asymmetrical in motion, 
which is also observed in the case of PBC, 
but the density is to be eventually localized at the boundary reflecting OBC.
Reference~\onlinecite{skin-EE} has demonstrated many-body HN model under OBC in the clean limit exhibits entanglement transition due to skin effect.
They have reported that entanglement entropy obeys logarithmic scaling, which is the same as the case of periodic boundary conditions.
However, the effective central charge is not equal to one.
They have analyzed this entanglement transition and shown that it originates from the skin effect.
Recently, both Refs.~\onlinecite{Skin-EE-transition,Skin-EE-transition2}  have investigated how disorder potential affects this entanglement transition.
Interestingly, they have reported that entanglement entropy exhibits non-monotonic behavior as a function of disorder strength in the case of OBC as well as that of PBC.
In the delocalized phase, entanglement entropy increases with the increase in disorder strength.
However, in the localized phase, entanglement entropy decreases with the increase in disorder strength.
While the origin of suppression of entanglement in the case of OBC is different from that of PBC, we consider that the quasi-particle picture and our discussion is still useful.
In the case of OBC, quasi-particle corresponds to skin mode, which is robust and localized at the boundary even if backscattering occurs.
However, as the disorder strength increases, the quasi-particle (skin mode) tends to move bidirectional rather than unidirectional motion due to backscattering as well as that of PBC, resulting in an increase in entanglement entropy.
Thus, disorder dependence of entanglement entropy is qualitatively independent of the choice of the boundary condition.

\section{Origin of the oscillatory behavior of $S_{\rm ent}$ in the weakly disordered regime}
\label{system-size}

\begin{figure}
\includegraphics[width=80mm]{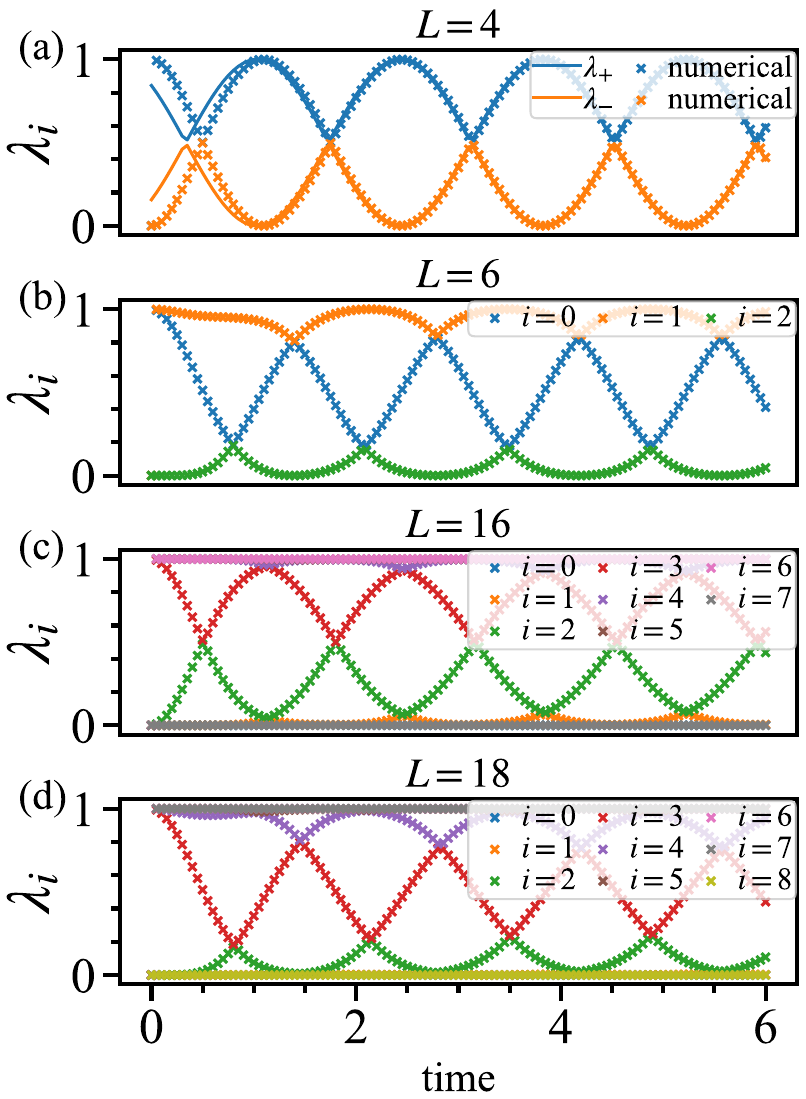}
\caption{
Time evolution of the eigenvalues $\lambda_k$
of correlation function~(\ref{correlation_append})
:
(a) $L=4$, (b) $L=6$, (c) $L=16$, and (d) $L=18$.
In panel (a), solid lines $\lambda_{\pm}=\frac{1}{2}\pm\frac{1}{2}\sqrt{\frac{1}{2}[\sin(4\cosh(g)t+\phi)+1]}$ are analytical solutions, where $\phi$ is a fitting parameter.
$g=0.5$.
}
\label{Fig-revise}
\end{figure}

Let us focus on the behavior of $S_{\rm ent}(t)$
in the weakly disordered regime
depicted in Fig.~\ref{Fig4}.
We have noticed that 
a small rapid oscillation is conspicuous
on top of its global tendency to saturate.
The oscillation is also rather conspicuous in the non-interacting case 
[Fig.~\ref{Fig4}, panel (a)], and
in the case of even number of particles $L/2=6, 8, 10$.
Here, we show that the oscillation stems from a 
two-fold degeneracy (in the imaginary part)
of the asymptotic state.

To simplify the argument, 
let us consider here
the non-interacting case, in which
$S_{\rm ent}$ is simply given by the
the eigenvalues $\lambda_i$
of the correlation function
in the subsystem (of size $\ell=L/2$)
\cite{conformal2}
\begin{eqnarray}
{\bf C}=
\begin{pmatrix} 
  \langle c_1^\dagger c_1\rangle& \dots  & \langle c_1^\dagger c_\ell\rangle \\
  \vdots & \ddots & \vdots \\
 \langle c_\ell^\dagger c_1 \rangle& \dots  & \langle c_\ell^\dagger c_\ell \rangle
\end{pmatrix},
\label{OPDM}    
\end{eqnarray}
as
\begin{eqnarray}
S_{\rm ent}=-\sum_{i=1}^{i=L/2}[\lambda_i\ln(\lambda_i)+(1-\lambda_i)\ln(1-\lambda_i)].
\label{EE-corr}    
\end{eqnarray}
The four panels of
Fig.~\ref{Fig-revise} shows
the behavior of numerically evaluated $\lambda_i$'s
in the cases of different number of particles $L/2$.
The plots show that 
there exists a qualitative difference in the behavior of $\lambda_i$'s
between the cases of $L/2$ even and odd;
%
in the case of $L/2$ even,
a pair of
$\lambda_i$'s appear 
symmetrically with respect to $\lambda=1/2$,
while
in the case of $L/2$ odd,
there exists no such a pairwise behavior.
Since
$\lambda_i$'s taking a value close to $\lambda=1/2$
gives the most relevant contribution to $S_{\rm ent}$,
one naturally expects that
such a pairwise behavior of $\lambda_i$'s
in the case of $L/2$ even
leads to
a conspicuous oscillation of $S_{\rm ent}$
in this case.

The reason why $S_{\rm ent}$ is oscillatory in the first place
may be understood in the following way.
In the asymptotic time regime $t\to\infty$
only the following two many-body states with
with a maximal imaginary part in the eigenenergy 
are relevant:
\begin{eqnarray}
&&|\Psi(t\to\infty)\rangle=\frac{1}{\sqrt{2}}(e^{-i\epsilon_{k=-\pi}t}c_{k=-\pi}^\dagger+e^{-i\epsilon_{k=0}t}c_{k=0}^\dagger)\nonumber\\
&&\times\Pi_{k=-2\pi/L\times(L/2-1)}^{k=-2\pi/L} e^{-i\epsilon_k t}c_k^\dagger|0\rangle.
\label{psit_append}    
\end{eqnarray}
One can estimate the
the correlation function (\ref{psit_append})
in this asymptotic regime
as
\begin{eqnarray}
&&{\bf C}_{n,m}=[\frac{1}{L}\sum e^{ik(n-m)}]+\frac{1}{2L}(e^{-i\pi(n-m)}+1)\nonumber\\
&&+\frac{1}{2L}(-1)^{L/2}(e^{-4i\cosh(g)t}e^{i\pi n}+e^{4i\cosh(g)t}e^{-i\pi m}).\nonumber\\
\label{correlation_append}    
\end{eqnarray}
This matrix can be easily diagonalized, e.g., in case of $L=4$.
In Fig.~\ref{Fig-revise} panel (a)
the analytic value of $\lambda_i$'s thus obtained are compared with
the ones found numerically.
The plots show that in most of the time regime considered in the figure
except the very early one around $t\simeq 0$
the two plots coincide, indicating that
the system is indeed controlled by the asymptotic state (\ref{psit_append}),
and
the two-fold degeneracy (in the imaginary part)
of the two relevant eigenstates
is the origin of the fast oscillation of $S_{\rm ent}$
in this case.

\section{Notes on the generalized Gibbs ensemble}
\label{Append-GGE}

In the limit of $t\to\infty$, $t_c(k)<t$ for all $k$ are satisfied, and then $|\Psi(t)\rangle$ reaches an equilibrium state.
Generally, we can obtain the corresponding statistical ensemble, assuming principle of maximum entropy under some constraint, such as expectation values of energy or total particles, using Lagrange multipliers.
In case of integrable systems, our target ($W=0$ and $V=0$), realized statistical ensemble is called generalized Gibbs ensemble (GGE), which forms maximum entropy under the constraint of $\hat{n}_k$.
GGE is defined as
\begin{equation}
\rho_{GGE}\equiv \frac{e^{-\sum_k \lambda_k\hat{n}_k}}{Z}
\label{GGE}
\end{equation}
where $Z=\Tr[e^{-\sum_k\lambda_k \hat{n}_k}]$, and $\lambda_k$ is the Lagrange multipliers that imposes constraint $\langle\Psi(t=0)|\hat{n}_k|\Psi(t=0)\rangle=\langle\Psi(\infty)|\hat{n}_k|\Psi(\infty)\rangle$.
GGE describes the expectation value of various quantities as well as the saturation value of $S_{\rm ent}$.
The statistical expectation value of $\hat{n}_k$ is defined by 
\begin{eqnarray}
\langle \hat{n}_k \rangle_{GGE}&&\equiv\Tr(\rho_{GGE}\hat{n}_k)\nonumber\\
&&=-\frac{\partial}{\partial \lambda_k }\log(Z)\nonumber\\
&&=\frac{1}{1+\exp(\lambda_k)}\nonumber\\
&&=\langle \Psi(t)| \hat{n}_k|\Psi(t)\rangle.
\label{GGE_nk}    
\end{eqnarray}
Additionally, the thermodynamic entropy of GGE is in accordance with the saturation value of $S_{\rm ent}$ in the thermodynamic limit, i.e.,
\begin{eqnarray}
S_{ent}(\infty)&=&\lim_{L\to\infty}S_{thermo}\nonumber\\
&\equiv&\lim_{L\to\infty}-\Tr\rho_{GGE}\ln\rho_{GGE}\nonumber\\
&=&\lim_{L\to\infty}\sum s_k\equiv\ell\int dk s(k).
\label{qpp_ee}
\end{eqnarray}
$ S_{\rm ent}(\infty)$ follows a volume-law ($S_{\rm ent}(\infty)\propto\ell$)
if most $\langle \hat{n}_k\rangle$ take neither 0 or 1, which is consistent to the fact that thermal entropy obeys volume-law.  
\begin{figure}
\includegraphics[width=85mm]{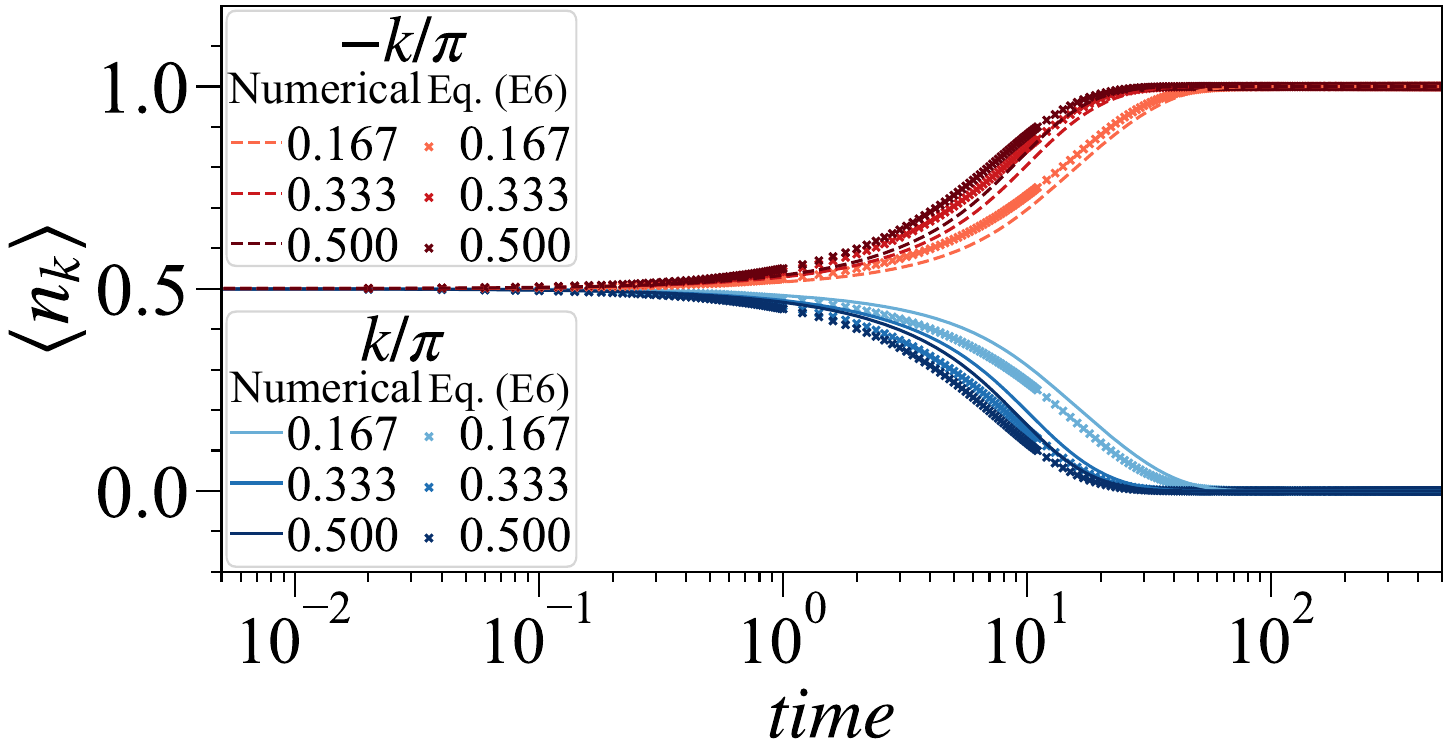}
\caption{Time evolution of $\langle \hat{n}_k\rangle$: $L=12$, $W=0$, and $V=0$.
The numerical result is obtained by averaging over $100$ different initial states (Eq.~(\ref{Eq:Append_A})).
Both solid and dashed line represent numerical result, while a scatter plots represents Eq.~(\ref{Eq:k,g-dependence-GGE2}).
}
\label{Fig11}
\end{figure}

\begin{figure}
\includegraphics[width=85mm]{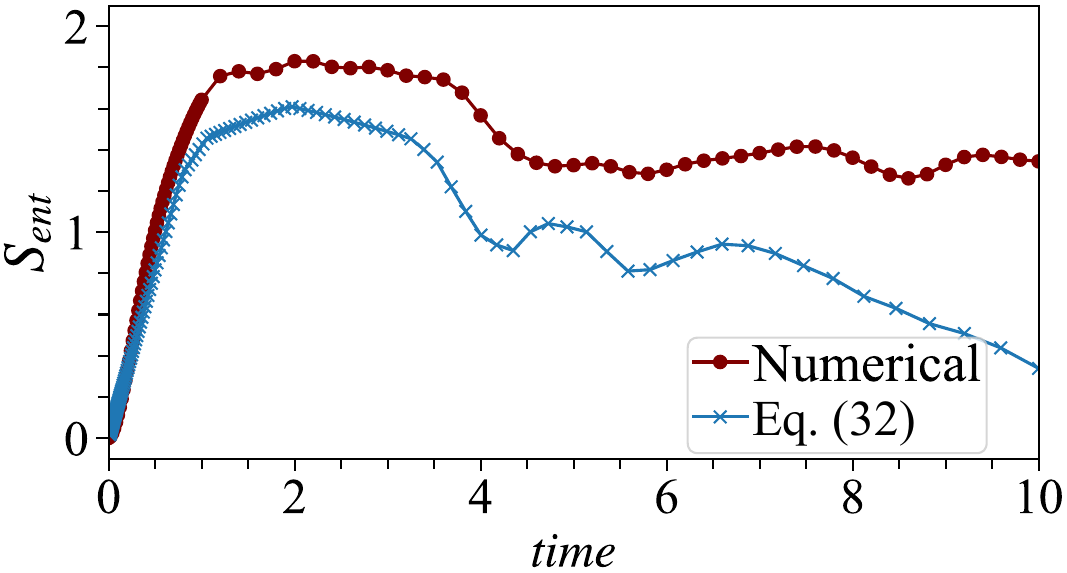}
\caption{Time evolution of $S_{\rm ent}$ (red solid line with circles) and 
the quasiparticle picture (Eq.~(\ref{qpp}), blue solid line with crosses): $L=16$, $W=0$, and, $V=0$.
We choose the DW state as the initial state.
We take into account the possibility of quasiparticles moving from the left (or right) end to the opposite end in the calculation Eq.~(\ref{qpp}) (for more details, refer to Ref.~\onlinecite{EE_revival}).
}
\label{Fig12}
\end{figure}



In the Hermitian case, since $\hat{n}_k$ is a conserved quantity, $\langle \Psi(t)|\hat{n}_k|\Psi(t)\rangle$ remains constant value over time, while $\langle \Psi(t)|\hat{n}_k|\Psi(t)\rangle$ varies during dynamics in the case of the many-body HN model due to the non-unitary time evolution $\partial_t \langle \hat{n}_k\rangle =i\langle\Psi(t)|H^\dagger\hat{n}_k-\hat{n}_kH|\Psi(t)\rangle\neq0$.
Let us focus on the relaxation of $\langle \Psi(t)|\hat{n}_k|\Psi(t)\rangle$, which is naively formulated by
\begin{eqnarray}
\langle\hat{n}_k\rangle &=& {\langle\Psi(0)|e^{iH^\dagger t}\hat{n}_ke^{-iHt}|\Psi(0)\rangle\over\langle\Psi(0)|e^{iH^\dagger t}e^{-iHt}|\Psi(0)\rangle}\nonumber\\
&=& {\Tr(|\Psi_{\{k\}}|^2\hat{n}_ke^{2\sum_k\rm{Im}(\epsilon_k)\hat{n}_kt})
\over\Tr(|\Psi_{\{k\}}|^2e^{2\sum_k\rm{Im}(\epsilon_k)\hat{n}_kt})},
\label{Eq:k,g-dependence}
\end{eqnarray}
where 
$\epsilon_k$ is a single particle eigenenergy (see Supplemental Material) and
$\Psi_{\{k\}}=\langle \{n_k\}|\Psi(0)\rangle=\langle n_{k_1}\cdots n_{k_L}|\Psi(0)\rangle$:
$|\{n_k\}\rangle$ represents the Fock space in momentum space. 
Here, we assume $|\Psi_{\{k\}}|^2$ is expressed as a GGE, so that
\begin{eqnarray}
\langle \hat{n}_k\rangle&=& {\Tr(|\Psi_{\{k\}}|^2\hat{n}_ke^{2\sum_k\rm{Im}(\epsilon_k)\hat{n}_kt})
\over\Tr(|\Psi_{\{k\}}|^2e^{2\sum_k\rm{Im}(\epsilon_k)\hat{n}_kt})}\nonumber\\
&\sim& {\Tr(\hat{n}_k e^{\sum_k (-\lambda_k+2\rm{Im}(\epsilon_k)t)\hat{n}_k}) \over Z},
\label{Eq:k,g-dependence-GGE}
\end{eqnarray}
where $Z=\Tr(e^{\sum_k(-\lambda_k+2\rm{Im}(\epsilon_k)t)\hat{n}_k})$,
and we assume superposition consists of various filling to use the knowledge of the grand canonical ensemble.
We can derives time dependent behavior of $\langle \hat{n}_k\rangle$,
which is defined as
\begin{eqnarray}
\langle \hat{n}_k\rangle
&&= {\Tr(\hat{n}_k e^{\sum_k (-\lambda_k+2\rm{Im}(\epsilon_k)t)\hat{n}_k}) \over Z}\nonumber\\
&&= -\frac{\partial}{\partial((\lambda_k-2\rm{Im}(\epsilon_k)t))}\log(Z)
\nonumber\\
&&=\frac{1}{1+e^{-2\rm{Im}(\epsilon_k)t}},
\label{Eq:k,g-dependence-GGE2}
\end{eqnarray}
where in the last line we take all $\lambda_k$ to be $0$, which is justified in case the initial state is prepared as DW-state.
Equation (\ref{Eq:k,g-dependence-GGE2}) implies that an imaginary eigenenergy either amplifies or decays a corresponding mode $\langle \hat{n}_k\rangle$ and this relaxation depends on the magnitude of Im$(\epsilon_k)$.



\section{Time dependence of $\langle \hat{n}_k\rangle$ for free-particle case}
\label{time-dependence nk}
In Sec.~\ref{Sec4}, we observed $\langle \hat{n}_k\rangle$ converge to stationary values more rapidly than Eq.~(\ref{Eq:k,g-dependence-GGE2}).
This discrepancy appears to arise from the fact that we assume $Q=\sum_i k_i$ takes the values ranging from $0$ to $L$ to derive an analytical expression of $\langle \hat{n}_k\rangle$, although we employ half-filling sector in actual numerical calculation.
To justify Eq.~(\ref{Eq:k,g-dependence-GGE2}),
we select the initial state as
\begin{eqnarray}
|\Psi(0)\rangle=\sum_{Q=0}^{Q=L}\frac{1}{\sqrt{L+1}}|\{n_k^Q\}\rangle,
\label{Eq:Append_A}
\end{eqnarray}
where $|\{n_k^Q\}\rangle$ is the Fock state that satisfies with $\sum_i k_i=Q$ and we randomly choose the Fock state $|\{n_k^Q\}\rangle$. 
Figure~\ref{Fig11} shows the time evolution of $\langle \hat{n}_k\rangle$ with the initial state given in Eq.~(\ref{Eq:Append_A}). The behavior of $\langle \hat{n}_k\rangle$ is closer to Eq.~(\ref{Eq:k,g-dependence-GGE2}) than the result shown in Fig.~\ref{Fig5}.

\begin{figure}
\includegraphics[width=85mm]{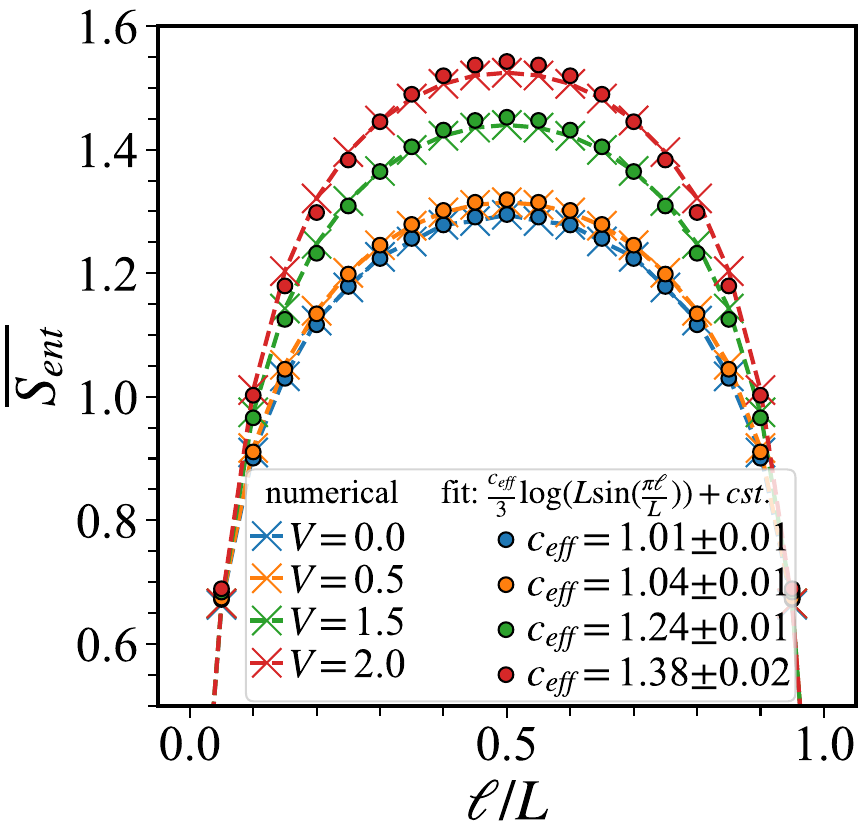}
\caption{
Scaling of $S_{\rm ent}$ as a function of $\ell/L$ with various values of $V$: $L=20$ and $W=0$.
A dashed line represents a numerical result, while a scatter plot represents a fitting curve.
}
\label{Fig13}
\end{figure}

\section{Quasiparticle picture for non-reciprocal system}
\label{Quasi}

In the HN model, quasiparticles decay or amplify under time evolution, leading to the question of when the quasiparticle picture becomes ill-defined. 
To address this question, we compare numerical results with the result suggested by the quasiparticle picture, as shown in Fig~\ref{Fig12}.
Initially, the result suggested by the quasiparticle picture agrees with the numerical result; however, as time evolves, it begins to converge to $0$, which differs from the numerical result.
This discrepancy stems from the assumption within the quasiparticle picture that $S_{\rm ent}$ behaves as thermal entropy, whereas in this case, $S_{\rm ent}$ actually characterizes quantum correlation.
Additionally, this discrepancy contrasts with a recent study in which the quasiparticle picture is used to describe the entanglement dynamics in the non-Hermitian system (PT-symmetric system).\cite{PT-quasi-particle}
As quasiparticles exhibit unidirectional motion in the HN model, the question of whether the quasiparticle picture quantitatively still describes entanglement dynamics is intriguing.
The quasiparticle picture can be compatible and generalized to many physical situations, such as an inhomogeneous initial state\cite{qpp-inhomo} and a state with no quasiparticle pair structure.\cite{no-pair}
Therefore, further study is necessary to generalize the quasiparticle picture to the HN model, which may become a framework for non-Hermitian GGE.

\section{The effect of interaction on the scaling of $S_{\rm ent}(\infty)$}
\label{central-charge}

Figure~\ref{Fig13} shows the saturation value of $S_{\rm ent}$ as a function $\ell/L$ with various values of $V$.
For weak $V$, a discrepancy between a numerical result (dashed line) and a fitting function (scatter plot), which is a form of Eq.~(\ref{log_sin_fit}), is negligible,
but it becomes more noticeable for large $V$.
Although a finite discrepancy exists for large $V$, a fitting function (Eq.~(\ref{log_sin_fit})) qualitatively characterizes numerical results, leading us to conclude that the scaling of $S_{\rm ent}$ is logarithmic.


\bibliography{ref_genko_221101}

\begin{thebibliography}{144}%
\makeatletter
\providecommand \@ifxundefined [1]{%
 \@ifx{#1\undefined}
}%
\providecommand \@ifnum [1]{%
 \ifnum #1\expandafter \@firstoftwo
 \else \expandafter \@secondoftwo
 \fi
}%
\providecommand \@ifx [1]{%
 \ifx #1\expandafter \@firstoftwo
 \else \expandafter \@secondoftwo
 \fi
}%
\providecommand \natexlab [1]{#1}%
\providecommand \enquote  [1]{``#1''}%
\providecommand \bibnamefont  [1]{#1}%
\providecommand \bibfnamefont [1]{#1}%
\providecommand \citenamefont [1]{#1}%
\providecommand \href@noop [0]{\@secondoftwo}%
\providecommand \href [0]{\begingroup \@sanitize@url \@href}%
\providecommand \@href[1]{\@@startlink{#1}\@@href}%
\providecommand \@@href[1]{\endgroup#1\@@endlink}%
\providecommand \@sanitize@url [0]{\catcode `\\12\catcode `\$12\catcode `\&12\catcode `\#12\catcode `\^12\catcode `\_12\catcode `\%12\relax}%
\providecommand \@@startlink[1]{}%
\providecommand \@@endlink[0]{}%
\providecommand \url  [0]{\begingroup\@sanitize@url \@url }%
\providecommand \@url [1]{\endgroup\@href {#1}{\urlprefix }}%
\providecommand \urlprefix  [0]{URL }%
\providecommand \Eprint [0]{\href }%
\providecommand \doibase [0]{http://dx.doi.org/}%
\providecommand \selectlanguage [0]{\@gobble}%
\providecommand \bibinfo  [0]{\@secondoftwo}%
\providecommand \bibfield  [0]{\@secondoftwo}%
\providecommand \translation [1]{[#1]}%
\providecommand \BibitemOpen [0]{}%
\providecommand \bibitemStop [0]{}%
\providecommand \bibitemNoStop [0]{.\EOS\space}%
\providecommand \EOS [0]{\spacefactor3000\relax}%
\providecommand \BibitemShut  [1]{\csname bibitem#1\endcsname}%
\let\auto@bib@innerbib\@empty
\bibitem [{\citenamefont {Einstein}\ \emph {et~al.}(1935)\citenamefont {Einstein}, \citenamefont {Podolsky},\ and\ \citenamefont {Rosen}}]{EPR}%
  \BibitemOpen
  \bibfield  {author} {\bibinfo {author} {\bibfnamefont {A.}~\bibnamefont {Einstein}}, \bibinfo {author} {\bibfnamefont {B.}~\bibnamefont {Podolsky}}, \ and\ \bibinfo {author} {\bibfnamefont {N.}~\bibnamefont {Rosen}},\ }\href {\doibase 10.1103/PhysRev.47.777} {\bibfield  {journal} {\bibinfo  {journal} {Phys. Rev.}\ }\textbf {\bibinfo {volume} {47}},\ \bibinfo {pages} {777} (\bibinfo {year} {1935})}\BibitemShut {NoStop}%
\bibitem [{\citenamefont {Freedman}\ and\ \citenamefont {Clauser}(1972)}]{nobel_c}%
  \BibitemOpen
  \bibfield  {author} {\bibinfo {author} {\bibfnamefont {S.~J.}\ \bibnamefont {Freedman}}\ and\ \bibinfo {author} {\bibfnamefont {J.~F.}\ \bibnamefont {Clauser}},\ }\href {\doibase 10.1103/PhysRevLett.28.938} {\bibfield  {journal} {\bibinfo  {journal} {Phys. Rev. Lett.}\ }\textbf {\bibinfo {volume} {28}},\ \bibinfo {pages} {938} (\bibinfo {year} {1972})}\BibitemShut {NoStop}%
\bibitem [{\citenamefont {Pan}\ \emph {et~al.}(1998)\citenamefont {Pan}, \citenamefont {Bouwmeester}, \citenamefont {Weinfurter},\ and\ \citenamefont {Zeilinger}}]{nobel_z}%
  \BibitemOpen
  \bibfield  {author} {\bibinfo {author} {\bibfnamefont {J.-W.}\ \bibnamefont {Pan}}, \bibinfo {author} {\bibfnamefont {D.}~\bibnamefont {Bouwmeester}}, \bibinfo {author} {\bibfnamefont {H.}~\bibnamefont {Weinfurter}}, \ and\ \bibinfo {author} {\bibfnamefont {A.}~\bibnamefont {Zeilinger}},\ }\href {\doibase 10.1103/PhysRevLett.80.3891} {\bibfield  {journal} {\bibinfo  {journal} {Phys. Rev. Lett.}\ }\textbf {\bibinfo {volume} {80}},\ \bibinfo {pages} {3891} (\bibinfo {year} {1998})}\BibitemShut {NoStop}%
\bibitem [{\citenamefont {Aspect}(1976)}]{aspect}%
  \BibitemOpen
  \bibfield  {author} {\bibinfo {author} {\bibfnamefont {A.}~\bibnamefont {Aspect}},\ }\href {\doibase 10.1103/PhysRevD.14.1944} {\bibfield  {journal} {\bibinfo  {journal} {Phys. Rev. D}\ }\textbf {\bibinfo {volume} {14}},\ \bibinfo {pages} {1944} (\bibinfo {year} {1976})}\BibitemShut {NoStop}%
\bibitem [{\citenamefont {Aspect}\ \emph {et~al.}(1982{\natexlab{a}})\citenamefont {Aspect}, \citenamefont {Dalibard},\ and\ \citenamefont {Roger}}]{aspect1?}%
  \BibitemOpen
  \bibfield  {author} {\bibinfo {author} {\bibfnamefont {A.}~\bibnamefont {Aspect}}, \bibinfo {author} {\bibfnamefont {J.}~\bibnamefont {Dalibard}}, \ and\ \bibinfo {author} {\bibfnamefont {G.}~\bibnamefont {Roger}},\ }\href {\doibase 10.1103/PhysRevLett.49.1804} {\bibfield  {journal} {\bibinfo  {journal} {Phys. Rev. Lett.}\ }\textbf {\bibinfo {volume} {49}},\ \bibinfo {pages} {1804} (\bibinfo {year} {1982}{\natexlab{a}})}\BibitemShut {NoStop}%
\bibitem [{\citenamefont {Aspect}\ \emph {et~al.}(1982{\natexlab{b}})\citenamefont {Aspect}, \citenamefont {Grangier},\ and\ \citenamefont {Roger}}]{aspect3}%
  \BibitemOpen
  \bibfield  {author} {\bibinfo {author} {\bibfnamefont {A.}~\bibnamefont {Aspect}}, \bibinfo {author} {\bibfnamefont {P.}~\bibnamefont {Grangier}}, \ and\ \bibinfo {author} {\bibfnamefont {G.}~\bibnamefont {Roger}},\ }\href {\doibase 10.1103/PhysRevLett.49.91} {\bibfield  {journal} {\bibinfo  {journal} {Phys. Rev. Lett.}\ }\textbf {\bibinfo {volume} {49}},\ \bibinfo {pages} {91} (\bibinfo {year} {1982}{\natexlab{b}})}\BibitemShut {NoStop}%
\bibitem [{\citenamefont {Deutsch}(1991)}]{ETH1}%
  \BibitemOpen
  \bibfield  {author} {\bibinfo {author} {\bibfnamefont {J.~M.}\ \bibnamefont {Deutsch}},\ }\href {\doibase 10.1103/PhysRevA.43.2046} {\bibfield  {journal} {\bibinfo  {journal} {Phys. Rev. A}\ }\textbf {\bibinfo {volume} {43}},\ \bibinfo {pages} {2046} (\bibinfo {year} {1991})}\BibitemShut {NoStop}%
\bibitem [{\citenamefont {Srednicki}(1994)}]{ETH2}%
  \BibitemOpen
  \bibfield  {author} {\bibinfo {author} {\bibfnamefont {M.}~\bibnamefont {Srednicki}},\ }\href {\doibase 10.1103/PhysRevE.50.888} {\bibfield  {journal} {\bibinfo  {journal} {Phys. Rev. E}\ }\textbf {\bibinfo {volume} {50}},\ \bibinfo {pages} {888} (\bibinfo {year} {1994})}\BibitemShut {NoStop}%
\bibitem [{\citenamefont {Rigol}\ \emph {et~al.}(2008)\citenamefont {Rigol}, \citenamefont {Dunjko},\ and\ \citenamefont {Olshanii}}]{ETH3}%
  \BibitemOpen
  \bibfield  {author} {\bibinfo {author} {\bibfnamefont {M.}~\bibnamefont {Rigol}}, \bibinfo {author} {\bibfnamefont {V.}~\bibnamefont {Dunjko}}, \ and\ \bibinfo {author} {\bibfnamefont {M.}~\bibnamefont {Olshanii}},\ }\href {\doibase 10.1038/nature06838} {\bibfield  {journal} {\bibinfo  {journal} {Nature}\ }\textbf {\bibinfo {volume} {452}},\ \bibinfo {pages} {854–858} (\bibinfo {year} {2008})}\BibitemShut {NoStop}%
\bibitem [{\citenamefont {Rigol}\ \emph {et~al.}(2007)\citenamefont {Rigol}, \citenamefont {Dunjko}, \citenamefont {Yurovsky},\ and\ \citenamefont {Olshanii}}]{GGE1}%
  \BibitemOpen
  \bibfield  {author} {\bibinfo {author} {\bibfnamefont {M.}~\bibnamefont {Rigol}}, \bibinfo {author} {\bibfnamefont {V.}~\bibnamefont {Dunjko}}, \bibinfo {author} {\bibfnamefont {V.}~\bibnamefont {Yurovsky}}, \ and\ \bibinfo {author} {\bibfnamefont {M.}~\bibnamefont {Olshanii}},\ }\href {\doibase 10.1103/PhysRevLett.98.050405} {\bibfield  {journal} {\bibinfo  {journal} {Phys. Rev. Lett.}\ }\textbf {\bibinfo {volume} {98}},\ \bibinfo {pages} {050405} (\bibinfo {year} {2007})}\BibitemShut {NoStop}%
\bibitem [{\citenamefont {Rigol}\ \emph {et~al.}(2006)\citenamefont {Rigol}, \citenamefont {Muramatsu},\ and\ \citenamefont {Olshanii}}]{GGE2}%
  \BibitemOpen
  \bibfield  {author} {\bibinfo {author} {\bibfnamefont {M.}~\bibnamefont {Rigol}}, \bibinfo {author} {\bibfnamefont {A.}~\bibnamefont {Muramatsu}}, \ and\ \bibinfo {author} {\bibfnamefont {M.}~\bibnamefont {Olshanii}},\ }\href {\doibase 10.1103/PhysRevA.74.053616} {\bibfield  {journal} {\bibinfo  {journal} {Phys. Rev. A}\ }\textbf {\bibinfo {volume} {74}},\ \bibinfo {pages} {053616} (\bibinfo {year} {2006})}\BibitemShut {NoStop}%
\bibitem [{\citenamefont {Iucci}\ and\ \citenamefont {Cazalilla}(2009)}]{GGE3}%
  \BibitemOpen
  \bibfield  {author} {\bibinfo {author} {\bibfnamefont {A.}~\bibnamefont {Iucci}}\ and\ \bibinfo {author} {\bibfnamefont {M.~A.}\ \bibnamefont {Cazalilla}},\ }\href {\doibase 10.1103/PhysRevA.80.063619} {\bibfield  {journal} {\bibinfo  {journal} {Phys. Rev. A}\ }\textbf {\bibinfo {volume} {80}},\ \bibinfo {pages} {063619} (\bibinfo {year} {2009})}\BibitemShut {NoStop}%
\bibitem [{\citenamefont {Vidmar}\ and\ \citenamefont {Rigol}(2016)}]{GGE4}%
  \BibitemOpen
  \bibfield  {author} {\bibinfo {author} {\bibfnamefont {L.}~\bibnamefont {Vidmar}}\ and\ \bibinfo {author} {\bibfnamefont {M.}~\bibnamefont {Rigol}},\ }\href {\doibase 10.1088/1742-5468/2016/06/064007} {\bibfield  {journal} {\bibinfo  {journal} {Journal of Statistical Mechanics: Theory and Experiment}\ }\textbf {\bibinfo {volume} {2016}},\ \bibinfo {pages} {064007} (\bibinfo {year} {2016})}\BibitemShut {NoStop}%
\bibitem [{\citenamefont {Calabrese}\ and\ \citenamefont {Cardy}(2005)}]{Calabrese1}%
  \BibitemOpen
  \bibfield  {author} {\bibinfo {author} {\bibfnamefont {P.}~\bibnamefont {Calabrese}}\ and\ \bibinfo {author} {\bibfnamefont {J.}~\bibnamefont {Cardy}},\ }\href {\doibase 10.1088/1742-5468/2005/04/p04010} {\bibfield  {journal} {\bibinfo  {journal} {Journal of Statistical Mechanics: Theory and Experiment}\ }\textbf {\bibinfo {volume} {2005}},\ \bibinfo {pages} {P04010} (\bibinfo {year} {2005})}\BibitemShut {NoStop}%
\bibitem [{\citenamefont {Alba}\ and\ \citenamefont {Calabrese}(2017)}]{quasi-particle1}%
  \BibitemOpen
  \bibfield  {author} {\bibinfo {author} {\bibfnamefont {V.}~\bibnamefont {Alba}}\ and\ \bibinfo {author} {\bibfnamefont {P.}~\bibnamefont {Calabrese}},\ }\href {\doibase 10.1073/pnas.1703516114} {\bibfield  {journal} {\bibinfo  {journal} {Proceedings of the National Academy of Sciences}\ }\textbf {\bibinfo {volume} {114}},\ \bibinfo {pages} {7947} (\bibinfo {year} {2017})}\BibitemShut {NoStop}%
\bibitem [{\citenamefont {Fagotti}\ and\ \citenamefont {Calabrese}(2008)}]{quasi-particle2}%
  \BibitemOpen
  \bibfield  {author} {\bibinfo {author} {\bibfnamefont {M.}~\bibnamefont {Fagotti}}\ and\ \bibinfo {author} {\bibfnamefont {P.}~\bibnamefont {Calabrese}},\ }\href {\doibase 10.1103/PhysRevA.78.010306} {\bibfield  {journal} {\bibinfo  {journal} {Phys. Rev. A}\ }\textbf {\bibinfo {volume} {78}},\ \bibinfo {pages} {010306} (\bibinfo {year} {2008})}\BibitemShut {NoStop}%
\bibitem [{\citenamefont {Chiara}\ \emph {et~al.}(2006)\citenamefont {Chiara}, \citenamefont {Montangero}, \citenamefont {Calabrese},\ and\ \citenamefont {Fazio}}]{quasi-particle3}%
  \BibitemOpen
  \bibfield  {author} {\bibinfo {author} {\bibfnamefont {G.~D.}\ \bibnamefont {Chiara}}, \bibinfo {author} {\bibfnamefont {S.}~\bibnamefont {Montangero}}, \bibinfo {author} {\bibfnamefont {P.}~\bibnamefont {Calabrese}}, \ and\ \bibinfo {author} {\bibfnamefont {R.}~\bibnamefont {Fazio}},\ }\href {\doibase 10.1088/1742-5468/2006/03/p03001} {\bibfield  {journal} {\bibinfo  {journal} {Journal of Statistical Mechanics: Theory and Experiment}\ }\textbf {\bibinfo {volume} {2006}},\ \bibinfo {pages} {P03001} (\bibinfo {year} {2006})}\BibitemShut {NoStop}%
\bibitem [{\citenamefont {Calabrese}(2018)}]{quasi-particle4}%
  \BibitemOpen
  \bibfield  {author} {\bibinfo {author} {\bibfnamefont {P.}~\bibnamefont {Calabrese}},\ }\href {\doibase https://doi.org/10.1016/j.physa.2017.10.011} {\bibfield  {journal} {\bibinfo  {journal} {Physica A: Statistical Mechanics and its Applications}\ }\textbf {\bibinfo {volume} {504}},\ \bibinfo {pages} {31} (\bibinfo {year} {2018})},\ \bibinfo {note} {lecture Notes of the 14th International Summer School on Fundamental Problems in Statistical Physics}\BibitemShut {NoStop}%
\bibitem [{\citenamefont {Nandkishore}\ and\ \citenamefont {Huse}(2015)}]{MBL_review}%
  \BibitemOpen
  \bibfield  {author} {\bibinfo {author} {\bibfnamefont {R.}~\bibnamefont {Nandkishore}}\ and\ \bibinfo {author} {\bibfnamefont {D.~A.}\ \bibnamefont {Huse}},\ }\href {\doibase 10.1146/annurev-conmatphys-031214-014726} {\bibfield  {journal} {\bibinfo  {journal} {Annual Review of Condensed Matter Physics}\ }\textbf {\bibinfo {volume} {6}},\ \bibinfo {pages} {15} (\bibinfo {year} {2015})}\BibitemShut {NoStop}%
\bibitem [{\citenamefont {Abanin}\ \emph {et~al.}(2019)\citenamefont {Abanin}, \citenamefont {Altman}, \citenamefont {Bloch},\ and\ \citenamefont {Serbyn}}]{MBL5}%
  \BibitemOpen
  \bibfield  {author} {\bibinfo {author} {\bibfnamefont {D.~A.}\ \bibnamefont {Abanin}}, \bibinfo {author} {\bibfnamefont {E.}~\bibnamefont {Altman}}, \bibinfo {author} {\bibfnamefont {I.}~\bibnamefont {Bloch}}, \ and\ \bibinfo {author} {\bibfnamefont {M.}~\bibnamefont {Serbyn}},\ }\href {\doibase 10.1103/RevModPhys.91.021001} {\bibfield  {journal} {\bibinfo  {journal} {Rev. Mod. Phys.}\ }\textbf {\bibinfo {volume} {91}},\ \bibinfo {pages} {021001} (\bibinfo {year} {2019})}\BibitemShut {NoStop}%
\bibitem [{\citenamefont {Page}(1993)}]{Pagesan}%
  \BibitemOpen
  \bibfield  {author} {\bibinfo {author} {\bibfnamefont {D.~N.}\ \bibnamefont {Page}},\ }\href {\doibase 10.1103/PhysRevLett.71.1291} {\bibfield  {journal} {\bibinfo  {journal} {Phys. Rev. Lett.}\ }\textbf {\bibinfo {volume} {71}},\ \bibinfo {pages} {1291} (\bibinfo {year} {1993})}\BibitemShut {NoStop}%
\bibitem [{\citenamefont {Turkeshi}\ and\ \citenamefont {Schir\'o}(2023)}]{Marco1}%
  \BibitemOpen
  \bibfield  {author} {\bibinfo {author} {\bibfnamefont {X.}~\bibnamefont {Turkeshi}}\ and\ \bibinfo {author} {\bibfnamefont {M.}~\bibnamefont {Schir\'o}},\ }\href {\doibase 10.1103/PhysRevB.107.L020403} {\bibfield  {journal} {\bibinfo  {journal} {Phys. Rev. B}\ }\textbf {\bibinfo {volume} {107}},\ \bibinfo {pages} {L020403} (\bibinfo {year} {2023})}\BibitemShut {NoStop}%
\bibitem [{\citenamefont {Skinner}\ \emph {et~al.}(2019)\citenamefont {Skinner}, \citenamefont {Ruhman},\ and\ \citenamefont {Nahum}}]{MIP}%
  \BibitemOpen
  \bibfield  {author} {\bibinfo {author} {\bibfnamefont {B.}~\bibnamefont {Skinner}}, \bibinfo {author} {\bibfnamefont {J.}~\bibnamefont {Ruhman}}, \ and\ \bibinfo {author} {\bibfnamefont {A.}~\bibnamefont {Nahum}},\ }\href {\doibase 10.1103/physrevx.9.031009} {\bibfield  {journal} {\bibinfo  {journal} {Physical Review X}\ }\textbf {\bibinfo {volume} {9}} (\bibinfo {year} {2019}),\ 10.1103/physrevx.9.031009}\BibitemShut {NoStop}%
\bibitem [{\citenamefont {Fuji}\ and\ \citenamefont {Ashida}(2020)}]{Fuji-Ashida}%
  \BibitemOpen
  \bibfield  {author} {\bibinfo {author} {\bibfnamefont {Y.}~\bibnamefont {Fuji}}\ and\ \bibinfo {author} {\bibfnamefont {Y.}~\bibnamefont {Ashida}},\ }\href {\doibase 10.1103/PhysRevB.102.054302} {\bibfield  {journal} {\bibinfo  {journal} {Phys. Rev. B}\ }\textbf {\bibinfo {volume} {102}},\ \bibinfo {pages} {054302} (\bibinfo {year} {2020})}\BibitemShut {NoStop}%
\bibitem [{\citenamefont {B\'acsi}\ and\ \citenamefont {D\'ora}(2021)}]{PT-quasi-particle}%
  \BibitemOpen
  \bibfield  {author} {\bibinfo {author} {\bibfnamefont {A.}~\bibnamefont {B\'acsi}}\ and\ \bibinfo {author} {\bibfnamefont {B.}~\bibnamefont {D\'ora}},\ }\href {\doibase 10.1103/PhysRevB.103.085137} {\bibfield  {journal} {\bibinfo  {journal} {Phys. Rev. B}\ }\textbf {\bibinfo {volume} {103}},\ \bibinfo {pages} {085137} (\bibinfo {year} {2021})}\BibitemShut {NoStop}%
\bibitem [{\citenamefont {Gal}\ \emph {et~al.}(2023)\citenamefont {Gal}, \citenamefont {Turkeshi},\ and\ \citenamefont {Schirò}}]{Marco2}%
  \BibitemOpen
  \bibfield  {author} {\bibinfo {author} {\bibfnamefont {Y.~L.}\ \bibnamefont {Gal}}, \bibinfo {author} {\bibfnamefont {X.}~\bibnamefont {Turkeshi}}, \ and\ \bibinfo {author} {\bibfnamefont {M.}~\bibnamefont {Schirò}},\ }\href {\doibase 10.21468/SciPostPhys.14.5.138} {\bibfield  {journal} {\bibinfo  {journal} {SciPost Phys.}\ }\textbf {\bibinfo {volume} {14}},\ \bibinfo {pages} {138} (\bibinfo {year} {2023})}\BibitemShut {NoStop}%
\bibitem [{\citenamefont {Li}\ \emph {et~al.}(2019)\citenamefont {Li}, \citenamefont {Chen},\ and\ \citenamefont {Fisher}}]{MIP1}%
  \BibitemOpen
  \bibfield  {author} {\bibinfo {author} {\bibfnamefont {Y.}~\bibnamefont {Li}}, \bibinfo {author} {\bibfnamefont {X.}~\bibnamefont {Chen}}, \ and\ \bibinfo {author} {\bibfnamefont {M.~P.~A.}\ \bibnamefont {Fisher}},\ }\href {\doibase 10.1103/PhysRevB.100.134306} {\bibfield  {journal} {\bibinfo  {journal} {Phys. Rev. B}\ }\textbf {\bibinfo {volume} {100}},\ \bibinfo {pages} {134306} (\bibinfo {year} {2019})}\BibitemShut {NoStop}%
\bibitem [{\citenamefont {Chan}\ \emph {et~al.}(2019)\citenamefont {Chan}, \citenamefont {Nandkishore}, \citenamefont {Pretko},\ and\ \citenamefont {Smith}}]{MIP2}%
  \BibitemOpen
  \bibfield  {author} {\bibinfo {author} {\bibfnamefont {A.}~\bibnamefont {Chan}}, \bibinfo {author} {\bibfnamefont {R.~M.}\ \bibnamefont {Nandkishore}}, \bibinfo {author} {\bibfnamefont {M.}~\bibnamefont {Pretko}}, \ and\ \bibinfo {author} {\bibfnamefont {G.}~\bibnamefont {Smith}},\ }\href {\doibase 10.1103/PhysRevB.99.224307} {\bibfield  {journal} {\bibinfo  {journal} {Phys. Rev. B}\ }\textbf {\bibinfo {volume} {99}},\ \bibinfo {pages} {224307} (\bibinfo {year} {2019})}\BibitemShut {NoStop}%
\bibitem [{\citenamefont {Choi}\ \emph {et~al.}(2020)\citenamefont {Choi}, \citenamefont {Bao}, \citenamefont {Qi},\ and\ \citenamefont {Altman}}]{MIP3}%
  \BibitemOpen
  \bibfield  {author} {\bibinfo {author} {\bibfnamefont {S.}~\bibnamefont {Choi}}, \bibinfo {author} {\bibfnamefont {Y.}~\bibnamefont {Bao}}, \bibinfo {author} {\bibfnamefont {X.-L.}\ \bibnamefont {Qi}}, \ and\ \bibinfo {author} {\bibfnamefont {E.}~\bibnamefont {Altman}},\ }\href {\doibase 10.1103/PhysRevLett.125.030505} {\bibfield  {journal} {\bibinfo  {journal} {Phys. Rev. Lett.}\ }\textbf {\bibinfo {volume} {125}},\ \bibinfo {pages} {030505} (\bibinfo {year} {2020})}\BibitemShut {NoStop}%
\bibitem [{\citenamefont {Zabalo}\ \emph {et~al.}(2020)\citenamefont {Zabalo}, \citenamefont {Gullans}, \citenamefont {Wilson}, \citenamefont {Gopalakrishnan}, \citenamefont {Huse},\ and\ \citenamefont {Pixley}}]{MIP4}%
  \BibitemOpen
  \bibfield  {author} {\bibinfo {author} {\bibfnamefont {A.}~\bibnamefont {Zabalo}}, \bibinfo {author} {\bibfnamefont {M.~J.}\ \bibnamefont {Gullans}}, \bibinfo {author} {\bibfnamefont {J.~H.}\ \bibnamefont {Wilson}}, \bibinfo {author} {\bibfnamefont {S.}~\bibnamefont {Gopalakrishnan}}, \bibinfo {author} {\bibfnamefont {D.~A.}\ \bibnamefont {Huse}}, \ and\ \bibinfo {author} {\bibfnamefont {J.~H.}\ \bibnamefont {Pixley}},\ }\href {\doibase 10.1103/PhysRevB.101.060301} {\bibfield  {journal} {\bibinfo  {journal} {Phys. Rev. B}\ }\textbf {\bibinfo {volume} {101}},\ \bibinfo {pages} {060301} (\bibinfo {year} {2020})}\BibitemShut {NoStop}%
\bibitem [{\citenamefont {Turkeshi}\ \emph {et~al.}(2020)\citenamefont {Turkeshi}, \citenamefont {Fazio},\ and\ \citenamefont {Dalmonte}}]{MIP5}%
  \BibitemOpen
  \bibfield  {author} {\bibinfo {author} {\bibfnamefont {X.}~\bibnamefont {Turkeshi}}, \bibinfo {author} {\bibfnamefont {R.}~\bibnamefont {Fazio}}, \ and\ \bibinfo {author} {\bibfnamefont {M.}~\bibnamefont {Dalmonte}},\ }\href {\doibase 10.1103/PhysRevB.102.014315} {\bibfield  {journal} {\bibinfo  {journal} {Phys. Rev. B}\ }\textbf {\bibinfo {volume} {102}},\ \bibinfo {pages} {014315} (\bibinfo {year} {2020})}\BibitemShut {NoStop}%
\bibitem [{\citenamefont {Sang}\ and\ \citenamefont {Hsieh}(2021)}]{MIP6}%
  \BibitemOpen
  \bibfield  {author} {\bibinfo {author} {\bibfnamefont {S.}~\bibnamefont {Sang}}\ and\ \bibinfo {author} {\bibfnamefont {T.~H.}\ \bibnamefont {Hsieh}},\ }\href {\doibase 10.1103/PhysRevResearch.3.023200} {\bibfield  {journal} {\bibinfo  {journal} {Phys. Rev. Res.}\ }\textbf {\bibinfo {volume} {3}},\ \bibinfo {pages} {023200} (\bibinfo {year} {2021})}\BibitemShut {NoStop}%
\bibitem [{\citenamefont {Yang}\ \emph {et~al.}(2023)\citenamefont {Yang}, \citenamefont {Mao},\ and\ \citenamefont {Jian}}]{MIP7}%
  \BibitemOpen
  \bibfield  {author} {\bibinfo {author} {\bibfnamefont {Z.}~\bibnamefont {Yang}}, \bibinfo {author} {\bibfnamefont {D.}~\bibnamefont {Mao}}, \ and\ \bibinfo {author} {\bibfnamefont {C.-M.}\ \bibnamefont {Jian}},\ }\href@noop {} {\enquote {\bibinfo {title} {Entanglement in one-dimensional critical state after measurements},}\ } (\bibinfo {year} {2023}),\ \Eprint {http://arxiv.org/abs/2301.08255} {arXiv:2301.08255 [quant-ph]} \BibitemShut {NoStop}%
\bibitem [{\citenamefont {Noel}\ \emph {et~al.}(2022)\citenamefont {Noel}, \citenamefont {Niroula}, \citenamefont {Zhu}, \citenamefont {Risinger}, \citenamefont {Egan}, \citenamefont {Biswas}, \citenamefont {Cetina}, \citenamefont {Gorshkov}, \citenamefont {Gullans}, \citenamefont {Huse},\ and\ \citenamefont {Monroe}}]{MIP_experiment}%
  \BibitemOpen
  \bibfield  {author} {\bibinfo {author} {\bibfnamefont {C.}~\bibnamefont {Noel}}, \bibinfo {author} {\bibfnamefont {P.}~\bibnamefont {Niroula}}, \bibinfo {author} {\bibfnamefont {D.}~\bibnamefont {Zhu}}, \bibinfo {author} {\bibfnamefont {A.}~\bibnamefont {Risinger}}, \bibinfo {author} {\bibfnamefont {L.}~\bibnamefont {Egan}}, \bibinfo {author} {\bibfnamefont {D.}~\bibnamefont {Biswas}}, \bibinfo {author} {\bibfnamefont {M.}~\bibnamefont {Cetina}}, \bibinfo {author} {\bibfnamefont {A.~V.}\ \bibnamefont {Gorshkov}}, \bibinfo {author} {\bibfnamefont {M.~J.}\ \bibnamefont {Gullans}}, \bibinfo {author} {\bibfnamefont {D.~A.}\ \bibnamefont {Huse}}, \ and\ \bibinfo {author} {\bibfnamefont {C.}~\bibnamefont {Monroe}},\ }\href {\doibase 10.1038/s41567-022-01619-7} {\bibfield  {journal} {\bibinfo  {journal} {Nature Physics}\ }\textbf {\bibinfo {volume} {18}},\ \bibinfo {pages} {760} (\bibinfo {year} {2022})}\BibitemShut {NoStop}%
\bibitem [{\citenamefont {Koh}\ \emph {et~al.}(2023)\citenamefont {Koh}, \citenamefont {Sun}, \citenamefont {Motta},\ and\ \citenamefont {Minnich}}]{MIP_exp2}%
  \BibitemOpen
  \bibfield  {author} {\bibinfo {author} {\bibfnamefont {J.~M.}\ \bibnamefont {Koh}}, \bibinfo {author} {\bibfnamefont {S.-N.}\ \bibnamefont {Sun}}, \bibinfo {author} {\bibfnamefont {M.}~\bibnamefont {Motta}}, \ and\ \bibinfo {author} {\bibfnamefont {A.~J.}\ \bibnamefont {Minnich}},\ }\href {\doibase 10.1038/s41567-023-02076-6} {\bibfield  {journal} {\bibinfo  {journal} {Nature Physics}\ } (\bibinfo {year} {2023}),\ 10.1038/s41567-023-02076-6}\BibitemShut {NoStop}%
\bibitem [{\citenamefont {Vasseur}\ \emph {et~al.}(2019)\citenamefont {Vasseur}, \citenamefont {Potter}, \citenamefont {You},\ and\ \citenamefont {Ludwig}}]{ET1}%
  \BibitemOpen
  \bibfield  {author} {\bibinfo {author} {\bibfnamefont {R.}~\bibnamefont {Vasseur}}, \bibinfo {author} {\bibfnamefont {A.~C.}\ \bibnamefont {Potter}}, \bibinfo {author} {\bibfnamefont {Y.-Z.}\ \bibnamefont {You}}, \ and\ \bibinfo {author} {\bibfnamefont {A.~W.~W.}\ \bibnamefont {Ludwig}},\ }\href {\doibase 10.1103/PhysRevB.100.134203} {\bibfield  {journal} {\bibinfo  {journal} {Phys. Rev. B}\ }\textbf {\bibinfo {volume} {100}},\ \bibinfo {pages} {134203} (\bibinfo {year} {2019})}\BibitemShut {NoStop}%
\bibitem [{\citenamefont {Gullans}\ and\ \citenamefont {Huse}(2020)}]{PuriT1}%
  \BibitemOpen
  \bibfield  {author} {\bibinfo {author} {\bibfnamefont {M.~J.}\ \bibnamefont {Gullans}}\ and\ \bibinfo {author} {\bibfnamefont {D.~A.}\ \bibnamefont {Huse}},\ }\href {\doibase 10.1103/PhysRevX.10.041020} {\bibfield  {journal} {\bibinfo  {journal} {Phys. Rev. X}\ }\textbf {\bibinfo {volume} {10}},\ \bibinfo {pages} {041020} (\bibinfo {year} {2020})}\BibitemShut {NoStop}%
\bibitem [{\citenamefont {Kuno}\ \emph {et~al.}(2022)\citenamefont {Kuno}, \citenamefont {Orito},\ and\ \citenamefont {Ichinose}}]{PuriT2}%
  \BibitemOpen
  \bibfield  {author} {\bibinfo {author} {\bibfnamefont {Y.}~\bibnamefont {Kuno}}, \bibinfo {author} {\bibfnamefont {T.}~\bibnamefont {Orito}}, \ and\ \bibinfo {author} {\bibfnamefont {I.}~\bibnamefont {Ichinose}},\ }\href {\doibase 10.1103/PhysRevB.106.214304} {\bibfield  {journal} {\bibinfo  {journal} {Phys. Rev. B}\ }\textbf {\bibinfo {volume} {106}},\ \bibinfo {pages} {214304} (\bibinfo {year} {2022})}\BibitemShut {NoStop}%
\bibitem [{\citenamefont {Hatano}\ and\ \citenamefont {Nelson}(1997)}]{HN_PRB97}%
  \BibitemOpen
  \bibfield  {author} {\bibinfo {author} {\bibfnamefont {N.}~\bibnamefont {Hatano}}\ and\ \bibinfo {author} {\bibfnamefont {D.~R.}\ \bibnamefont {Nelson}},\ }\href {\doibase 10.1103/PhysRevB.56.8651} {\bibfield  {journal} {\bibinfo  {journal} {Phys. Rev. B}\ }\textbf {\bibinfo {volume} {56}},\ \bibinfo {pages} {8651} (\bibinfo {year} {1997})}\BibitemShut {NoStop}%
\bibitem [{\citenamefont {Hatano}\ and\ \citenamefont {Nelson}(1998)}]{HN_PRB98}%
  \BibitemOpen
  \bibfield  {author} {\bibinfo {author} {\bibfnamefont {N.}~\bibnamefont {Hatano}}\ and\ \bibinfo {author} {\bibfnamefont {D.~R.}\ \bibnamefont {Nelson}},\ }\href {\doibase 10.1103/PhysRevB.58.8384} {\bibfield  {journal} {\bibinfo  {journal} {Phys. Rev. B}\ }\textbf {\bibinfo {volume} {58}},\ \bibinfo {pages} {8384} (\bibinfo {year} {1998})}\BibitemShut {NoStop}%
\bibitem [{\citenamefont {Hatano}\ and\ \citenamefont {Nelson}(1996)}]{HN_PRL}%
  \BibitemOpen
  \bibfield  {author} {\bibinfo {author} {\bibfnamefont {N.}~\bibnamefont {Hatano}}\ and\ \bibinfo {author} {\bibfnamefont {D.~R.}\ \bibnamefont {Nelson}},\ }\href {\doibase 10.1103/PhysRevLett.77.570} {\bibfield  {journal} {\bibinfo  {journal} {Phys. Rev. Lett.}\ }\textbf {\bibinfo {volume} {77}},\ \bibinfo {pages} {570} (\bibinfo {year} {1996})}\BibitemShut {NoStop}%
\bibitem [{\citenamefont {Longhi}(2021)}]{Longhi_AA}%
  \BibitemOpen
  \bibfield  {author} {\bibinfo {author} {\bibfnamefont {S.}~\bibnamefont {Longhi}},\ }\href {\doibase 10.1103/PhysRevB.103.054203} {\bibfield  {journal} {\bibinfo  {journal} {Phys. Rev. B}\ }\textbf {\bibinfo {volume} {103}},\ \bibinfo {pages} {054203} (\bibinfo {year} {2021})}\BibitemShut {NoStop}%
\bibitem [{Note1()}]{Note1}%
  \BibitemOpen
  \bibinfo {note} {See Supplemental Material at [URL will be inserted by the publisher] for a numerical demonstration and detailed explanation of wave-packet dynamics. The Supplemental Material also contains Refs.~\protect \rev@citealpnum {SM-revise1,SM-revise2,thouless}.}\BibitemShut {Stop}%
\bibitem [{\citenamefont {Orito}\ and\ \citenamefont {Imura}(2022)}]{OI22A}%
  \BibitemOpen
  \bibfield  {author} {\bibinfo {author} {\bibfnamefont {T.}~\bibnamefont {Orito}}\ and\ \bibinfo {author} {\bibfnamefont {K.-I.}\ \bibnamefont {Imura}},\ }\href {\doibase 10.1103/PhysRevB.105.024303} {\bibfield  {journal} {\bibinfo  {journal} {Phys. Rev. B}\ }\textbf {\bibinfo {volume} {105}},\ \bibinfo {pages} {024303} (\bibinfo {year} {2022})}\BibitemShut {NoStop}%
\bibitem [{\citenamefont {Saad}(1992)}]{Arnoldi}%
  \BibitemOpen
  \bibfield  {author} {\bibinfo {author} {\bibfnamefont {Y.}~\bibnamefont {Saad}},\ }\href {\doibase 10.1137/0729014} {\bibfield  {journal} {\bibinfo  {journal} {SIAM Journal on Numerical Analysis}\ }\textbf {\bibinfo {volume} {29}},\ \bibinfo {pages} {209} (\bibinfo {year} {1992})},\ \Eprint {http://arxiv.org/abs/https://doi.org/10.1137/0729014} {https://doi.org/10.1137/0729014} \BibitemShut {NoStop}%
\bibitem [{\citenamefont {Aubry}\ and\ \citenamefont {André}(1980)}]{AA}%
  \BibitemOpen
  \bibfield  {author} {\bibinfo {author} {\bibfnamefont {S.}~\bibnamefont {Aubry}}\ and\ \bibinfo {author} {\bibfnamefont {G.}~\bibnamefont {André}},\ }\href@noop {} {\bibfield  {journal} {\bibinfo  {journal} {Ann. Israel Phys. Soc.}\ }\textbf {\bibinfo {volume} {3}},\ \bibinfo {pages} {133} (\bibinfo {year} {1980})}\BibitemShut {NoStop}%
\bibitem [{\citenamefont {Gong}\ \emph {et~al.}(2018)\citenamefont {Gong}, \citenamefont {Ashida}, \citenamefont {Kawabata}, \citenamefont {Takasan}, \citenamefont {Higashikawa},\ and\ \citenamefont {Ueda}}]{PhysRevX_Gong}%
  \BibitemOpen
  \bibfield  {author} {\bibinfo {author} {\bibfnamefont {Z.}~\bibnamefont {Gong}}, \bibinfo {author} {\bibfnamefont {Y.}~\bibnamefont {Ashida}}, \bibinfo {author} {\bibfnamefont {K.}~\bibnamefont {Kawabata}}, \bibinfo {author} {\bibfnamefont {K.}~\bibnamefont {Takasan}}, \bibinfo {author} {\bibfnamefont {S.}~\bibnamefont {Higashikawa}}, \ and\ \bibinfo {author} {\bibfnamefont {M.}~\bibnamefont {Ueda}},\ }\href {\doibase 10.1103/PhysRevX.8.031079} {\bibfield  {journal} {\bibinfo  {journal} {Phys. Rev. X}\ }\textbf {\bibinfo {volume} {8}},\ \bibinfo {pages} {031079} (\bibinfo {year} {2018})}\BibitemShut {NoStop}%
\bibitem [{\citenamefont {Yao}\ and\ \citenamefont {Wang}(2018)}]{YW}%
  \BibitemOpen
  \bibfield  {author} {\bibinfo {author} {\bibfnamefont {S.}~\bibnamefont {Yao}}\ and\ \bibinfo {author} {\bibfnamefont {Z.}~\bibnamefont {Wang}},\ }\href {\doibase 10.1103/PhysRevLett.121.086803} {\bibfield  {journal} {\bibinfo  {journal} {Phys. Rev. Lett.}\ }\textbf {\bibinfo {volume} {121}},\ \bibinfo {pages} {086803} (\bibinfo {year} {2018})}\BibitemShut {NoStop}%
\bibitem [{\citenamefont {Yokomizo}\ and\ \citenamefont {Murakami}(2019)}]{YM}%
  \BibitemOpen
  \bibfield  {author} {\bibinfo {author} {\bibfnamefont {K.}~\bibnamefont {Yokomizo}}\ and\ \bibinfo {author} {\bibfnamefont {S.}~\bibnamefont {Murakami}},\ }\href {\doibase 10.1103/PhysRevLett.123.066404} {\bibfield  {journal} {\bibinfo  {journal} {Phys. Rev. Lett.}\ }\textbf {\bibinfo {volume} {123}},\ \bibinfo {pages} {066404} (\bibinfo {year} {2019})}\BibitemShut {NoStop}%
\bibitem [{\citenamefont {Imura}\ and\ \citenamefont {Takane}(2019)}]{imura1}%
  \BibitemOpen
  \bibfield  {author} {\bibinfo {author} {\bibfnamefont {K.-I.}\ \bibnamefont {Imura}}\ and\ \bibinfo {author} {\bibfnamefont {Y.}~\bibnamefont {Takane}},\ }\href {\doibase 10.1103/PhysRevB.100.165430} {\bibfield  {journal} {\bibinfo  {journal} {Phys. Rev. B}\ }\textbf {\bibinfo {volume} {100}},\ \bibinfo {pages} {165430} (\bibinfo {year} {2019})}\BibitemShut {NoStop}%
\bibitem [{\citenamefont {Imura}\ and\ \citenamefont {Takane}(2020)}]{imura2}%
  \BibitemOpen
  \bibfield  {author} {\bibinfo {author} {\bibfnamefont {K.-I.}\ \bibnamefont {Imura}}\ and\ \bibinfo {author} {\bibfnamefont {Y.}~\bibnamefont {Takane}},\ }\href {\doibase 10.1093/ptep/ptaa100} {\bibfield  {journal} {\bibinfo  {journal} {Progress of Theoretical and Experimental Physics}\ }\textbf {\bibinfo {volume} {2020}} (\bibinfo {year} {2020}),\ 10.1093/ptep/ptaa100},\ \bibinfo {note} {12A103},\ \Eprint {http://arxiv.org/abs/https://academic.oup.com/ptep/article-pdf/2020/12/12A103/35611802/ptaa100.pdf} {https://academic.oup.com/ptep/article-pdf/2020/12/12A103/35611802/ptaa100.pdf} \BibitemShut {NoStop}%
\bibitem [{\citenamefont {Ashida}\ \emph {et~al.}(2020)\citenamefont {Ashida}, \citenamefont {Gong},\ and\ \citenamefont {Ueda}}]{Ashida_2020}%
  \BibitemOpen
  \bibfield  {author} {\bibinfo {author} {\bibfnamefont {Y.}~\bibnamefont {Ashida}}, \bibinfo {author} {\bibfnamefont {Z.}~\bibnamefont {Gong}}, \ and\ \bibinfo {author} {\bibfnamefont {M.}~\bibnamefont {Ueda}},\ }\href {\doibase 10.1080/00018732.2021.1876991} {\bibfield  {journal} {\bibinfo  {journal} {Advances in Physics}\ }\textbf {\bibinfo {volume} {69}},\ \bibinfo {pages} {249} (\bibinfo {year} {2020})}\BibitemShut {NoStop}%
\bibitem [{\citenamefont {Yoshida}\ and\ \citenamefont {Hatsugai}(2022)}]{HN_topo1}%
  \BibitemOpen
  \bibfield  {author} {\bibinfo {author} {\bibfnamefont {T.}~\bibnamefont {Yoshida}}\ and\ \bibinfo {author} {\bibfnamefont {Y.}~\bibnamefont {Hatsugai}},\ }\href {\doibase 10.1103/PhysRevB.106.205147} {\bibfield  {journal} {\bibinfo  {journal} {Phys. Rev. B}\ }\textbf {\bibinfo {volume} {106}},\ \bibinfo {pages} {205147} (\bibinfo {year} {2022})}\BibitemShut {NoStop}%
\bibitem [{\citenamefont {Kawabata}\ \emph {et~al.}(2022)\citenamefont {Kawabata}, \citenamefont {Shiozaki},\ and\ \citenamefont {Ryu}}]{HN_topo3}%
  \BibitemOpen
  \bibfield  {author} {\bibinfo {author} {\bibfnamefont {K.}~\bibnamefont {Kawabata}}, \bibinfo {author} {\bibfnamefont {K.}~\bibnamefont {Shiozaki}}, \ and\ \bibinfo {author} {\bibfnamefont {S.}~\bibnamefont {Ryu}},\ }\href {\doibase 10.1103/PhysRevB.105.165137} {\bibfield  {journal} {\bibinfo  {journal} {Phys. Rev. B}\ }\textbf {\bibinfo {volume} {105}},\ \bibinfo {pages} {165137} (\bibinfo {year} {2022})}\BibitemShut {NoStop}%
\bibitem [{\citenamefont {Zhang}\ \emph {et~al.}(2022)\citenamefont {Zhang}, \citenamefont {Denner}, \citenamefont {Bzdu\ifmmode~\check{s}\else \v{s}\fi{}ek}, \citenamefont {Sentef},\ and\ \citenamefont {Neupert}}]{HN-skin2}%
  \BibitemOpen
  \bibfield  {author} {\bibinfo {author} {\bibfnamefont {S.-B.}\ \bibnamefont {Zhang}}, \bibinfo {author} {\bibfnamefont {M.~M.}\ \bibnamefont {Denner}}, \bibinfo {author} {\bibfnamefont {T.~c.~v.}\ \bibnamefont {Bzdu\ifmmode~\check{s}\else \v{s}\fi{}ek}}, \bibinfo {author} {\bibfnamefont {M.~A.}\ \bibnamefont {Sentef}}, \ and\ \bibinfo {author} {\bibfnamefont {T.}~\bibnamefont {Neupert}},\ }\href {\doibase 10.1103/PhysRevB.106.L121102} {\bibfield  {journal} {\bibinfo  {journal} {Phys. Rev. B}\ }\textbf {\bibinfo {volume} {106}},\ \bibinfo {pages} {L121102} (\bibinfo {year} {2022})}\BibitemShut {NoStop}%
\bibitem [{\citenamefont {Banerjee}\ \emph {et~al.}(2023)\citenamefont {Banerjee}, \citenamefont {Sarkar}, \citenamefont {Dey},\ and\ \citenamefont {Narayan}}]{HN_topo2}%
  \BibitemOpen
  \bibfield  {author} {\bibinfo {author} {\bibfnamefont {A.}~\bibnamefont {Banerjee}}, \bibinfo {author} {\bibfnamefont {R.}~\bibnamefont {Sarkar}}, \bibinfo {author} {\bibfnamefont {S.}~\bibnamefont {Dey}}, \ and\ \bibinfo {author} {\bibfnamefont {A.}~\bibnamefont {Narayan}},\ }\href {\doibase 10.1088/1361-648X/acd1cb} {\bibfield  {journal} {\bibinfo  {journal} {Journal of Physics: Condensed Matter}\ }\textbf {\bibinfo {volume} {35}},\ \bibinfo {pages} {333001} (\bibinfo {year} {2023})}\BibitemShut {NoStop}%
\bibitem [{\citenamefont {Yoshida}\ and\ \citenamefont {Hatsugai}(2023)}]{YH2}%
  \BibitemOpen
  \bibfield  {author} {\bibinfo {author} {\bibfnamefont {T.}~\bibnamefont {Yoshida}}\ and\ \bibinfo {author} {\bibfnamefont {Y.}~\bibnamefont {Hatsugai}},\ }\href {\doibase 10.1103/PhysRevB.107.075118} {\bibfield  {journal} {\bibinfo  {journal} {Phys. Rev. B}\ }\textbf {\bibinfo {volume} {107}},\ \bibinfo {pages} {075118} (\bibinfo {year} {2023})}\BibitemShut {NoStop}%
\bibitem [{\citenamefont {Molignini}\ \emph {et~al.}(2023)\citenamefont {Molignini}, \citenamefont {Arandes},\ and\ \citenamefont {Bergholtz}}]{HN-skin1}%
  \BibitemOpen
  \bibfield  {author} {\bibinfo {author} {\bibfnamefont {P.}~\bibnamefont {Molignini}}, \bibinfo {author} {\bibfnamefont {O.}~\bibnamefont {Arandes}}, \ and\ \bibinfo {author} {\bibfnamefont {E.~J.}\ \bibnamefont {Bergholtz}},\ }\href {\doibase 10.1103/PhysRevResearch.5.033058} {\bibfield  {journal} {\bibinfo  {journal} {Phys. Rev. Res.}\ }\textbf {\bibinfo {volume} {5}},\ \bibinfo {pages} {033058} (\bibinfo {year} {2023})}\BibitemShut {NoStop}%
\bibitem [{\citenamefont {Qi}\ \emph {et~al.}(2023)\citenamefont {Qi}, \citenamefont {Cao},\ and\ \citenamefont {Jiang}}]{HN-localization}%
  \BibitemOpen
  \bibfield  {author} {\bibinfo {author} {\bibfnamefont {R.}~\bibnamefont {Qi}}, \bibinfo {author} {\bibfnamefont {J.}~\bibnamefont {Cao}}, \ and\ \bibinfo {author} {\bibfnamefont {X.-P.}\ \bibnamefont {Jiang}},\ }\href@noop {} {\enquote {\bibinfo {title} {Localization and mobility edges in non-hermitian disorder-free lattices},}\ } (\bibinfo {year} {2023}),\ \Eprint {http://arxiv.org/abs/2306.03807} {arXiv:2306.03807 [cond-mat.dis-nn]} \BibitemShut {NoStop}%
\bibitem [{\citenamefont {Zhai}\ \emph {et~al.}(2022)\citenamefont {Zhai}, \citenamefont {Huang},\ and\ \citenamefont {Yin}}]{HN-localization2}%
  \BibitemOpen
  \bibfield  {author} {\bibinfo {author} {\bibfnamefont {L.-J.}\ \bibnamefont {Zhai}}, \bibinfo {author} {\bibfnamefont {G.-Y.}\ \bibnamefont {Huang}}, \ and\ \bibinfo {author} {\bibfnamefont {S.}~\bibnamefont {Yin}},\ }\href {\doibase 10.1103/PhysRevB.106.014204} {\bibfield  {journal} {\bibinfo  {journal} {Phys. Rev. B}\ }\textbf {\bibinfo {volume} {106}},\ \bibinfo {pages} {014204} (\bibinfo {year} {2022})}\BibitemShut {NoStop}%
\bibitem [{\citenamefont {Suthar}\ \emph {et~al.}(2022)\citenamefont {Suthar}, \citenamefont {Wang}, \citenamefont {Huang}, \citenamefont {Jen},\ and\ \citenamefont {You}}]{NonHMBL1}%
  \BibitemOpen
  \bibfield  {author} {\bibinfo {author} {\bibfnamefont {K.}~\bibnamefont {Suthar}}, \bibinfo {author} {\bibfnamefont {Y.-C.}\ \bibnamefont {Wang}}, \bibinfo {author} {\bibfnamefont {Y.-P.}\ \bibnamefont {Huang}}, \bibinfo {author} {\bibfnamefont {H.~H.}\ \bibnamefont {Jen}}, \ and\ \bibinfo {author} {\bibfnamefont {J.-S.}\ \bibnamefont {You}},\ }\href {\doibase 10.1103/PhysRevB.106.064208} {\bibfield  {journal} {\bibinfo  {journal} {Phys. Rev. B}\ }\textbf {\bibinfo {volume} {106}},\ \bibinfo {pages} {064208} (\bibinfo {year} {2022})}\BibitemShut {NoStop}%
\bibitem [{\citenamefont {Liu}\ and\ \citenamefont {Xu}(2023)}]{NonHMBL2}%
  \BibitemOpen
  \bibfield  {author} {\bibinfo {author} {\bibfnamefont {J.}~\bibnamefont {Liu}}\ and\ \bibinfo {author} {\bibfnamefont {Z.}~\bibnamefont {Xu}},\ }\href@noop {} {\enquote {\bibinfo {title} {From ergodicity to many-body localization in a one-dimensional interacting non-hermitian stark system},}\ } (\bibinfo {year} {2023}),\ \Eprint {http://arxiv.org/abs/2305.13636} {arXiv:2305.13636 [cond-mat.dis-nn]} \BibitemShut {NoStop}%
\bibitem [{\citenamefont {Zhai}\ \emph {et~al.}(2020)\citenamefont {Zhai}, \citenamefont {Yin},\ and\ \citenamefont {Huang}}]{nonHMBL-AA}%
  \BibitemOpen
  \bibfield  {author} {\bibinfo {author} {\bibfnamefont {L.-J.}\ \bibnamefont {Zhai}}, \bibinfo {author} {\bibfnamefont {S.}~\bibnamefont {Yin}}, \ and\ \bibinfo {author} {\bibfnamefont {G.-Y.}\ \bibnamefont {Huang}},\ }\href {\doibase 10.1103/PhysRevB.102.064206} {\bibfield  {journal} {\bibinfo  {journal} {Phys. Rev. B}\ }\textbf {\bibinfo {volume} {102}},\ \bibinfo {pages} {064206} (\bibinfo {year} {2020})}\BibitemShut {NoStop}%
\bibitem [{\citenamefont {Cheng}\ \emph {et~al.}(2023)\citenamefont {Cheng}, \citenamefont {Yin},\ and\ \citenamefont {Yao}}]{GL-non-herm}%
  \BibitemOpen
  \bibfield  {author} {\bibinfo {author} {\bibfnamefont {J.-Q.}\ \bibnamefont {Cheng}}, \bibinfo {author} {\bibfnamefont {S.}~\bibnamefont {Yin}}, \ and\ \bibinfo {author} {\bibfnamefont {D.-X.}\ \bibnamefont {Yao}},\ }\href@noop {} {\enquote {\bibinfo {title} {Dynamical localization transition in the non-hermitian $\mathbb{Z}_2$ gauge theory},}\ } (\bibinfo {year} {2023}),\ \Eprint {http://arxiv.org/abs/2307.08750} {arXiv:2307.08750 [cond-mat.dis-nn]} \BibitemShut {NoStop}%
\bibitem [{\citenamefont {Mák}\ \emph {et~al.}(2023)\citenamefont {Mák}, \citenamefont {Bhaseen},\ and\ \citenamefont {Pal}}]{non-Hermitian-MBL-arXiv}%
  \BibitemOpen
  \bibfield  {author} {\bibinfo {author} {\bibfnamefont {J.}~\bibnamefont {Mák}}, \bibinfo {author} {\bibfnamefont {M.~J.}\ \bibnamefont {Bhaseen}}, \ and\ \bibinfo {author} {\bibfnamefont {A.}~\bibnamefont {Pal}},\ }\href@noop {} {\enquote {\bibinfo {title} {Statics and dynamics of non-hermitian many-body localization},}\ } (\bibinfo {year} {2023}),\ \Eprint {http://arxiv.org/abs/2301.01763} {arXiv:2301.01763 [cond-mat.dis-nn]} \BibitemShut {NoStop}%
\bibitem [{\citenamefont {Slevin}\ and\ \citenamefont {Ohtsuki}(1999)}]{transfer_matrix1-Herm}%
  \BibitemOpen
  \bibfield  {author} {\bibinfo {author} {\bibfnamefont {K.}~\bibnamefont {Slevin}}\ and\ \bibinfo {author} {\bibfnamefont {T.}~\bibnamefont {Ohtsuki}},\ }\href {\doibase 10.1103/PhysRevLett.82.382} {\bibfield  {journal} {\bibinfo  {journal} {Phys. Rev. Lett.}\ }\textbf {\bibinfo {volume} {82}},\ \bibinfo {pages} {382} (\bibinfo {year} {1999})}\BibitemShut {NoStop}%
\bibitem [{\citenamefont {Slevin}\ and\ \citenamefont {Ohtsuki}(2014)}]{transfer_matrix2-Herm}%
  \BibitemOpen
  \bibfield  {author} {\bibinfo {author} {\bibfnamefont {K.}~\bibnamefont {Slevin}}\ and\ \bibinfo {author} {\bibfnamefont {T.}~\bibnamefont {Ohtsuki}},\ }\href {\doibase 10.1088/1367-2630/16/1/015012} {\bibfield  {journal} {\bibinfo  {journal} {New Journal of Physics}\ }\textbf {\bibinfo {volume} {16}},\ \bibinfo {pages} {015012} (\bibinfo {year} {2014})}\BibitemShut {NoStop}%
\bibitem [{\citenamefont {Kawabata}\ and\ \citenamefont {Ryu}(2021)}]{transfer_matrix-nonH1}%
  \BibitemOpen
  \bibfield  {author} {\bibinfo {author} {\bibfnamefont {K.}~\bibnamefont {Kawabata}}\ and\ \bibinfo {author} {\bibfnamefont {S.}~\bibnamefont {Ryu}},\ }\href {\doibase 10.1103/PhysRevLett.126.166801} {\bibfield  {journal} {\bibinfo  {journal} {Phys. Rev. Lett.}\ }\textbf {\bibinfo {volume} {126}},\ \bibinfo {pages} {166801} (\bibinfo {year} {2021})}\BibitemShut {NoStop}%
\bibitem [{\citenamefont {Luo}\ \emph {et~al.}(2021)\citenamefont {Luo}, \citenamefont {Ohtsuki},\ and\ \citenamefont {Shindou}}]{transfer_matrix-nonH2}%
  \BibitemOpen
  \bibfield  {author} {\bibinfo {author} {\bibfnamefont {X.}~\bibnamefont {Luo}}, \bibinfo {author} {\bibfnamefont {T.}~\bibnamefont {Ohtsuki}}, \ and\ \bibinfo {author} {\bibfnamefont {R.}~\bibnamefont {Shindou}},\ }\href {\doibase 10.1103/PhysRevB.104.104203} {\bibfield  {journal} {\bibinfo  {journal} {Phys. Rev. B}\ }\textbf {\bibinfo {volume} {104}},\ \bibinfo {pages} {104203} (\bibinfo {year} {2021})}\BibitemShut {NoStop}%
\bibitem [{\citenamefont {Heu\ss{}en}\ \emph {et~al.}(2021)\citenamefont {Heu\ss{}en}, \citenamefont {White},\ and\ \citenamefont {Refael}}]{Gil}%
  \BibitemOpen
  \bibfield  {author} {\bibinfo {author} {\bibfnamefont {S.}~\bibnamefont {Heu\ss{}en}}, \bibinfo {author} {\bibfnamefont {C.~D.}\ \bibnamefont {White}}, \ and\ \bibinfo {author} {\bibfnamefont {G.}~\bibnamefont {Refael}},\ }\href {\doibase 10.1103/PhysRevB.103.064201} {\bibfield  {journal} {\bibinfo  {journal} {Phys. Rev. B}\ }\textbf {\bibinfo {volume} {103}},\ \bibinfo {pages} {064201} (\bibinfo {year} {2021})}\BibitemShut {NoStop}%
\bibitem [{Note2()}]{Note2}%
  \BibitemOpen
  \bibinfo {note} {Inserting a real flux $\Phi $ (Hermitian system) changes a localized eigenstate $\psi (j,\Phi =0)\sim \exp (-\protect \frac {|j|}{\xi })$ to $\psi (j, \Phi \protect \neq 0)\sim \exp (-\protect \frac {|j|}{\xi }+i\Phi j)$. In contrast, inserting an imaginary flux $ig$ (HN-model) modifies a localized eigenstate $\psi (j,g=0)\sim \exp (-\protect \frac {|j|}{\xi })$ to \protect \[ \begin {aligned}[t] \psi (j,g\protect \neq 0)\sim &\exp \left (-\protect \frac {(1-g\xi )|j|}{\xi }\right ) & \protect \text {if } & j<0 \\ \psi (j,g\protect \neq 0)\sim &\exp \left (-\protect \frac {(1+g\xi )|j|}{\xi }\right ) & \protect \text {if } & j\geq 0. \end {aligned} \protect \] One can easily observe that the delocalization transition is induced by $ig$ and occurs at $g=\xi ^{-1}$, from which we can determine the localization length.}\BibitemShut {Stop}%
\bibitem [{\citenamefont {Panda}\ and\ \citenamefont {Banerjee}(2020)}]{panda}%
  \BibitemOpen
  \bibfield  {author} {\bibinfo {author} {\bibfnamefont {A.}~\bibnamefont {Panda}}\ and\ \bibinfo {author} {\bibfnamefont {S.}~\bibnamefont {Banerjee}},\ }\href {\doibase 10.1103/PhysRevB.101.184201} {\bibfield  {journal} {\bibinfo  {journal} {Phys. Rev. B}\ }\textbf {\bibinfo {volume} {101}},\ \bibinfo {pages} {184201} (\bibinfo {year} {2020})}\BibitemShut {NoStop}%
\bibitem [{\citenamefont {Brody}(2013)}]{Brody_2013}%
  \BibitemOpen
  \bibfield  {author} {\bibinfo {author} {\bibfnamefont {D.~C.}\ \bibnamefont {Brody}},\ }\href {\doibase 10.1088/1751-8113/47/3/035305} {\bibfield  {journal} {\bibinfo  {journal} {Journal of Physics A: Mathematical and Theoretical}\ }\textbf {\bibinfo {volume} {47}},\ \bibinfo {pages} {035305} (\bibinfo {year} {2013})}\BibitemShut {NoStop}%
\bibitem [{Note3()}]{Note3}%
  \BibitemOpen
  \bibinfo {note} {In the Hermitian case, the total probability: $\DOTSB \sum@ \slimits@ _\alpha |c_\alpha (t)|^2 (=1)$ is, of course, conserved. Here, in the non-Hermitian case, the quantity, $\DOTSB \sum@ \slimits@ _\alpha |c_\alpha (t)|^2$ itself does not have much meaning, since $\langle \Psi (t)|\Psi (t)\rangle \protect \neq \DOTSB \sum@ \slimits@ _\alpha |c_\alpha (t)|^2$. If one expands $\langle \Psi (t)|$ into contributions from different {\protect \it left} eigenmodes as, $\langle \Psi (t)|=\DOTSB \sum@ \slimits@ _\alpha b_\alpha (t)\langle \langle \alpha |$, and uses the biorthogonal relation (\ref {biortho}), then one finds, $\langle \Psi (t)|\Psi (t)\rangle =\DOTSB \sum@ \slimits@ _\alpha b_\alpha (t) c_\alpha (t)$.}\BibitemShut {Stop}%
\bibitem [{\citenamefont {Lindblad}(1976)}]{GKSL1}%
  \BibitemOpen
  \bibfield  {author} {\bibinfo {author} {\bibfnamefont {G.}~\bibnamefont {Lindblad}},\ }\href {\doibase 10.1007/BF01608499} {\bibfield  {journal} {\bibinfo  {journal} {Commun. Math. Phys.}\ }\textbf {\bibinfo {volume} {48}},\ \bibinfo {pages} {119} (\bibinfo {year} {1976})}\BibitemShut {NoStop}%
\bibitem [{\citenamefont {Gorini}\ \emph {et~al.}(1976)\citenamefont {Gorini}, \citenamefont {Kossakowski},\ and\ \citenamefont {Sudarshan}}]{GKSL2}%
  \BibitemOpen
  \bibfield  {author} {\bibinfo {author} {\bibfnamefont {V.}~\bibnamefont {Gorini}}, \bibinfo {author} {\bibfnamefont {A.}~\bibnamefont {Kossakowski}}, \ and\ \bibinfo {author} {\bibfnamefont {E.~C.~G.}\ \bibnamefont {Sudarshan}},\ }\href {\doibase 10.1063/1.522979} {\bibfield  {journal} {\bibinfo  {journal} {J. Math. Phys.}\ }\textbf {\bibinfo {volume} {17}},\ \bibinfo {pages} {821} (\bibinfo {year} {1976})}\BibitemShut {NoStop}%
\bibitem [{Note4()}]{Note4}%
  \BibitemOpen
  \bibinfo {note} {Later we will encounter the case in which some largest Im($E_{\alpha })$'s are quasi-degenerate: ${\protect \rm Im}(E_{\alpha _1})\simeq {\protect \rm Im}(E_{\alpha _2}) \simeq \protect \cdots $, and contribute equally to $|\Psi (t\to \infty )\rangle $. Such degeneracy in the imaginary part becomes indeed relevant in the long-time dynamics of the non-interacting case; see Secs.~\ref {Sec3} and~\ref {Sec5} for details.}\BibitemShut {Stop}%
\bibitem [{\citenamefont {Hamazaki}\ \emph {et~al.}(2019)\citenamefont {Hamazaki}, \citenamefont {Kawabata},\ and\ \citenamefont {Ueda}}]{Hamaz}%
  \BibitemOpen
  \bibfield  {author} {\bibinfo {author} {\bibfnamefont {R.}~\bibnamefont {Hamazaki}}, \bibinfo {author} {\bibfnamefont {K.}~\bibnamefont {Kawabata}}, \ and\ \bibinfo {author} {\bibfnamefont {M.}~\bibnamefont {Ueda}},\ }\href {\doibase 10.1103/PhysRevLett.123.090603} {\bibfield  {journal} {\bibinfo  {journal} {Phys. Rev. Lett.}\ }\textbf {\bibinfo {volume} {123}},\ \bibinfo {pages} {090603} (\bibinfo {year} {2019})}\BibitemShut {NoStop}%
\bibitem [{\citenamefont {Qin}\ and\ \citenamefont {Li}(2023)}]{MBskin-dynamics}%
  \BibitemOpen
  \bibfield  {author} {\bibinfo {author} {\bibfnamefont {Y.}~\bibnamefont {Qin}}\ and\ \bibinfo {author} {\bibfnamefont {L.}~\bibnamefont {Li}},\ }\href@noop {} {\enquote {\bibinfo {title} {Pairing-dependent particle separation in non-hermitian fermionic systems},}\ } (\bibinfo {year} {2023}),\ \Eprint {http://arxiv.org/abs/2307.07964} {arXiv:2307.07964 [cond-mat.quant-gas]} \BibitemShut {NoStop}%
\bibitem [{Note5()}]{Note5}%
  \BibitemOpen
  \bibinfo {note} {In the thermodynamic and clean limit, the evolution of $\langle n_j\rangle $ exhibits algebraic decay with oscillation and ultimately reaches the homogenous state. This tendency is consistent with the feature of delocalization.}\BibitemShut {Stop}%
\bibitem [{\citenamefont {\ifmmode \check{Z}\else \v{Z}\fi{}nidari\ifmmode~\check{c}\else \v{c}\fi{}}\ \emph {et~al.}(2008)\citenamefont {\ifmmode \check{Z}\else \v{Z}\fi{}nidari\ifmmode~\check{c}\else \v{c}\fi{}}, \citenamefont {Prosen},\ and\ \citenamefont {Prelov\ifmmode~\check{s}\else \v{s}\fi{}ek}}]{logt0}%
  \BibitemOpen
  \bibfield  {author} {\bibinfo {author} {\bibfnamefont {M.}~\bibnamefont {\ifmmode \check{Z}\else \v{Z}\fi{}nidari\ifmmode~\check{c}\else \v{c}\fi{}}}, \bibinfo {author} {\bibfnamefont {T.~c.~v.}\ \bibnamefont {Prosen}}, \ and\ \bibinfo {author} {\bibfnamefont {P.}~\bibnamefont {Prelov\ifmmode~\check{s}\else \v{s}\fi{}ek}},\ }\href {\doibase 10.1103/PhysRevB.77.064426} {\bibfield  {journal} {\bibinfo  {journal} {Phys. Rev. B}\ }\textbf {\bibinfo {volume} {77}},\ \bibinfo {pages} {064426} (\bibinfo {year} {2008})}\BibitemShut {NoStop}%
\bibitem [{\citenamefont {Bardarson}\ \emph {et~al.}(2012)\citenamefont {Bardarson}, \citenamefont {Pollmann},\ and\ \citenamefont {Moore}}]{logt1}%
  \BibitemOpen
  \bibfield  {author} {\bibinfo {author} {\bibfnamefont {J.~H.}\ \bibnamefont {Bardarson}}, \bibinfo {author} {\bibfnamefont {F.}~\bibnamefont {Pollmann}}, \ and\ \bibinfo {author} {\bibfnamefont {J.~E.}\ \bibnamefont {Moore}},\ }\href {\doibase 10.1103/PhysRevLett.109.017202} {\bibfield  {journal} {\bibinfo  {journal} {Phys. Rev. Lett.}\ }\textbf {\bibinfo {volume} {109}},\ \bibinfo {pages} {017202} (\bibinfo {year} {2012})}\BibitemShut {NoStop}%
\bibitem [{\citenamefont {Serbyn}\ \emph {et~al.}(2013{\natexlab{a}})\citenamefont {Serbyn}, \citenamefont {Papi\ifmmode~\acute{c}\else \'{c}\fi{}},\ and\ \citenamefont {Abanin}}]{logt2}%
  \BibitemOpen
  \bibfield  {author} {\bibinfo {author} {\bibfnamefont {M.}~\bibnamefont {Serbyn}}, \bibinfo {author} {\bibfnamefont {Z.}~\bibnamefont {Papi\ifmmode~\acute{c}\else \'{c}\fi{}}}, \ and\ \bibinfo {author} {\bibfnamefont {D.~A.}\ \bibnamefont {Abanin}},\ }\href {\doibase 10.1103/PhysRevLett.110.260601} {\bibfield  {journal} {\bibinfo  {journal} {Phys. Rev. Lett.}\ }\textbf {\bibinfo {volume} {110}},\ \bibinfo {pages} {260601} (\bibinfo {year} {2013}{\natexlab{a}})}\BibitemShut {NoStop}%
\bibitem [{\citenamefont {Serbyn}\ \emph {et~al.}(2013{\natexlab{b}})\citenamefont {Serbyn}, \citenamefont {Papi\ifmmode~\acute{c}\else \'{c}\fi{}},\ and\ \citenamefont {Abanin}}]{logt3}%
  \BibitemOpen
  \bibfield  {author} {\bibinfo {author} {\bibfnamefont {M.}~\bibnamefont {Serbyn}}, \bibinfo {author} {\bibfnamefont {Z.}~\bibnamefont {Papi\ifmmode~\acute{c}\else \'{c}\fi{}}}, \ and\ \bibinfo {author} {\bibfnamefont {D.~A.}\ \bibnamefont {Abanin}},\ }\href {\doibase 10.1103/PhysRevLett.111.127201} {\bibfield  {journal} {\bibinfo  {journal} {Phys. Rev. Lett.}\ }\textbf {\bibinfo {volume} {111}},\ \bibinfo {pages} {127201} (\bibinfo {year} {2013}{\natexlab{b}})}\BibitemShut {NoStop}%
\bibitem [{\citenamefont {Huse}\ \emph {et~al.}(2014)\citenamefont {Huse}, \citenamefont {Nandkishore},\ and\ \citenamefont {Oganesyan}}]{Huse_FMBL}%
  \BibitemOpen
  \bibfield  {author} {\bibinfo {author} {\bibfnamefont {D.~A.}\ \bibnamefont {Huse}}, \bibinfo {author} {\bibfnamefont {R.}~\bibnamefont {Nandkishore}}, \ and\ \bibinfo {author} {\bibfnamefont {V.}~\bibnamefont {Oganesyan}},\ }\href {\doibase 10.1103/PhysRevB.90.174202} {\bibfield  {journal} {\bibinfo  {journal} {Phys. Rev. B}\ }\textbf {\bibinfo {volume} {90}},\ \bibinfo {pages} {174202} (\bibinfo {year} {2014})}\BibitemShut {NoStop}%
\bibitem [{\citenamefont {Luitz}\ \emph {et~al.}(2016)\citenamefont {Luitz}, \citenamefont {Laflorencie},\ and\ \citenamefont {Alet}}]{MBL-EEdisorder1}%
  \BibitemOpen
  \bibfield  {author} {\bibinfo {author} {\bibfnamefont {D.~J.}\ \bibnamefont {Luitz}}, \bibinfo {author} {\bibfnamefont {N.}~\bibnamefont {Laflorencie}}, \ and\ \bibinfo {author} {\bibfnamefont {F.}~\bibnamefont {Alet}},\ }\href {\doibase 10.1103/PhysRevB.93.060201} {\bibfield  {journal} {\bibinfo  {journal} {Phys. Rev. B}\ }\textbf {\bibinfo {volume} {93}},\ \bibinfo {pages} {060201} (\bibinfo {year} {2016})}\BibitemShut {NoStop}%
\bibitem [{\citenamefont {Doggen}\ \emph {et~al.}(2018)\citenamefont {Doggen}, \citenamefont {Schindler}, \citenamefont {Tikhonov}, \citenamefont {Mirlin}, \citenamefont {Neupert}, \citenamefont {Polyakov},\ and\ \citenamefont {Gornyi}}]{MBL-EEdisorder2}%
  \BibitemOpen
  \bibfield  {author} {\bibinfo {author} {\bibfnamefont {E.~V.~H.}\ \bibnamefont {Doggen}}, \bibinfo {author} {\bibfnamefont {F.}~\bibnamefont {Schindler}}, \bibinfo {author} {\bibfnamefont {K.~S.}\ \bibnamefont {Tikhonov}}, \bibinfo {author} {\bibfnamefont {A.~D.}\ \bibnamefont {Mirlin}}, \bibinfo {author} {\bibfnamefont {T.}~\bibnamefont {Neupert}}, \bibinfo {author} {\bibfnamefont {D.~G.}\ \bibnamefont {Polyakov}}, \ and\ \bibinfo {author} {\bibfnamefont {I.~V.}\ \bibnamefont {Gornyi}},\ }\href {\doibase 10.1103/PhysRevB.98.174202} {\bibfield  {journal} {\bibinfo  {journal} {Phys. Rev. B}\ }\textbf {\bibinfo {volume} {98}},\ \bibinfo {pages} {174202} (\bibinfo {year} {2018})}\BibitemShut {NoStop}%
\bibitem [{\citenamefont {Mazza}\ \emph {et~al.}(2016)\citenamefont {Mazza}, \citenamefont {St{\'{e} }phan}, \citenamefont {Canovi}, \citenamefont {Alba}, \citenamefont {Brockmann},\ and\ \citenamefont {Haque}}]{nk=0.5}%
  \BibitemOpen
  \bibfield  {author} {\bibinfo {author} {\bibfnamefont {P.~P.}\ \bibnamefont {Mazza}}, \bibinfo {author} {\bibfnamefont {J.-M.}\ \bibnamefont {St{\'{e} }phan}}, \bibinfo {author} {\bibfnamefont {E.}~\bibnamefont {Canovi}}, \bibinfo {author} {\bibfnamefont {V.}~\bibnamefont {Alba}}, \bibinfo {author} {\bibfnamefont {M.}~\bibnamefont {Brockmann}}, \ and\ \bibinfo {author} {\bibfnamefont {M.}~\bibnamefont {Haque}},\ }\href {\doibase 10.1088/1742-5468/2016/01/013104} {\bibfield  {journal} {\bibinfo  {journal} {Journal of Statistical Mechanics: Theory and Experiment}\ }\textbf {\bibinfo {volume} {2016}},\ \bibinfo {pages} {013104} (\bibinfo {year} {2016})}\BibitemShut {NoStop}%
\bibitem [{Note6()}]{Note6}%
  \BibitemOpen
  \bibinfo {note} {That is, $\langle \protect \hat {n}_k\rangle $ suggested by GGE (see Appendix~\ref {Append-GGE} for more details) converges either to 0 or to 1 more slowly than Eq.~(\ref {nk_tilde}). This discrepancy is because we assume superposition consists of various filling to derive Eq.~(\ref {Eq:k,g-dependence-GGE2}), whereas in the actual numerical calculation, we consider the half-filling case. If the initial state is prepared as a superposition consists various filling $Q=\DOTSB \sum@ \slimits@ _i k_i/L$, the time dependence of $\langle \protect \hat {n}_k\rangle $ is akin to Eq.~(\ref {Eq:k,g-dependence-GGE2}) (see Appendix~\ref {time-dependence nk}).}\BibitemShut {Stop}%
\bibitem [{\citenamefont {Avella}\ and\ \citenamefont {Mancini}(2013)}]{Avella2013}%
  \BibitemOpen
  \bibfield  {author} {\bibinfo {author} {\bibfnamefont {A.}~\bibnamefont {Avella}}\ and\ \bibinfo {author} {\bibfnamefont {F.}~\bibnamefont {Mancini}},\ }\href {https://books.google.co.jp/books?id=Be4\_AAAAQBAJ} {\emph {\bibinfo {title} {Strongly Correlated Systems: Numerical Methods}}},\ Springer Series in Solid-State Sciences\ (\bibinfo  {publisher} {Springer Berlin Heidelberg},\ \bibinfo {year} {2013})\BibitemShut {NoStop}%
\bibitem [{Note7()}]{Note7}%
  \BibitemOpen
  \bibinfo {note} {Strictly speaking, since we treat a finite system, each quasiparticles can be located in the same subsystem due to the boundary effect (we later comment on this effect), leading to decay in the $S_{\protect \rm ent}$. Although this effect can be non-negligible in a finite system, it is already known that the less important this effect, the larger the system size we treat;\cite {EE_revival} therefore, we can interpret $t_c(k)$ as a characteristic time scale.}\BibitemShut {Stop}%
\bibitem [{Note8()}]{Note8}%
  \BibitemOpen
  \bibinfo {note} {In numerical calculation, we choose the length of subsystem size $\ell $ to be small because it may be the simplest way to realize the non-monotonic behavior of $S_{\protect \rm ent}$. Since $S_{\protect \rm ent}(t\to \infty )$ decreases with a decrease of $\ell $, in case of small $\ell $, the condition $S_{\protect \rm ent}(t)> S_{\protect \rm ent}(t\to \infty )$, which is required to realize such a behavior, becomes easier to achieve.}\BibitemShut {Stop}%
\bibitem [{\citenamefont {Modak}\ \emph {et~al.}(2020)\citenamefont {Modak}, \citenamefont {Alba},\ and\ \citenamefont {Calabrese}}]{EE_revival}%
  \BibitemOpen
  \bibfield  {author} {\bibinfo {author} {\bibfnamefont {R.}~\bibnamefont {Modak}}, \bibinfo {author} {\bibfnamefont {V.}~\bibnamefont {Alba}}, \ and\ \bibinfo {author} {\bibfnamefont {P.}~\bibnamefont {Calabrese}},\ }\href {\doibase 10.1088/1742-5468/aba9d9} {\bibfield  {journal} {\bibinfo  {journal} {Journal of Statistical Mechanics: Theory and Experiment}\ }\textbf {\bibinfo {volume} {2020}},\ \bibinfo {pages} {083110} (\bibinfo {year} {2020})}\BibitemShut {NoStop}%
\bibitem [{\citenamefont {Orito}\ and\ \citenamefont {Imura}(2023)}]{OI2023}%
  \BibitemOpen
  \bibfield  {author} {\bibinfo {author} {\bibfnamefont {T.}~\bibnamefont {Orito}}\ and\ \bibinfo {author} {\bibfnamefont {K.-I.}\ \bibnamefont {Imura}},\ }\href {\doibase 10.7566/jpscp.38.011187} {\bibfield  {journal} {\bibinfo  {journal} {Proceedings of the 29th International Conference on Low Temperature Physics ({LT}29)}\ } (\bibinfo {year} {2023}),\ 10.7566/jpscp.38.011187}\BibitemShut {NoStop}%
\bibitem [{\citenamefont {Calabrese}\ and\ \citenamefont {Cardy}(2009)}]{conformal1}%
  \BibitemOpen
  \bibfield  {author} {\bibinfo {author} {\bibfnamefont {P.}~\bibnamefont {Calabrese}}\ and\ \bibinfo {author} {\bibfnamefont {J.}~\bibnamefont {Cardy}},\ }\href {\doibase 10.1088/1751-8113/42/50/504005} {\bibfield  {journal} {\bibinfo  {journal} {Journal of Physics A: Mathematical and Theoretical}\ }\textbf {\bibinfo {volume} {42}},\ \bibinfo {pages} {504005} (\bibinfo {year} {2009})}\BibitemShut {NoStop}%
\bibitem [{\citenamefont {Peschel}(2003)}]{conformal2}%
  \BibitemOpen
  \bibfield  {author} {\bibinfo {author} {\bibfnamefont {I.}~\bibnamefont {Peschel}},\ }\href {\doibase 10.1088/0305-4470/36/14/101} {\bibfield  {journal} {\bibinfo  {journal} {Journal of Physics A: Mathematical and General}\ }\textbf {\bibinfo {volume} {36}},\ \bibinfo {pages} {L205} (\bibinfo {year} {2003})}\BibitemShut {NoStop}%
\bibitem [{\citenamefont {Calabrese}\ and\ \citenamefont {Cardy}(2004)}]{conformal3}%
  \BibitemOpen
  \bibfield  {author} {\bibinfo {author} {\bibfnamefont {P.}~\bibnamefont {Calabrese}}\ and\ \bibinfo {author} {\bibfnamefont {J.}~\bibnamefont {Cardy}},\ }\href {\doibase 10.1088/1742-5468/2004/06/P06002} {\bibfield  {journal} {\bibinfo  {journal} {Journal of Statistical Mechanics: Theory and Experiment}\ }\textbf {\bibinfo {volume} {2004}},\ \bibinfo {pages} {P06002} (\bibinfo {year} {2004})}\BibitemShut {NoStop}%
\bibitem [{\citenamefont {Holzhey}\ \emph {et~al.}(1994)\citenamefont {Holzhey}, \citenamefont {Larsen},\ and\ \citenamefont {Wilczek}}]{conformal4}%
  \BibitemOpen
  \bibfield  {author} {\bibinfo {author} {\bibfnamefont {C.}~\bibnamefont {Holzhey}}, \bibinfo {author} {\bibfnamefont {F.}~\bibnamefont {Larsen}}, \ and\ \bibinfo {author} {\bibfnamefont {F.}~\bibnamefont {Wilczek}},\ }\href {\doibase https://doi.org/10.1016/0550-3213(94)90402-2} {\bibfield  {journal} {\bibinfo  {journal} {Nuclear Physics B}\ }\textbf {\bibinfo {volume} {424}},\ \bibinfo {pages} {443} (\bibinfo {year} {1994})}\BibitemShut {NoStop}%
\bibitem [{\citenamefont {Vidal}\ \emph {et~al.}(2003)\citenamefont {Vidal}, \citenamefont {Latorre}, \citenamefont {Rico},\ and\ \citenamefont {Kitaev}}]{conformal5}%
  \BibitemOpen
  \bibfield  {author} {\bibinfo {author} {\bibfnamefont {G.}~\bibnamefont {Vidal}}, \bibinfo {author} {\bibfnamefont {J.~I.}\ \bibnamefont {Latorre}}, \bibinfo {author} {\bibfnamefont {E.}~\bibnamefont {Rico}}, \ and\ \bibinfo {author} {\bibfnamefont {A.}~\bibnamefont {Kitaev}},\ }\href {\doibase 10.1103/PhysRevLett.90.227902} {\bibfield  {journal} {\bibinfo  {journal} {Phys. Rev. Lett.}\ }\textbf {\bibinfo {volume} {90}},\ \bibinfo {pages} {227902} (\bibinfo {year} {2003})}\BibitemShut {NoStop}%
\bibitem [{\citenamefont {Furukawa}\ \emph {et~al.}(2009)\citenamefont {Furukawa}, \citenamefont {Pasquier},\ and\ \citenamefont {Shiraishi}}]{conformal6}%
  \BibitemOpen
  \bibfield  {author} {\bibinfo {author} {\bibfnamefont {S.}~\bibnamefont {Furukawa}}, \bibinfo {author} {\bibfnamefont {V.}~\bibnamefont {Pasquier}}, \ and\ \bibinfo {author} {\bibfnamefont {J.}~\bibnamefont {Shiraishi}},\ }\href {\doibase 10.1103/PhysRevLett.102.170602} {\bibfield  {journal} {\bibinfo  {journal} {Phys. Rev. Lett.}\ }\textbf {\bibinfo {volume} {102}},\ \bibinfo {pages} {170602} (\bibinfo {year} {2009})}\BibitemShut {NoStop}%
\bibitem [{\citenamefont {Fisher}\ and\ \citenamefont {Hartwig}(1969)}]{fisherhartwig1}%
  \BibitemOpen
  \bibfield  {author} {\bibinfo {author} {\bibfnamefont {M.~E.}\ \bibnamefont {Fisher}}\ and\ \bibinfo {author} {\bibfnamefont {R.~E.}\ \bibnamefont {Hartwig}},\ }\href {\doibase 10.1002/9780470143605.ch18} {\bibfield  {journal} {\bibinfo  {journal} {Advances in Chemical Physics}\ }\textbf {\bibinfo {volume} {15}},\ \bibinfo {pages} {333} (\bibinfo {year} {1969})}\BibitemShut {NoStop}%
\bibitem [{\citenamefont {Jin}\ and\ \citenamefont {Korepin}(2004)}]{fisherhartwig2}%
  \BibitemOpen
  \bibfield  {author} {\bibinfo {author} {\bibfnamefont {B.-Q.}\ \bibnamefont {Jin}}\ and\ \bibinfo {author} {\bibfnamefont {V.~E.}\ \bibnamefont {Korepin}},\ }\href {\doibase 10.1023/b:joss.0000037230.37166.42} {\bibfield  {journal} {\bibinfo  {journal} {Journal of Statistical Physics}\ }\textbf {\bibinfo {volume} {116}},\ \bibinfo {pages} {79} (\bibinfo {year} {2004})}\BibitemShut {NoStop}%
\bibitem [{\citenamefont {Nishimoto}(2011)}]{Nishimoto_2011}%
  \BibitemOpen
  \bibfield  {author} {\bibinfo {author} {\bibfnamefont {S.}~\bibnamefont {Nishimoto}},\ }\href {\doibase 10.1103/physrevb.84.195108} {\bibfield  {journal} {\bibinfo  {journal} {Physical Review B}\ }\textbf {\bibinfo {volume} {84}} (\bibinfo {year} {2011}),\ 10.1103/physrevb.84.195108}\BibitemShut {NoStop}%
\bibitem [{\citenamefont {Mao}\ \emph {et~al.}(2023)\citenamefont {Mao}, \citenamefont {Hao},\ and\ \citenamefont {Pan}}]{Bethe2}%
  \BibitemOpen
  \bibfield  {author} {\bibinfo {author} {\bibfnamefont {L.}~\bibnamefont {Mao}}, \bibinfo {author} {\bibfnamefont {Y.}~\bibnamefont {Hao}}, \ and\ \bibinfo {author} {\bibfnamefont {L.}~\bibnamefont {Pan}},\ }\href {\doibase 10.1103/PhysRevA.107.043315} {\bibfield  {journal} {\bibinfo  {journal} {Phys. Rev. A}\ }\textbf {\bibinfo {volume} {107}},\ \bibinfo {pages} {043315} (\bibinfo {year} {2023})}\BibitemShut {NoStop}%
\bibitem [{\citenamefont {Ishiguro}\ \emph {et~al.}(2023)\citenamefont {Ishiguro}, \citenamefont {Sato},\ and\ \citenamefont {Nishinari}}]{Bethe3}%
  \BibitemOpen
  \bibfield  {author} {\bibinfo {author} {\bibfnamefont {Y.}~\bibnamefont {Ishiguro}}, \bibinfo {author} {\bibfnamefont {J.}~\bibnamefont {Sato}}, \ and\ \bibinfo {author} {\bibfnamefont {K.}~\bibnamefont {Nishinari}},\ }\href@noop {} {\enquote {\bibinfo {title} {Asymmetry-induced delocalization transition in the integrable non-hermitian spin chain},}\ } (\bibinfo {year} {2023}),\ \Eprint {http://arxiv.org/abs/2101.10647} {arXiv:2101.10647 [cond-mat.stat-mech]} \BibitemShut {NoStop}%
\bibitem [{\citenamefont {Yamamoto}\ \emph {et~al.}(2022)\citenamefont {Yamamoto}, \citenamefont {Nakagawa}, \citenamefont {Tezuka}, \citenamefont {Ueda},\ and\ \citenamefont {Kawakami}}]{DMRG3andBethe}%
  \BibitemOpen
  \bibfield  {author} {\bibinfo {author} {\bibfnamefont {K.}~\bibnamefont {Yamamoto}}, \bibinfo {author} {\bibfnamefont {M.}~\bibnamefont {Nakagawa}}, \bibinfo {author} {\bibfnamefont {M.}~\bibnamefont {Tezuka}}, \bibinfo {author} {\bibfnamefont {M.}~\bibnamefont {Ueda}}, \ and\ \bibinfo {author} {\bibfnamefont {N.}~\bibnamefont {Kawakami}},\ }\href {\doibase 10.1103/PhysRevB.105.205125} {\bibfield  {journal} {\bibinfo  {journal} {Phys. Rev. B}\ }\textbf {\bibinfo {volume} {105}},\ \bibinfo {pages} {205125} (\bibinfo {year} {2022})}\BibitemShut {NoStop}%
\bibitem [{\citenamefont {White}(1992)}]{DMRG1}%
  \BibitemOpen
  \bibfield  {author} {\bibinfo {author} {\bibfnamefont {S.~R.}\ \bibnamefont {White}},\ }\href {\doibase 10.1103/PhysRevLett.69.2863} {\bibfield  {journal} {\bibinfo  {journal} {Phys. Rev. Lett.}\ }\textbf {\bibinfo {volume} {69}},\ \bibinfo {pages} {2863} (\bibinfo {year} {1992})}\BibitemShut {NoStop}%
\bibitem [{\citenamefont {Schollwöck}(2011)}]{DMRG2}%
  \BibitemOpen
  \bibfield  {author} {\bibinfo {author} {\bibfnamefont {U.}~\bibnamefont {Schollwöck}},\ }\href {\doibase https://doi.org/10.1016/j.aop.2010.09.012} {\bibfield  {journal} {\bibinfo  {journal} {Annals of Physics}\ }\textbf {\bibinfo {volume} {326}},\ \bibinfo {pages} {96} (\bibinfo {year} {2011})}\BibitemShut {NoStop}%
\bibitem [{\citenamefont {Orito}\ \emph {et~al.}(2021)\citenamefont {Orito}, \citenamefont {Kuno},\ and\ \citenamefont {Ichinose}}]{tensornet3}%
  \BibitemOpen
  \bibfield  {author} {\bibinfo {author} {\bibfnamefont {T.}~\bibnamefont {Orito}}, \bibinfo {author} {\bibfnamefont {Y.}~\bibnamefont {Kuno}}, \ and\ \bibinfo {author} {\bibfnamefont {I.}~\bibnamefont {Ichinose}},\ }\href {\doibase 10.1103/PhysRevB.103.L060301} {\bibfield  {journal} {\bibinfo  {journal} {Phys. Rev. B}\ }\textbf {\bibinfo {volume} {103}},\ \bibinfo {pages} {L060301} (\bibinfo {year} {2021})}\BibitemShut {NoStop}%
\bibitem [{\citenamefont {Sierant}\ and\ \citenamefont {Zakrzewski}(2022)}]{tensor}%
  \BibitemOpen
  \bibfield  {author} {\bibinfo {author} {\bibfnamefont {P.}~\bibnamefont {Sierant}}\ and\ \bibinfo {author} {\bibfnamefont {J.}~\bibnamefont {Zakrzewski}},\ }\href {\doibase 10.1103/PhysRevB.105.224203} {\bibfield  {journal} {\bibinfo  {journal} {Phys. Rev. B}\ }\textbf {\bibinfo {volume} {105}},\ \bibinfo {pages} {224203} (\bibinfo {year} {2022})}\BibitemShut {NoStop}%
\bibitem [{\citenamefont {Blankenbecler}\ \emph {et~al.}(1981)\citenamefont {Blankenbecler}, \citenamefont {Scalapino},\ and\ \citenamefont {Sugar}}]{QMC1}%
  \BibitemOpen
  \bibfield  {author} {\bibinfo {author} {\bibfnamefont {R.}~\bibnamefont {Blankenbecler}}, \bibinfo {author} {\bibfnamefont {D.~J.}\ \bibnamefont {Scalapino}}, \ and\ \bibinfo {author} {\bibfnamefont {R.~L.}\ \bibnamefont {Sugar}},\ }\href {\doibase 10.1103/PhysRevD.24.2278} {\bibfield  {journal} {\bibinfo  {journal} {Phys. Rev. D}\ }\textbf {\bibinfo {volume} {24}},\ \bibinfo {pages} {2278} (\bibinfo {year} {1981})}\BibitemShut {NoStop}%
\bibitem [{\citenamefont {Duane}\ \emph {et~al.}(1987)\citenamefont {Duane}, \citenamefont {Kennedy}, \citenamefont {Pendleton},\ and\ \citenamefont {Roweth}}]{QMC2}%
  \BibitemOpen
  \bibfield  {author} {\bibinfo {author} {\bibfnamefont {S.}~\bibnamefont {Duane}}, \bibinfo {author} {\bibfnamefont {A.}~\bibnamefont {Kennedy}}, \bibinfo {author} {\bibfnamefont {B.~J.}\ \bibnamefont {Pendleton}}, \ and\ \bibinfo {author} {\bibfnamefont {D.}~\bibnamefont {Roweth}},\ }\href {\doibase https://doi.org/10.1016/0370-2693(87)91197-X} {\bibfield  {journal} {\bibinfo  {journal} {Physics Letters B}\ }\textbf {\bibinfo {volume} {195}},\ \bibinfo {pages} {216} (\bibinfo {year} {1987})}\BibitemShut {NoStop}%
\bibitem [{\citenamefont {Hayata}\ and\ \citenamefont {Yamamoto}(2021)}]{QMC3}%
  \BibitemOpen
  \bibfield  {author} {\bibinfo {author} {\bibfnamefont {T.}~\bibnamefont {Hayata}}\ and\ \bibinfo {author} {\bibfnamefont {A.}~\bibnamefont {Yamamoto}},\ }\href {\doibase 10.1103/PhysRevB.104.125102} {\bibfield  {journal} {\bibinfo  {journal} {Phys. Rev. B}\ }\textbf {\bibinfo {volume} {104}},\ \bibinfo {pages} {125102} (\bibinfo {year} {2021})}\BibitemShut {NoStop}%
\bibitem [{\citenamefont {Hu}\ \emph {et~al.}(2023)\citenamefont {Hu}, \citenamefont {Fu},\ and\ \citenamefont {Zhang}}]{QMC4}%
  \BibitemOpen
  \bibfield  {author} {\bibinfo {author} {\bibfnamefont {S.-X.}\ \bibnamefont {Hu}}, \bibinfo {author} {\bibfnamefont {Y.}~\bibnamefont {Fu}}, \ and\ \bibinfo {author} {\bibfnamefont {Y.}~\bibnamefont {Zhang}},\ }\href@noop {} {\enquote {\bibinfo {title} {Nontrivial worldline winding in non-hermitian quantum systems},}\ } (\bibinfo {year} {2023}),\ \Eprint {http://arxiv.org/abs/2307.01260} {arXiv:2307.01260 [quant-ph]} \BibitemShut {NoStop}%
\bibitem [{\citenamefont {Li}\ \emph {et~al.}(2023{\natexlab{a}})\citenamefont {Li}, \citenamefont {Liu},\ and\ \citenamefont {Xu}}]{Skin-EE-transition}%
  \BibitemOpen
  \bibfield  {author} {\bibinfo {author} {\bibfnamefont {K.}~\bibnamefont {Li}}, \bibinfo {author} {\bibfnamefont {Z.-C.}\ \bibnamefont {Liu}}, \ and\ \bibinfo {author} {\bibfnamefont {Y.}~\bibnamefont {Xu}},\ }\href@noop {} {\enquote {\bibinfo {title} {Disorder-induced entanglement phase transitions in non-hermitian systems with skin effects},}\ } (\bibinfo {year} {2023}{\natexlab{a}}),\ \Eprint {http://arxiv.org/abs/2305.12342} {arXiv:2305.12342 [quant-ph]} \BibitemShut {NoStop}%
\bibitem [{Note9()}]{Note9}%
  \BibitemOpen
  \bibinfo {note} {We note that we can also use $S_{\protect \rm ent}(\infty )$ as a quantity to characterize the delocalization-localization transition. We expect that $S_{\protect \rm ent}(\infty )$ obeys the volume-law in the delocalized phase and the area-law in the delocalized phase, reflecting the property of $|\alpha _1\rangle $.}\BibitemShut {Stop}%
\bibitem [{\citenamefont {De~Roeck}\ and\ \citenamefont {Huveneers}(2017)}]{Avalanche}%
  \BibitemOpen
  \bibfield  {author} {\bibinfo {author} {\bibfnamefont {W.}~\bibnamefont {De~Roeck}}\ and\ \bibinfo {author} {\bibfnamefont {F.~m.~c.}\ \bibnamefont {Huveneers}},\ }\href {\doibase 10.1103/PhysRevB.95.155129} {\bibfield  {journal} {\bibinfo  {journal} {Phys. Rev. B}\ }\textbf {\bibinfo {volume} {95}},\ \bibinfo {pages} {155129} (\bibinfo {year} {2017})}\BibitemShut {NoStop}%
\bibitem [{\citenamefont {Thiery}\ \emph {et~al.}(2018)\citenamefont {Thiery}, \citenamefont {Huveneers}, \citenamefont {M\"uller},\ and\ \citenamefont {De~Roeck}}]{Avalanche2}%
  \BibitemOpen
  \bibfield  {author} {\bibinfo {author} {\bibfnamefont {T.}~\bibnamefont {Thiery}}, \bibinfo {author} {\bibfnamefont {F.~m.~c.}\ \bibnamefont {Huveneers}}, \bibinfo {author} {\bibfnamefont {M.}~\bibnamefont {M\"uller}}, \ and\ \bibinfo {author} {\bibfnamefont {W.}~\bibnamefont {De~Roeck}},\ }\href {\doibase 10.1103/PhysRevLett.121.140601} {\bibfield  {journal} {\bibinfo  {journal} {Phys. Rev. Lett.}\ }\textbf {\bibinfo {volume} {121}},\ \bibinfo {pages} {140601} (\bibinfo {year} {2018})}\BibitemShut {NoStop}%
\bibitem [{\citenamefont {Kiefer-Emmanouilidis}\ \emph {et~al.}(2020)\citenamefont {Kiefer-Emmanouilidis}, \citenamefont {Unanyan}, \citenamefont {Fleischhauer},\ and\ \citenamefont {Sirker}}]{thermal}%
  \BibitemOpen
  \bibfield  {author} {\bibinfo {author} {\bibfnamefont {M.}~\bibnamefont {Kiefer-Emmanouilidis}}, \bibinfo {author} {\bibfnamefont {R.}~\bibnamefont {Unanyan}}, \bibinfo {author} {\bibfnamefont {M.}~\bibnamefont {Fleischhauer}}, \ and\ \bibinfo {author} {\bibfnamefont {J.}~\bibnamefont {Sirker}},\ }\href {\doibase 10.1103/PhysRevLett.124.243601} {\bibfield  {journal} {\bibinfo  {journal} {Phys. Rev. Lett.}\ }\textbf {\bibinfo {volume} {124}},\ \bibinfo {pages} {243601} (\bibinfo {year} {2020})}\BibitemShut {NoStop}%
\bibitem [{\citenamefont {Evers}\ and\ \citenamefont {Bera}(2023)}]{thermal2}%
  \BibitemOpen
  \bibfield  {author} {\bibinfo {author} {\bibfnamefont {F.}~\bibnamefont {Evers}}\ and\ \bibinfo {author} {\bibfnamefont {S.}~\bibnamefont {Bera}},\ }\href@noop {} {\enquote {\bibinfo {title} {The internal clock of many-body (de-)localization},}\ } (\bibinfo {year} {2023}),\ \Eprint {http://arxiv.org/abs/2302.11384} {arXiv:2302.11384 [cond-mat.dis-nn]} \BibitemShut {NoStop}%
\bibitem [{\citenamefont {Mac\'e}\ \emph {et~al.}(2019)\citenamefont {Mac\'e}, \citenamefont {Alet},\ and\ \citenamefont {Laflorencie}}]{ETH-MBL1}%
  \BibitemOpen
  \bibfield  {author} {\bibinfo {author} {\bibfnamefont {N.}~\bibnamefont {Mac\'e}}, \bibinfo {author} {\bibfnamefont {F.}~\bibnamefont {Alet}}, \ and\ \bibinfo {author} {\bibfnamefont {N.}~\bibnamefont {Laflorencie}},\ }\href {\doibase 10.1103/PhysRevLett.123.180601} {\bibfield  {journal} {\bibinfo  {journal} {Phys. Rev. Lett.}\ }\textbf {\bibinfo {volume} {123}},\ \bibinfo {pages} {180601} (\bibinfo {year} {2019})}\BibitemShut {NoStop}%
\bibitem [{\citenamefont {Laflorencie}\ \emph {et~al.}(2020)\citenamefont {Laflorencie}, \citenamefont {Lemari\'e},\ and\ \citenamefont {Mac\'e}}]{ETH-MBL2}%
  \BibitemOpen
  \bibfield  {author} {\bibinfo {author} {\bibfnamefont {N.}~\bibnamefont {Laflorencie}}, \bibinfo {author} {\bibfnamefont {G.}~\bibnamefont {Lemari\'e}}, \ and\ \bibinfo {author} {\bibfnamefont {N.}~\bibnamefont {Mac\'e}},\ }\href {\doibase 10.1103/PhysRevResearch.2.042033} {\bibfield  {journal} {\bibinfo  {journal} {Phys. Rev. Res.}\ }\textbf {\bibinfo {volume} {2}},\ \bibinfo {pages} {042033} (\bibinfo {year} {2020})}\BibitemShut {NoStop}%
\bibitem [{\citenamefont {Roy}\ and\ \citenamefont {Logan}(2020)}]{ETH-MBL3}%
  \BibitemOpen
  \bibfield  {author} {\bibinfo {author} {\bibfnamefont {S.}~\bibnamefont {Roy}}\ and\ \bibinfo {author} {\bibfnamefont {D.~E.}\ \bibnamefont {Logan}},\ }\href {\doibase 10.1103/PhysRevB.101.134202} {\bibfield  {journal} {\bibinfo  {journal} {Phys. Rev. B}\ }\textbf {\bibinfo {volume} {101}},\ \bibinfo {pages} {134202} (\bibinfo {year} {2020})}\BibitemShut {NoStop}%
\bibitem [{\citenamefont {De~Tomasi}\ \emph {et~al.}(2021)\citenamefont {De~Tomasi}, \citenamefont {Khaymovich}, \citenamefont {Pollmann},\ and\ \citenamefont {Warzel}}]{ETH-MBL4}%
  \BibitemOpen
  \bibfield  {author} {\bibinfo {author} {\bibfnamefont {G.}~\bibnamefont {De~Tomasi}}, \bibinfo {author} {\bibfnamefont {I.~M.}\ \bibnamefont {Khaymovich}}, \bibinfo {author} {\bibfnamefont {F.}~\bibnamefont {Pollmann}}, \ and\ \bibinfo {author} {\bibfnamefont {S.}~\bibnamefont {Warzel}},\ }\href {\doibase 10.1103/PhysRevB.104.024202} {\bibfield  {journal} {\bibinfo  {journal} {Phys. Rev. B}\ }\textbf {\bibinfo {volume} {104}},\ \bibinfo {pages} {024202} (\bibinfo {year} {2021})}\BibitemShut {NoStop}%
\bibitem [{\citenamefont {Bahovadinov}\ \emph {et~al.}(2022)\citenamefont {Bahovadinov}, \citenamefont {Buijsman}, \citenamefont {Fedorov}, \citenamefont {Gritsev},\ and\ \citenamefont {Kurlov}}]{ETH-MBL5}%
  \BibitemOpen
  \bibfield  {author} {\bibinfo {author} {\bibfnamefont {M.~S.}\ \bibnamefont {Bahovadinov}}, \bibinfo {author} {\bibfnamefont {W.}~\bibnamefont {Buijsman}}, \bibinfo {author} {\bibfnamefont {A.~K.}\ \bibnamefont {Fedorov}}, \bibinfo {author} {\bibfnamefont {V.}~\bibnamefont {Gritsev}}, \ and\ \bibinfo {author} {\bibfnamefont {D.~V.}\ \bibnamefont {Kurlov}},\ }\href {\doibase 10.1103/PhysRevB.106.224205} {\bibfield  {journal} {\bibinfo  {journal} {Phys. Rev. B}\ }\textbf {\bibinfo {volume} {106}},\ \bibinfo {pages} {224205} (\bibinfo {year} {2022})}\BibitemShut {NoStop}%
\bibitem [{\citenamefont {Morningstar}\ \emph {et~al.}(2020)\citenamefont {Morningstar}, \citenamefont {Huse},\ and\ \citenamefont {Imbrie}}]{insta_MBL1}%
  \BibitemOpen
  \bibfield  {author} {\bibinfo {author} {\bibfnamefont {A.}~\bibnamefont {Morningstar}}, \bibinfo {author} {\bibfnamefont {D.~A.}\ \bibnamefont {Huse}}, \ and\ \bibinfo {author} {\bibfnamefont {J.~Z.}\ \bibnamefont {Imbrie}},\ }\href {\doibase 10.1103/PhysRevB.102.125134} {\bibfield  {journal} {\bibinfo  {journal} {Phys. Rev. B}\ }\textbf {\bibinfo {volume} {102}},\ \bibinfo {pages} {125134} (\bibinfo {year} {2020})}\BibitemShut {NoStop}%
\bibitem [{\citenamefont {Morningstar}\ \emph {et~al.}(2022)\citenamefont {Morningstar}, \citenamefont {Colmenarez}, \citenamefont {Khemani}, \citenamefont {Luitz},\ and\ \citenamefont {Huse}}]{insta_MBL2}%
  \BibitemOpen
  \bibfield  {author} {\bibinfo {author} {\bibfnamefont {A.}~\bibnamefont {Morningstar}}, \bibinfo {author} {\bibfnamefont {L.}~\bibnamefont {Colmenarez}}, \bibinfo {author} {\bibfnamefont {V.}~\bibnamefont {Khemani}}, \bibinfo {author} {\bibfnamefont {D.~J.}\ \bibnamefont {Luitz}}, \ and\ \bibinfo {author} {\bibfnamefont {D.~A.}\ \bibnamefont {Huse}},\ }\href {\doibase 10.1103/PhysRevB.105.174205} {\bibfield  {journal} {\bibinfo  {journal} {Phys. Rev. B}\ }\textbf {\bibinfo {volume} {105}},\ \bibinfo {pages} {174205} (\bibinfo {year} {2022})}\BibitemShut {NoStop}%
\bibitem [{\citenamefont {Kawabata}\ \emph {et~al.}(2023)\citenamefont {Kawabata}, \citenamefont {Numasawa},\ and\ \citenamefont {Ryu}}]{skin-EE}%
  \BibitemOpen
  \bibfield  {author} {\bibinfo {author} {\bibfnamefont {K.}~\bibnamefont {Kawabata}}, \bibinfo {author} {\bibfnamefont {T.}~\bibnamefont {Numasawa}}, \ and\ \bibinfo {author} {\bibfnamefont {S.}~\bibnamefont {Ryu}},\ }\href {\doibase 10.1103/PhysRevX.13.021007} {\bibfield  {journal} {\bibinfo  {journal} {Phys. Rev. X}\ }\textbf {\bibinfo {volume} {13}},\ \bibinfo {pages} {021007} (\bibinfo {year} {2023})}\BibitemShut {NoStop}%
\bibitem [{\citenamefont {Weinberg}\ and\ \citenamefont {Bukov}(2017)}]{Quspin1}%
  \BibitemOpen
  \bibfield  {author} {\bibinfo {author} {\bibfnamefont {P.}~\bibnamefont {Weinberg}}\ and\ \bibinfo {author} {\bibfnamefont {M.}~\bibnamefont {Bukov}},\ }\href {\doibase 10.21468/SciPostPhys.2.1.003} {\bibfield  {journal} {\bibinfo  {journal} {SciPost Phys.}\ }\textbf {\bibinfo {volume} {2}},\ \bibinfo {pages} {003} (\bibinfo {year} {2017})}\BibitemShut {NoStop}%
\bibitem [{\citenamefont {Weinberg}\ and\ \citenamefont {Bukov}(2019)}]{Quspin2}%
  \BibitemOpen
  \bibfield  {author} {\bibinfo {author} {\bibfnamefont {P.}~\bibnamefont {Weinberg}}\ and\ \bibinfo {author} {\bibfnamefont {M.}~\bibnamefont {Bukov}},\ }\href {\doibase 10.21468/SciPostPhys.7.2.020} {\bibfield  {journal} {\bibinfo  {journal} {SciPost Phys.}\ }\textbf {\bibinfo {volume} {7}},\ \bibinfo {pages} {20} (\bibinfo {year} {2019})}\BibitemShut {NoStop}%
\bibitem [{\citenamefont {Daley}(2014)}]{Daley_2014}%
  \BibitemOpen
  \bibfield  {author} {\bibinfo {author} {\bibfnamefont {A.~J.}\ \bibnamefont {Daley}},\ }\href {\doibase 10.1080/00018732.2014.933502} {\bibfield  {journal} {\bibinfo  {journal} {Advances in Physics}\ }\textbf {\bibinfo {volume} {63}},\ \bibinfo {pages} {77} (\bibinfo {year} {2014})}\BibitemShut {NoStop}%
\bibitem [{\citenamefont {Plenio}\ and\ \citenamefont {Knight}(1998)}]{quantum-jump}%
  \BibitemOpen
  \bibfield  {author} {\bibinfo {author} {\bibfnamefont {M.~B.}\ \bibnamefont {Plenio}}\ and\ \bibinfo {author} {\bibfnamefont {P.~L.}\ \bibnamefont {Knight}},\ }\href {\doibase 10.1103/RevModPhys.70.101} {\bibfield  {journal} {\bibinfo  {journal} {Rev. Mod. Phys.}\ }\textbf {\bibinfo {volume} {70}},\ \bibinfo {pages} {101} (\bibinfo {year} {1998})}\BibitemShut {NoStop}%
\bibitem [{\citenamefont {Dalibard}\ \emph {et~al.}(1992)\citenamefont {Dalibard}, \citenamefont {Castin},\ and\ \citenamefont {M\o{}lmer}}]{first-monte-1}%
  \BibitemOpen
  \bibfield  {author} {\bibinfo {author} {\bibfnamefont {J.}~\bibnamefont {Dalibard}}, \bibinfo {author} {\bibfnamefont {Y.}~\bibnamefont {Castin}}, \ and\ \bibinfo {author} {\bibfnamefont {K.}~\bibnamefont {M\o{}lmer}},\ }\href {\doibase 10.1103/PhysRevLett.68.580} {\bibfield  {journal} {\bibinfo  {journal} {Phys. Rev. Lett.}\ }\textbf {\bibinfo {volume} {68}},\ \bibinfo {pages} {580} (\bibinfo {year} {1992})}\BibitemShut {NoStop}%
\bibitem [{\citenamefont {Dum}\ \emph {et~al.}(1992)\citenamefont {Dum}, \citenamefont {Zoller},\ and\ \citenamefont {Ritsch}}]{first-monte-2}%
  \BibitemOpen
  \bibfield  {author} {\bibinfo {author} {\bibfnamefont {R.}~\bibnamefont {Dum}}, \bibinfo {author} {\bibfnamefont {P.}~\bibnamefont {Zoller}}, \ and\ \bibinfo {author} {\bibfnamefont {H.}~\bibnamefont {Ritsch}},\ }\href {\doibase 10.1103/PhysRevA.45.4879} {\bibfield  {journal} {\bibinfo  {journal} {Phys. Rev. A}\ }\textbf {\bibinfo {volume} {45}},\ \bibinfo {pages} {4879} (\bibinfo {year} {1992})}\BibitemShut {NoStop}%
\bibitem [{\citenamefont {Bianchini}\ \emph {et~al.}(2015)\citenamefont {Bianchini}, \citenamefont {Castro-Alvaredo},\ and\ \citenamefont {Doyon}}]{non-unitary-CFT1}%
  \BibitemOpen
  \bibfield  {author} {\bibinfo {author} {\bibfnamefont {D.}~\bibnamefont {Bianchini}}, \bibinfo {author} {\bibfnamefont {O.~A.}\ \bibnamefont {Castro-Alvaredo}}, \ and\ \bibinfo {author} {\bibfnamefont {B.}~\bibnamefont {Doyon}},\ }\href {\doibase 10.1016/j.nuclphysb.2015.05.013} {\bibfield  {journal} {\bibinfo  {journal} {Nuclear Physics B}\ }\textbf {\bibinfo {volume} {896}},\ \bibinfo {pages} {835} (\bibinfo {year} {2015})}\BibitemShut {NoStop}%
\bibitem [{\citenamefont {Couvreur}\ \emph {et~al.}(2017)\citenamefont {Couvreur}, \citenamefont {Jacobsen},\ and\ \citenamefont {Saleur}}]{non-unitary-CFT2}%
  \BibitemOpen
  \bibfield  {author} {\bibinfo {author} {\bibfnamefont {R.}~\bibnamefont {Couvreur}}, \bibinfo {author} {\bibfnamefont {J.~L.}\ \bibnamefont {Jacobsen}}, \ and\ \bibinfo {author} {\bibfnamefont {H.}~\bibnamefont {Saleur}},\ }\href {\doibase 10.1103/PhysRevLett.119.040601} {\bibfield  {journal} {\bibinfo  {journal} {Phys. Rev. Lett.}\ }\textbf {\bibinfo {volume} {119}},\ \bibinfo {pages} {040601} (\bibinfo {year} {2017})}\BibitemShut {NoStop}%
\bibitem [{\citenamefont {Tu}\ \emph {et~al.}(2022)\citenamefont {Tu}, \citenamefont {Tzeng},\ and\ \citenamefont {Chang}}]{negative-c}%
  \BibitemOpen
  \bibfield  {author} {\bibinfo {author} {\bibfnamefont {Y.-T.}\ \bibnamefont {Tu}}, \bibinfo {author} {\bibfnamefont {Y.-C.}\ \bibnamefont {Tzeng}}, \ and\ \bibinfo {author} {\bibfnamefont {P.-Y.}\ \bibnamefont {Chang}},\ }\href {\doibase 10.21468/SciPostPhys.12.6.194} {\bibfield  {journal} {\bibinfo  {journal} {SciPost Phys.}\ }\textbf {\bibinfo {volume} {12}},\ \bibinfo {pages} {194} (\bibinfo {year} {2022})}\BibitemShut {NoStop}%
\bibitem [{\citenamefont {Agarwal}\ \emph {et~al.}(2023)\citenamefont {Agarwal}, \citenamefont {Konar}, \citenamefont {Lakkaraju},\ and\ \citenamefont {De}}]{EE-herm-like}%
  \BibitemOpen
  \bibfield  {author} {\bibinfo {author} {\bibfnamefont {K.~D.}\ \bibnamefont {Agarwal}}, \bibinfo {author} {\bibfnamefont {T.~K.}\ \bibnamefont {Konar}}, \bibinfo {author} {\bibfnamefont {L.~G.~C.}\ \bibnamefont {Lakkaraju}}, \ and\ \bibinfo {author} {\bibfnamefont {A.~S.}\ \bibnamefont {De}},\ }\href@noop {} {\enquote {\bibinfo {title} {Recognizing critical lines via entanglement in non-hermitian systems},}\ } (\bibinfo {year} {2023}),\ \Eprint {http://arxiv.org/abs/2305.08374} {arXiv:2305.08374 [quant-ph]} \BibitemShut {NoStop}%
\bibitem [{\citenamefont {Li}\ \emph {et~al.}(2023{\natexlab{b}})\citenamefont {Li}, \citenamefont {Yu},\ and\ \citenamefont {Li}}]{Skin-EE-transition2}%
  \BibitemOpen
  \bibfield  {author} {\bibinfo {author} {\bibfnamefont {S.-Z.}\ \bibnamefont {Li}}, \bibinfo {author} {\bibfnamefont {X.-J.}\ \bibnamefont {Yu}}, \ and\ \bibinfo {author} {\bibfnamefont {Z.}~\bibnamefont {Li}},\ }\href@noop {} {\enquote {\bibinfo {title} {Emergent entanglement phase transitions in non-hermitian aubry-andr\'e-harper chains},}\ } (\bibinfo {year} {2023}{\natexlab{b}}),\ \Eprint {http://arxiv.org/abs/2309.03546} {arXiv:2309.03546 [cond-mat.dis-nn]} \BibitemShut {NoStop}%
\bibitem [{\citenamefont {Alba}(2018)}]{qpp-inhomo}%
  \BibitemOpen
  \bibfield  {author} {\bibinfo {author} {\bibfnamefont {V.}~\bibnamefont {Alba}},\ }\href {\doibase 10.1103/PhysRevB.97.245135} {\bibfield  {journal} {\bibinfo  {journal} {Phys. Rev. B}\ }\textbf {\bibinfo {volume} {97}},\ \bibinfo {pages} {245135} (\bibinfo {year} {2018})}\BibitemShut {NoStop}%
\bibitem [{\citenamefont {Bertini}\ \emph {et~al.}(2018)\citenamefont {Bertini}, \citenamefont {Tartaglia},\ and\ \citenamefont {Calabrese}}]{no-pair}%
  \BibitemOpen
  \bibfield  {author} {\bibinfo {author} {\bibfnamefont {B.}~\bibnamefont {Bertini}}, \bibinfo {author} {\bibfnamefont {E.}~\bibnamefont {Tartaglia}}, \ and\ \bibinfo {author} {\bibfnamefont {P.}~\bibnamefont {Calabrese}},\ }\href {\doibase 10.1088/1742-5468/aac73f} {\bibfield  {journal} {\bibinfo  {journal} {Journal of Statistical Mechanics: Theory and Experiment}\ }\textbf {\bibinfo {volume} {2018}},\ \bibinfo {pages} {063104} (\bibinfo {year} {2018})}\BibitemShut {NoStop}%
\bibitem [{\citenamefont {Eichelkraut}\ \emph {et~al.}(2013)\citenamefont {Eichelkraut}, \citenamefont {Heilmann}, \citenamefont {Weimann}, \citenamefont {Stützer}, \citenamefont {Dreisow}, \citenamefont {Christodoulides}, \citenamefont {Nolte},\ and\ \citenamefont {Szameit}}]{SM-revise1}%
  \BibitemOpen
  \bibfield  {author} {\bibinfo {author} {\bibfnamefont {T.}~\bibnamefont {Eichelkraut}}, \bibinfo {author} {\bibfnamefont {R.}~\bibnamefont {Heilmann}}, \bibinfo {author} {\bibfnamefont {S.}~\bibnamefont {Weimann}}, \bibinfo {author} {\bibfnamefont {S.}~\bibnamefont {Stützer}}, \bibinfo {author} {\bibfnamefont {F.}~\bibnamefont {Dreisow}}, \bibinfo {author} {\bibfnamefont {D.~N.}\ \bibnamefont {Christodoulides}}, \bibinfo {author} {\bibfnamefont {S.}~\bibnamefont {Nolte}}, \ and\ \bibinfo {author} {\bibfnamefont {A.}~\bibnamefont {Szameit}},\ }\href {\doibase https://doi.org/10.1038/ncomms3533} {\bibfield  {journal} {\bibinfo  {journal} {Nature Communications}\ }\textbf {\bibinfo {volume} {4}},\ \bibinfo {pages} {2533} (\bibinfo {year} {2013})}\BibitemShut {NoStop}%
\bibitem [{\citenamefont {Longhi}(2019)}]{SM-revise2}%
  \BibitemOpen
  \bibfield  {author} {\bibinfo {author} {\bibfnamefont {S.}~\bibnamefont {Longhi}},\ }\href {\doibase 10.1103/PhysRevResearch.1.023013} {\bibfield  {journal} {\bibinfo  {journal} {Phys. Rev. Res.}\ }\textbf {\bibinfo {volume} {1}},\ \bibinfo {pages} {023013} (\bibinfo {year} {2019})}\BibitemShut {NoStop}%
\bibitem [{\citenamefont {Thouless}(1983)}]{thouless}%
  \BibitemOpen
  \bibfield  {author} {\bibinfo {author} {\bibfnamefont {D.~J.}\ \bibnamefont {Thouless}},\ }\href {\doibase 10.1103/PhysRevB.28.4272} {\bibfield  {journal} {\bibinfo  {journal} {Phys. Rev. B}\ }\textbf {\bibinfo {volume} {28}},\ \bibinfo {pages} {4272} (\bibinfo {year} {1983})}\BibitemShut {NoStop}%
\end{thebibliography}%

\onecolumngrid
\clearpage

{\centering
    \large{\textbf{{Supplemental Material}}}
\par}

\bigskip

\maketitle
\section*{Single-particle dynamics in the Hatano-Nelson model}
In this 
Supplemental Material,
we provide a brief overview of the properties of the Hatano-Nelson model and its single-particle dynamics.
The Hatano-Nelson model is a one-dimensional disordered tight-binding system with non-reciprocal hopping,\cite{HN_PRB97,HN_PRB98,HN_PRL} as defined by
\begin{eqnarray}
H&=&-\sum_{j=0}^{L-1}
\Big(
\Gamma_R
|j+1\rangle\langle j|
+\Gamma_L
|j\rangle\langle j+1|
\Big)
+
\sum_{j=0}^{L-1}
W_j
|j\rangle\langle j|,
\label{ham_sp}
\end{eqnarray}
where $|j\rangle$ represents a particle located at site $j$ and the notation is the same as Eq.~(1) in the main text.
The non-Hermiticity of this model is determined by the parameter $g$ in
\begin{equation}
\Gamma_L=e^g\Gamma_0,\ \
\Gamma_R=e^{-g}\Gamma_0.
\end{equation}
The static properties of this model ($g\neq0$) are essentially different from the Hermitian case ($g=0$).

\subsection{Static and dynamical properties (single-particle)}

In the Hermitian case: $\Gamma_R=\Gamma_L=\Gamma_0$ 
(i.e., in the case of symmetric hopping)
and under the periodic boundary,
the plane waves:
\begin{equation}
\label{pw}
|k\rangle=\sum_j e^{ikj} |j\rangle
\end{equation}
are eigenstates of the tight-binding model (\ref{ham_sp}) in the clean limit $W=0$,
and the corresponding eigenenergies are $-2\Gamma_0 \cos k$.
In the non-Hermitian case: $\Gamma_R\neq \Gamma_L$:
under the periodic boundary,
the plane waves  (\ref{pw}) are still eigenstates of the
Hamiltonian (\ref{ham_sp}) in the clean limit $W=0$,
but the asymmetry in hopping makes
the corresponding eigenenergy $\epsilon_k$ 
complex:
\begin{equation}
\label{spec}
\epsilon_k=-2\Gamma_0 (\cosh g \cos k + i \sinh g \sin k).
\end{equation}
Under the open boundaries,
the eigenstates of Eq.~(\ref{ham_sp}) are no longer the
simple sine function composed of the plane waves (\ref{pw})
even in the clean limit,\cite{PhysRevX_Gong}
but
becomes a skin-effect wave function
either exponentially damping or amplifying,
i.e., the system shows 
the so-called non-Hermitian skin effect in which
the eigen wave functions are localized
at the neighborhood of
either of the two open boundaries.
The corresponding eigenenergies are, on contrary,
real.
Thus, 
both under periodic and open boundaries,
a specific non-Hermitian feature
such as the complex spectrum or the skin effect
appears
either in the eigenvalues or
in the eigen wave functions.

\begin{figure*}

\includegraphics[width=170mm]{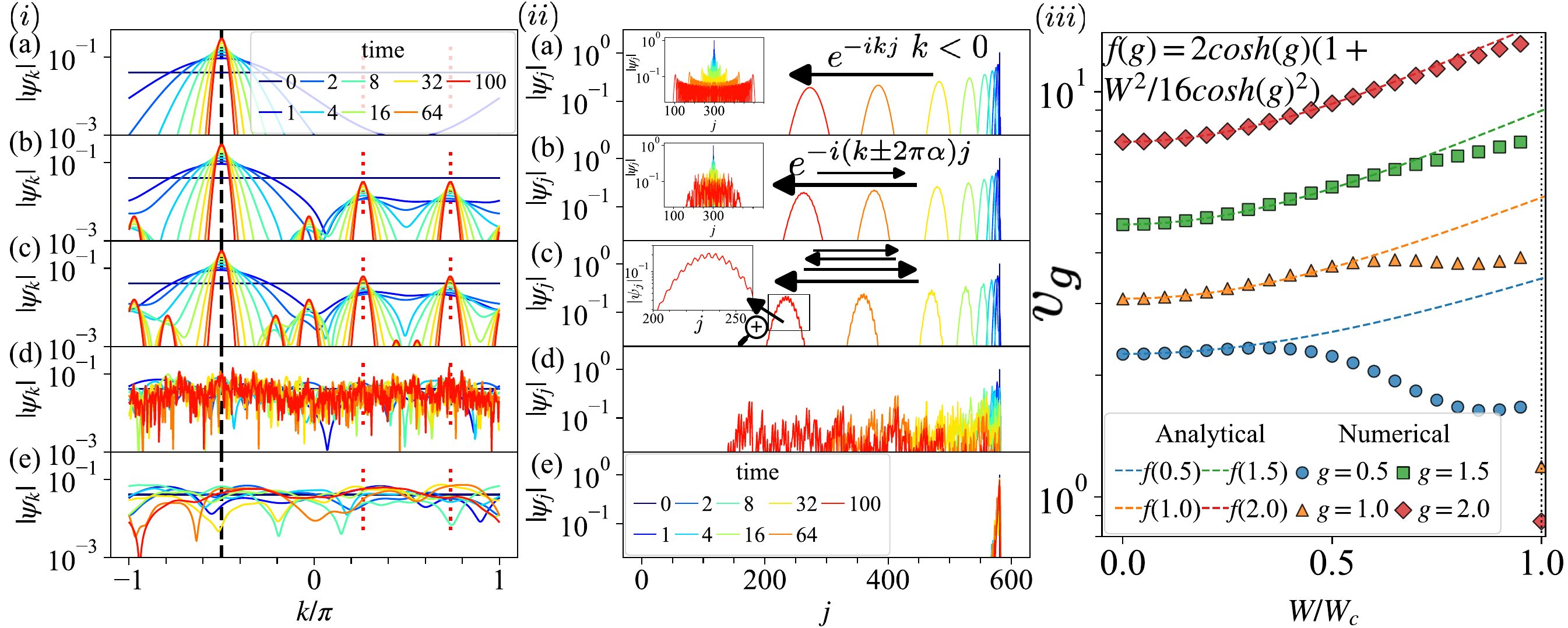}
\caption{Time evolution of wave-packets dynamics in the $k$ space: 
(i-a) $W=0$; (i-b), $W=1.0$; (i-c), $W=5.4$; (i-d), $W=7.0$; (i-e), $W=7.0$, in the real space:
(ii-a) $W=0$; (ii-b), $W=1.0$; (ii-c), $W=5.4$; (ii-d), $W=7.0$; (ii-e), $W=7.0$, for $g=1.0$ and its velocity: $v_g=\frac{\partial \langle x\rangle}{\partial t}$ as a function of $W/W_c$, where $W_c=2\exp(g)$, for various values $g$.
Black dashed lines in panel (i) represent $k=-\frac{\pi}{2}$, and red dashed lines in panel (i) represent $k=(-\frac{\pi}{2}\pm2\pi\alpha)\ {\rm mod}\ (2\pi)$.
These data are obtained with $j_0=581$, $L=601$ except for the inset panels.
The inset panels, including panels (ii-a) and (ii-b), correspond to the Hermitian case with the same disorder strength and different initial state $j_0=301$. 
}
\label{Fig14}
\end{figure*}

\subsection{Unidirectional motion and effect of disorder}
Let us choose, for simplicity, that
the initial wave packet $|\psi (0)\rangle$ is localized at a single site: $|\psi (0)\rangle=|j_0\rangle$.
Here, we use the notation that $|j\rangle$ represents a real space basis;
$|j\rangle$ represents a state localized at site $j$. 
At time $t$ this evolves as
\begin{equation}
\label{super}
|\psi (t)\rangle=\sum_k \psi_k e^{-i\epsilon_k t}|k\rangle,
\end{equation}
where $\psi_k=\langle k|j_0\rangle=e^{-ikj_0}$.
$|k\rangle$ represents a (crystal) momentum space ($k$-space) basis
or a plane wave eigenstate (\ref{pw}).
Here, we have the clean limit: $W=0$ in mind.
In the more generic case of $W\neq 0$,
\begin{equation}
|\psi (t)=\sum_{n} c_n e^{-i\epsilon_n t} |n\rangle,
\label{evo_sp}
\end{equation}
where
$|n\rangle$ represents the $n$th single-particle eigenstate of the Hamiltonian (\ref{ham_sp})
with an eigenenergy $\epsilon_n$; $H |n\rangle=\epsilon_n |n\rangle$,
while
$c_n =\langle\langle n|\psi (t=0)\rangle$.
Here, 
$\langle\langle n|$ represents the {\it left} eigenstate corresponding to the eigenenergy $\epsilon_n$:
$\langle\langle n|H=\epsilon_n\langle\langle n|$
and not $|n\rangle^\dagger$;
$\langle\langle n|\neq |n\rangle^\dagger$.
In Eq.~(\ref{super})
the initial state $|\psi (0)\rangle$
is expressed as a superposition of plane-wave eigenstates.
In the Hermitian case,
in the interference of such many plane waves,
those satisfying the stationary phase condition 
survive
and contribute to forming the shape of the wave front (see the insets of panels (ii-a) and (ii-b) of Fig.~\ref{Fig14}).
In the non-Hermitian case, on the other hand,
the imaginary part of the eigenenergy (\ref{spec}) plays instead a decisive role
in forming the shape of the wave front.
In case of $g\neq 0$,
the eigenenergy $\epsilon_n$ is typically complex, so that
the time-evolved wave packet 
$|\psi (t)\rangle$
literally as given in Eq.~(\ref{ham_sp})
tends to either decay or decay; its norm $\langle \psi (t)|\psi (t)\rangle$ is not conserved
due to the contribution from states with Im $\epsilon_n\neq0$. 
In the actual computation, 
we, therefore, rescale (renormalize) $|\psi (t)\rangle$.
We employed the Krylov subspace method with $\delta t=0.2$ and $M=15$ (see Eq.~(17) in the main text).
In the time evolution
a plane-wave eigenstate $|k\rangle$
with a maximal imaginary part Im$(\epsilon_k)$, i.e., $|k=-\pi/2\rangle$
in the superposition (\ref{super}).
Thus,
the initial wave packet $|\psi (0)\rangle$
formed as a superposition of many eigenstate:
$|\psi (0)\rangle=\sum_k \psi_k |k\rangle$
tends to evolve into a single eigenstate $|k=-\pi/2\rangle$.

Such an evolution can be indeed seen in the
numerical simulation of the density profile $\psi_k (t)$
in the crystal momentum space.
Different panels of
Fig.~\ref{Fig14} show such evolution 
at different strength of disorder $W$.
In the clean limit: $W=0$ [panel (i-a)]
and at weak disorder;
e.g., at $W=1.0$ [panel (i-b)],
one can see that
as time evolves
contribution from
the $k=-\pi/2$ component with a maximal Im$(\epsilon_k)$ [see Eq.~(\ref{spec})]
tends to become dominant.
In case of finite $W$ [panel (i-b) and (i-c)]
subdominant peaks associated with the quasi-periodic potential
appear.
The dominant and subdominant peaks disappear near the localization transition point $W\sim W_c$, and $|\psi_k|$ rapidly oscillates (panel (i-d)).
Once a quantum state localizes, initially given momentum distribution almost sustains (panel (i-e)).

As for the specific non-Hermitian characteristics in the real-space dynamics, the remarkable feature is the unidirectional motion of the wave packet, as shown in panel (ii-a) of Fig.~\ref{Fig14}. 
Once the quasi-periodic potential is introduced, the wave packet exhibits a weak modulation, reflecting the subdominant peak of $\psi_k$(panel (ii-b)).
As $W$ increases, this modulation develops into a more complex structure, reflecting the higher-order perturbation process (panel (ii-c)).
Near the localization-delocalization transition point $W\sim W_c$, a wave packet exhibits cascade-like spreading (panel (ii-d)).
Once $W$ exceeds the $W_c$, the quantum state localizes (panel (ii-e)). 
As observed, the wave packet spreading in the HN model is immune to disorder potential.
On the contrary, the velocity of wave packet $v_g=\frac{\partial \langle x\rangle}{\partial t}$ increases as $W$ increases, indicating that disorder enhances wave packet spreading.
Through the utilization of the second-order perturbation theory, we can elucidate this peculiar behavior (see the next subsection for detail).
Panel (iii) of Fig.~\ref{Fig14} shows $v_g$ as a function of $W/W_c$ with various values of $g$. For large $g$, analytical (perturbative) results exhibit good agreement with numerical results.

\subsection{Perturbative effect on non-reciprocal wave-packet dynamics}
\label{perturbation}
In the previous subsection,
we have examined the single particle dynamics of the HN model, where a wave packet exhibits unidirectional motion, and the velocity of unidirectional motion (sliding velocity) increases as $W$ increases for large $g$. 
To understand this peculiar behavior, we investigate the effect of the disorder using perturbation theory, considering the quasi-periodic potential as a perturbation term to free-particle dynamics.
Before discussing the effect of disorder, we initially provide an intuitive reason why the wave packet exhibits a unidirectional motion in the clean limit.
An essential factor is the role of Im$(E)$, which can either amplify or decay the corresponding eigenstate (plane wave) during dynamics.
Consequently, as time passes, the quantum state converges to the eigenstate whose Im$(E)$ is the maximal. In the presence case, Im$(E)$ becomes maximal at $k=-\frac{\pi}{2}$.
This tendency has been observed in panels (i) of Fig.~\ref{Fig14} and indicates that imaginary parts of eigenenergy act as a filter to determine to what extent the contribution of eigenstate to $|\Psi(t)\rangle$ remains after the time evolution, which is different from the Hermitian case, where interference plays one of the important roles in wave packet dynamics.
In fact, Refs.~\onlinecite{SM-revise1,SM-revise2} have focused on the role of the imaginary part of eigenenergy and obtained an analytical solution that agrees with numerical and experimental results.
That is, the wave packet dynamics of the non-Hermitian system are unique even in the clean limit, as the key factor that dominates the wave packet dynamics of the non-Hermitian system is different from the Hermitian system.
In our previous study,\cite{OI22A} we derived the trial function, which captures the unidirectional motion of the wave packet, as given by
\begin{eqnarray}
|\psi(t)\rangle&=&\sum_{j}|j\rangle\int^{2\pi}_{0} dk \frac{1}{\sqrt{2\pi}}e^{2i\cos(k-ig)t+ikj}\nonumber\\
&\simeq&\sum_{j}|j\rangle\int^{-\frac{\pi}{2}+\delta k}_{-\frac{\pi}{2}-\delta k} dk \frac{1}{\sqrt{2\pi}}e^{2i\cosh(g)(k+\frac{\pi}{2})t}\times e^{2\sinh(g)(1-\frac{1}{2}(k+\frac{\pi}{2})^2)t+i(k+\frac{\pi}{2})j}\nonumber\\
&\simeq&\sum_{j}|j\rangle\int^{\infty}_{-\infty} dk \frac{1}{\sqrt{2\pi}}e^{2i\cosh(g)kt}\times e^{2\sinh(g)(1-\frac{k^2}{2})t+ikj}\nonumber\\
&=&\sum_{j}|j\rangle \exp(-\frac{(j+2\cosh(g)t)^2}{4\sinh(g)t})\times e^{2\sinh(g)t}/\sqrt{4\sinh(g)t}.
\label{sup_k_int1}
\end{eqnarray}
Here, we impose the normalization condition on $|\Psi(t)\rangle$ as
\begin{eqnarray}
|\psi(t)\rangle
&=&\sum_{j}|j\rangle \frac{\exp(-\frac{(j+2\cosh(g)t)^2}{4\sinh(g)t})}{(2\pi\sinh(g)t)^{\frac{1}{4}}},
\label{trial-function}
\end{eqnarray}
where $2\cosh(g)={\rm Re}(\frac{\partial\epsilon_k}{\partial k})|_{k=-\frac{\pi}{2}}$, $2\sinh(g)={\rm Im}(\epsilon_k)|_{k=-\frac{\pi}{2}}$, and $j_0=0$.
Using these definitions, we obtained analytical expressions of the sliding velocity $v_g=\frac{\partial \langle x\rangle}{\partial t}$ and mean square displacement $\sigma=\langle x^2\rangle-\langle x\rangle^2$.
Next, we introduce the quasi-periodic disorder and employ a second-order perturbation theory.
Since the eigenstates are plane waves, we transform the quasi-periodic potential into momentum space and obtain the perturbation terms given by
\begin{eqnarray}
&&\sum_{j,k,k'} |k\rangle\langle k|W\cos(2\pi\alpha j )|j\rangle\langle j||k'\rangle\langle k'|=\sum_{k,\beta=\pm2\pi\alpha} \frac{W}{2}|k\rangle\langle k+\beta|+h.c..
\label{pertur}
\end{eqnarray}
In this calculation, we approximate the irrational number $\alpha$ as the rational number.\cite{Longhi_AA,thouless}
Indeed, $\alpha$ is defined as the limit of the ratio of consecutive Fibonacci numbers: $\alpha=\lim_{n\to\infty} \frac{f_n}{f_{n+1}}$,
where $f_{n+1}=f_{n}+f_{n-1}$ and $f_0=f_1=1$. 
We obtain perturbation energy and states as follows:
\begin{eqnarray}
&&E_n^1=\langle\psi_n^0|(\sum_{k,\beta=\pm2\pi\alpha} \frac{W}{2}|k\rangle\langle k+\beta|+h.c.)|\psi_n^0\rangle=0\nonumber,\\
&&|\psi_n^1\rangle=\frac{W}{2}(\frac{|\psi_{n+2\pi\alpha}^0\rangle}{E_n^0-E_{n+2\pi\alpha}^0}+\frac{|\psi_{n-2\pi\alpha}^0\rangle}{E_n^0-E_{n-2\pi\alpha}^0})\nonumber,\\
&&E_n^2=\frac{W^2}{4}(\frac{1}{E_n^0-E_{n+2\pi\alpha}^0}+\frac{1}{E_n^0-E_{n-2\pi\alpha}^0}),
\label{pertubation_enevec}
\end{eqnarray}
where the superscript and subscript represent $n$-th order of perturbation and eigenstate (or eigenenergy), respectively.
For large $g$, $(E_n- E_{n\pm2\pi\alpha})^{-1}$ is approximated by
\begin{eqnarray}
&&(E_n^0-E_{n\pm2\pi\alpha}^0)^{-1}=(2\cosh(g)(\cos(k_n\pm2\pi\alpha)-\cos(k_n))+2i\sinh(g)(\sin(k_n\pm2\pi\alpha)-\sin(k_n)))^{-1}\nonumber\\
&&=(-4\cosh(g)\sin(k_n\pm\pi\alpha)\sin(\pm\pi\alpha)-4i\sinh(g)\cos(k_n\pm\pi\alpha)\sin(\pm\pi\alpha))^{-1}\nonumber\\
&&=\frac{(-4\sin(\pm\pi\alpha))^{-1}}{\cosh(g)\sin(k_n\pm\pi\alpha)+i\sinh(g)\cos(k_n\pm\pi\alpha)}\nonumber\\
&&=\frac{-1}{4\sin(\pi\alpha)}\times\frac{\cosh(g)\sin(k_n\pm\pi\alpha)-i\sinh(g)\cos(k_n\pm\pi\alpha)}{\cosh^2(g)\sin^2(k_n\pm\pi\alpha)+\sinh^2(g)\cos^2(k_n\pm\pi\alpha)}\nonumber\\
&&=-\frac{\cosh(g)\sin(k_n\pm\pi\alpha)-i\sinh(g)\cos(k_n\pm\pi\alpha)}{4\sin(\pm\pi\alpha)(\cosh^2(g)-\cos^2(k_n\pm\pi\alpha))}\nonumber\\
&&\sim-\frac{\cosh(g)\sin(k_n\pm\pi\alpha)-i\sinh(g)\cos(k_n\pm\pi\alpha)}{4\sin(\pm\pi\alpha)\cosh^2(g)},
\label{pertubation_ene_c}
\end{eqnarray}
and thus nth-eigenenergy becomes
\begin{eqnarray}
E_n&\sim& E_n^0+E_n^1+E_n^2
\sim 2\cosh(g)(1+\frac{W^2}{16\cosh^2(g)})\cos(k_n)+2i\sinh(g)(1-\frac{W^2}{16\sinh^2(g)})\sin(k_n).
\label{pertubation_ene}
\end{eqnarray}
We note that the left eigenvector is the transpose of the right eigenvector multiplied by a constant factor: $\langle\langle n^L|\propto |n^R\rangle^T\equiv(|\psi^0_n\rangle+|\psi^1_n\rangle)^T$.
Additionally, the normalization condition $\langle\langle n^L|n^R\rangle=1$ is almost satisfied for large $g$.
Using eigenenergies and eigenvectors, we can derive $|\Psi(t)\rangle$, which is given by
\begin{eqnarray}
|\Psi(t)\rangle&&=\sum_ne^{-iE_nt}|n^R\rangle\langle\langle n^L|\Psi(0)\rangle\nonumber\\
&&\sim\sum_ne^{-i(E_n^0+E_n^2)t}(
(|\psi_n^0\rangle+\sum_{\beta=\pm2\pi\alpha}\frac{W}{2(E_n^0-E_{n+\beta}^0)}|\psi_{n+\beta}^0\rangle)(\langle\langle\psi_n^0|+\sum_{\beta=\pm2\pi\alpha}\frac{W}{2(E_n^0-E_{n+\beta}^0)}
\langle\langle\psi_{n+\beta}^0|))|\Psi(0)\rangle\nonumber\\
&&=\frac{1}{\sqrt{L}}\sum_{n,j}|j\rangle\langle j|e^{-i(E_n^0+E_n^2)t}(
(|\psi_n^0\rangle+\sum_{\beta=\pm2\pi\alpha}\frac{W}{2(E_n^0-E_{n+\beta}^0)}|\psi_{n+\beta}^0\rangle)(1+\sum_{\beta=\pm2\pi\alpha}\frac{W}{2(E_n^0-E_{n+\beta}^0)}))\nonumber\\
&&=\frac{1}{L}\sum_{n,j}|j\rangle e^{-i(E_n^0+E_n^2)t}
e^{ik_nj}(1+\sum_{\beta=\pm2\pi\alpha}\frac{W}{2(E_n^0-E_{n+\beta}^0)}e^{ik_{\beta}j})(1+\sum_{\beta=\pm2\pi\alpha}\frac{W}{2(E_n^0-E_{n+\beta}^0)}
))\nonumber\\
&&=\frac{1}{L}\sum_{n,j}|j\rangle e^{-i(E_n^0+E_n^2)t}
e^{ik_nj}(1+\sum_{\beta=\pm2\pi\alpha}\frac{W}{2(E_n^0-E_{n+\beta}^0)}e^{ik_{\beta}j}+\sum_{\beta=\pm2\pi\alpha}\frac{W}{2(E_n^0-E_{n+\beta}^0)})
))+\mathcal{O}(\cosh^{-2}(g))\nonumber\\
&&\sim\frac{\exp(-\frac{(j+2\cosh(g)(1+\frac{W^2}{16\cosh^2(g)})t)^2}{4\sinh(g)(1-\frac{W^2}{16\sinh^2(g)})t})}{(2\pi\sinh(g)(1-\frac{W^2}{16\sinh^2(g)})t)^{\frac{1}{4}}}\times
(1+\sum_{\beta=\pm2\pi\alpha}\frac{We^{ik_{\beta}j}}{2(E_n^0-E_{n+\beta}^0)}
|_{k_n=\frac{-\pi}{2}}+\sum_{\beta=\pm2\pi\alpha}\frac{W}{2(E_n^0-E_{n+\beta}^0)}|_{k_n=\frac{-\pi}{2}})\nonumber\\
&&=\frac{\exp(-\frac{(j+2\cosh(g)(1+\frac{W^2}{16\cosh^2(g)})t)^2}{4\sinh(g)(1-\frac{W^2}{16\sinh^2(g)})t})}{(2\pi\sinh(g)(1-\frac{W^2}{16\sinh^2(g)})t)^{\frac{1}{4}}}\times
(1+i\frac{\sinh(g)\sin(\pi\alpha)(1+\cos(2\pi\alpha j))+\cosh(g)\cos(\pi\alpha)\sin(2\pi\alpha j)}{4\sin(\pi\alpha)\cosh(g)^2} ).\nonumber\\
\label{time-psit}
\end{eqnarray}
We can extract a distinctive characteristic of $|\Psi(t)\rangle$ from Eq.~(\ref{time-psit}).
As the second term of Eq.~(\ref{time-psit}) in the last line consists solely of imaginary numbers, its contribution to $\langle x \rangle$ and $\langle x^2 \rangle$ become $\mathcal{O}(\cosh^{-2}(g))$ and can be considered negligible.
Therefore, we can obtain the perturbative solution by adding $\epsilon_n^2$ to $\epsilon_n^0$.
The perturbative effect of the eigenenergy on the wave packet dynamics differs from that of the eigenstate.
The perturbative correction in the eigenenergy leads to an increase in $v_g$ as a function of $W$, while that in the eigenstate is irrelevant for dynamics since non-interference occurs, resulting in unidirectional motion immune to the disorder potential.
Once interference becomes non-negligible, a cascade-like spreading emerges.

\end{document}